# Factorisation of 3d $\mathcal{N} = 4$ twisted indices and the geometry of vortex moduli space

**Samuel Crew, Nick Dorey and Daniel Zhang**

*Department of Applied Mathematics and Theoretical Physics, University of Cambridge, Wilberforce Road, Cambridge, CB3 0WA, U.K.*

*E-mail:* s.c.crew@damtp.cam.ac.uk, n.dorey@damtp.cam.ac.uk, d.zhang@damtp.cam.ac.uk

Abstract: We study the twisted indices of $\mathcal{N} = 4$ supersymmetric gauge theories in three dimensions on spatial $S^2$ with an angular momentum refinement. We demonstrate factorisation of the index into holomorphic blocks for the $T[\mathrm{SU}(N)]$ theory in the presence of generic fluxes and fugacities. We also investigate the relation between the twisted index, Hilbert series and the moduli space of vortices. In particular, we show that each holomorphic block coincides with a generating function for the $\chi_t$ genera of the moduli spaces of "local" vortices. The twisted index itself coincides with a corresponding generating function for the $\chi_t$ genera of moduli spaces of "global" vortices in agreement with a proposal of Bullimore et al. We generalise this geometric interpretation of the twisted index to include fluxes and Chern-Simons levels. For the $T[\mathrm{SU}(N)]$ theory, the relevant moduli spaces are the local and global versions of Laumon space respectively and we demonstrate the proposed agreements explicitly using results from the mathematical literature. Finally, we exhibit a precise relation between the Coulomb branch Hilbert series and the Poincaré polynomials of the corresponding vortex moduli spaces.

## Contents

# 1 Introduction and summary

Supersymmetric indices are powerful tools for gaining non-perturbative information on supersymmetric quantum field theories. In particular, topologically twisted indices [1] of three-dimensional theories with $\mathcal{N} \geq 2$ supersymmetry have interesting applications which include microstate counting for black holes in $AdS_4$ [2], the topology of 3-manifolds and the representation theory of chiral algebras [3, 4]. These indices have remarkable properties including holomorphic factorisation, modular invariance, wall-crossing and large-rank saddle points. Protected indices in supersymmetric QFT often have an interpretation in terms of the geometry of an appropriate moduli space of classical solutions. In recent work, Bullimore et al. [5, 6], have suggested such an interpretation for the twisted index. In this paper we will confirm this proposal and extend it in several ways.

We will focus on theories with $\mathcal{N} = 4$ supersymmetry in three-dimensions. These theories have a moduli space of vacua with Higgs and Coulomb branches characterised by distinct SU(2) R-symmetries. Correspondingly one can define two partial twists on $S^2$, denoted $A$ and $B$ in [7], which mix the U(1) of spatial rotation with the Cartan subgroup of the Higgs and Coulomb branch R-symmetry. Theories may come in pairs related by *mirror symmetry* which interchanges Higgs and Coulomb branches and thus interchanges the $A$- and $B$-twisted indices. As we review below, these indices naturally depend on fugacities for the conserved charges corresponding to global and topological symmetries as well as on background magnetic fluxes associated to these symmetries. An index for such a theory defined on spatial $S^2$ may also be refined further by introducing a fugacity for angular momentum. In the following we will study the properties of the $A$- and $B$-twisted $S^2$ indices retaining the full dependence on these parameters. Building on earlier work [8], we exhibit the factorisation of the topological index into holomorphic blocks, in this general case, for $T[\mathrm{SU}(N)]$ and $\mathrm{SQCD}[k, N]$ theories.

We also investigate the geometrical interpretation of the $\mathcal{N} = 4$ twisted index and its blocks. The blocks have an expansion in integer powers of the fugacity for topological charges corresponding to the contributions of BPS vortices in the three-dimensional gauge theory. Hence we expect to reproduce these contributions from an appropriate computation in supersymmetric quantum mechanics on the vortex moduli space. As each block corresponds to a particular vacuum state of the three-dimensional theory and provides a boundary condition at infinity for the vortex solution, we refer to these configurations as "local" vortices. As we review below, local vortices have an equivalent IR description as *based quasi-maps* from $\mathbb{P}^1$ to the Higgs branch of the gauge theory. We find that the vortex contribution to the block coincides with a particular $\chi_t$ genus of the moduli space. For theories of type $T_\rho[\mathrm{SU}(N)]$, the moduli space of local vortices can be realised explicitly as a Handsaw quiver variety. These spaces are Kähler cones and admit a $\mathfrak{u}(1,1|2)$ superconformal quantum mechanics of the type recently studied by two of the authors in [9], with a superconformal index which coincides with the equivariant $\chi_t$ genus[1] of (a smooth resolution of) the cone. Thus the vortex contribution to the block coincides with the superconformal index of quantum mechanics on the moduli space of local vortices.

[1]We recap the definition of the $\chi_t$ genus in section 4.1. The parameter $t$ which counts cohomological degree in the $\chi_t$ genus corresponds to a fugacity for an $R$-symmetry of the $\mathcal{N} = 4$ theory.

As mentioned above, Bullimore et al. [5, 6] have proposed a geometrical interpretation of the twisted index of $\mathcal{N}=4$ theories in terms of $\chi_t$ genera of vortex moduli spaces. In this case the relevant configurations are "global" vortices corresponding in the IR to unbased quasi-maps from $\mathbb{P}^1$ to the Higgs branch. We confirm this interpretation in the case of the $T[\mathrm{SU}(N)]$ theory and show that it is equivalent to our result for the blocks in terms of the contributions of "local" vortices as described above. In this case, the relevant moduli spaces are the global and local versions of Laumon space and we find that their $\chi_t$ genera are related via equivariant localisation in a way which is precisely consistent with the factorisation (1.7) of the twisted index into products of holomorphic blocks.

Another remarkable feature of the $A$- and $B$- twisted indices defined in [7] is that, in the absence of flux, they reduce to the Hilbert series counting holomorphic functions on the Coulomb and Higgs branches respectively.

Combining this insight with our results above, we provide a new formula for the Coulomb branch Hilbert series in terms of the Poincaré polynomials of Laumon spaces. This is consistent with the philosophy of Nakajima's mathematical construction of the Coulomb branch algebra [10, 11] in terms of the homology of a quasi-map space. We also perform some new tests of mirror symmetry in the presence of fluxes. The main results of the paper are described in the remainder of this introductory section.

**The A and B twisted index.** Three dimensional $\mathcal{N}=4$ theories, $T$, have an $\mathrm{SU}(2)_H\times\mathrm{SU}(2)_C$ R-symmetry with the two SU(2) factors acting non-trivially on the Higgs and Coulomb branches respectively, denoted $\mathcal{M}_H(T)$ and $\mathcal{M}_C(T)$. The $\mathcal{N}=4$ twisted index is a special case of the $\mathcal{N}=2$ twisted index, defined in [1], selected by picking a $\mathcal{N}=2$ subalgebra with abelian R-symmetry corresponding to either $\mathrm{U}(1)_{\mathcal{R}}=2\mathrm{U}(1)_H$ (A twist) or $\mathrm{U}(1)_{\mathcal{R}}=2\mathrm{U}(1)_C$ (B twist). Working on spatial $S^2$, a background gauge field corresponding to a unit of flux for the distinguished R symmetry is turned on to cancel the spin connection in the Killing spinor equations. From the point of view of $\mathcal{N}=2$ supersymmetry, the "other" $\mathcal{N}=4$ R-symmetry is a distinguished global symmetry. In particular we define the combination $\mathrm{U}(1)_t=2[\mathrm{U}(1)_H-\mathrm{U}(1)_C]$ and a corresponding fugacity $t$. The theory may have additional global symmetries $\{Q_i\}$, with fugacities $\{y_i\}$, which act non-trivially on the Higgs branch as well as topological symmetries $\{\tilde{Q}_a\}$, with fugacities $\{\xi_a\}$, which act on the Coulomb branch. We may also introduce background fluxes on $S^2$, denoted $\{\mathfrak{n}_i\}$ and $\{\tilde{\mathfrak{n}}_a\}$ respectively. Finally we also introduce a fugacity $q$ for angular momentum $J$ on $S^2$. Defining a Hilbert space $\mathcal{H}[X;\mathfrak{n},\tilde{\mathfrak{n}}]$ for the $\mathcal{N}=4$ theory $T$ on $S^2$ with $X=A$ or $B$ labelling the two choices of twist, the two twisted indices we will consider in this paper correspond to,

$$Z^X[T]\,(t,q,y,\xi,\mathfrak{n},\tilde{\mathfrak{n}})=\mathrm{Tr}_{\mathcal{H}[X;\mathfrak{n},\tilde{\mathfrak{n}}]}\left((-1)^F\,q^J t^{Q_t}\prod_i y_i^{Q_i}\prod_a \xi_a^{\tilde{Q}_a}\right) \tag{1.1}$$

with $X=A$ or $X=B$.

As usual, we may also compute the index by computing an appropriate Euclidean path integral. The $A$ and $B$ twists are implemented by introducing a background gauge field for the corresponding $R$ symmetry. Introducing the fugacity $q = e^{i\varsigma/2}$ for angular momentum is accomplished by placing the theory on an $\Omega$-background corresponding to the space-time $S^2 \times_q S^1$ with metric,

$$ds^2 = R^2(d\theta^2 + \sin^2\theta\,(d\phi - \varsigma dt)^2) + \beta dt^2 \tag{1.2}$$

In the Lagrangian formulation, fugacities for global symmetries correspond to mass terms while those for topological symmetries are similarly related to Fayet-Illiopoulos (F.I.) parameters.

Localising the path integral on the Coulomb branch [1], the twisted indices can be computed via residues of a certain meromorphic form, $Z_{\mathfrak{m}}$, which is determined by the field content of the theory:

$$Z^X_{S^2\times_q S^1}(t,q,y,\xi,\mathfrak{n},\tilde{\mathfrak{n}}) = \frac{1}{|W|}\sum_{\mathfrak{m}}\oint_{\text{JK}}\frac{dx}{2\pi i x} Z^X_{\mathfrak{m}}(x,q,t,y,\xi,\mathfrak{n},\tilde{\mathfrak{n}})\,, \tag{1.3}$$

here $x = e^{iu}$ where $u = A_t + i\beta\sigma$ is the complexified holonomy of the gauge field. $A_t$ is the holonomy of the gauge field around $S^1$, $\sigma$ the adjoint-valued scalar living in the $\mathcal{N} = 2$ vector multiplet and $\beta$ is the $S^1$ radius. The sum is over magnetic flux sectors of the gauge field, where $\mathfrak{m}$ lies in the co-root lattice of the gauge group $G$. The integral is divided by the Weyl group of the gauge group, to account for gauge-equivalent configurations. The JK subscript refers to the Jeffrey-Kirwan residue procedure [12], which prescribes a contour enclosing a particular set of poles specified by the data of the theory. We compute these residue integrals for the $T[\mathrm{SU}(N)]$ and $\mathrm{SQCD}[k,N]$ theories (which are defined in the next paragraph) in section 2, and describe the JK residue procedure therein.

In this work we will perform explicit calculations for theories of type $T_\rho[\mathrm{SU}(N)]$. These are $\mathcal{N} = 4$ theories described by the quiver diagram in figure 1. They are specified by a partition:

$$\rho = (\rho_1, \rho_2, \ldots, \rho_L) \equiv (N_1, N_2 - N_1, \ldots, N - N_{L-1}) \tag{1.4}$$

where $\rho_s \geq \rho_{s+1}$. This is a 'good' theory in the terminology of [13] and is expected to flow in the IR to an interacting CFT when all dimensionful parameters are set to zero. These theories have product gauge group $G = \mathrm{U}(N_1) \times \ldots \times \mathrm{U}(N_{L-1})$, with a ($\mathcal{N} = 4$) vector multiplet for each gauge node, a hypermultiplet in the bifundamental of $\mathrm{U}(N_s) \times \mathrm{U}(N_{s+1})$ for each $s = 1, \ldots, L-2$, and $N$ hypermultiplets in the fundamental of $\mathrm{U}(N_{L-1})$. The Higgs branch of the theory, $\mathcal{M}_H$, is $T^*\mathcal{B}(N_1, \ldots, N_L)$ where $\mathcal{B}(N_1, \ldots, N_L)$ is the space of flags of signature $(N_1, \ldots, N_{L-1})$ in $\mathbb{C}^N$. When $\rho$ is the trivial partition, $\rho = (1, \ldots, 1)$, the theory is denoted $T[\mathrm{SU}(N)]$ and we use the shorthand $\mathcal{M}_H = T^*\mathcal{B}_N$ for the Higgs branch. The $T[\mathrm{SU}(N)]$ theory is self mirror dual and thus the Coulomb branch is isomorphic to the Higgs branch. The other main example studied in this work is the $\mathrm{SQCD}[k,N]$ theory. This is a $T_\rho[\mathrm{SU}(N)]$ theory with $\rho = (k, N-k)$ and the Higgs branch is the cotangent bundle to the Grassmanian: $\mathcal{M}_H = T^*Gr(k,N)$.

**3d mirror symmetry and Hilbert series.** 3d $\mathcal{N} = 4$ theories enjoy a mirror duality, which is an IR duality of two gauge theories [14], composed with the mirror automorphism of the $\mathcal{N} = 4$ superalgebra, given by exchanging $\mathrm{SU}(2)_H$ and $\mathrm{SU}(2)_C$. In our context the duality acts by interchanging the A and the B twisted index for a pair of mirror dual theories. If $T$ and $\check{T}$ are mirror dual then the expected duality is:

$$Z^A_{S^2\times_q S^1}[T]\,(t,q,\{y,\xi\},\{\mathfrak{n},\tilde{\mathfrak{n}}\}) \longleftrightarrow Z^B_{S^2\times_q S^1}[\check{T}]\,\left(t^{-1},q,\{\xi,y\},\{\tilde{\mathfrak{n}},\mathfrak{n}\}\right) \tag{1.5}$$

where fugacities and fluxes for the hypermultiplet flavour symmetry and topological symmetry are exchanged.

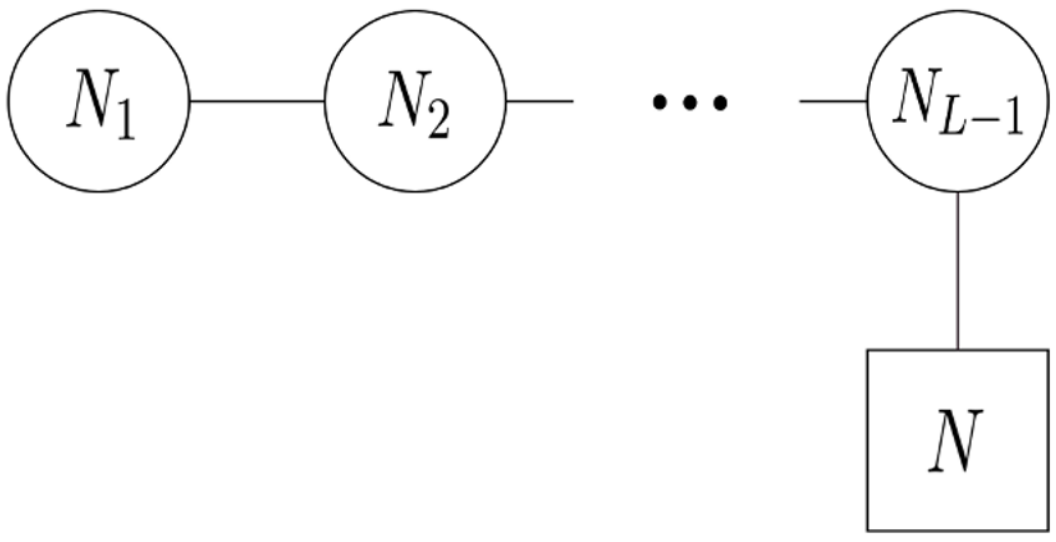

**Figure 1**. The quiver diagram for the $T_\rho[\mathrm{SU}(N)]$ quiver theory.

We explicitly demonstrate mirror symmetry in the presence of generic fluxes for the angular momentum refined twisted index of the Abelian $T[\mathrm{SU}(2)]$ and SQED[1] theories in section 3. Although a full check is much harder for higher rank gauge groups, we also demonstrate mirror symmetry for the $T[\mathrm{SU}(N)]$ theory in the $t \to 1$ limit.

In the absence of fluxes and angular momentum refinement,[2] it is argued in [7] that the A and B twisted index coincide with the Hilbert series of the Coulomb and the Higgs branch respectively. The Higgs branch Hilbert series can be computed using the standard Molien integral techniques [15] while the Coulomb branch Hilbert series is given by the formula of Hanany et al. [16] which counts the monopole operators. The agreement of the Higgs and Coulomb branch expressions for mirror pairs has been checked in many cases. In fact, it is easy to see that the contour integral representation of the B twist under certain assumptions coincides with the Molien integral representation of the Coulomb branch Hilbert series[3] implied by mirror symmetry. In section 5, we discuss how these arguments are modified in the presence of background fluxes and the angular momentum refinement. We use 3d mirror symmetry to give a mirror dual interpretation of the Coulomb branch Hilbert series with background flux in terms of the geometry of quasi-maps to the Higgs branch.

**Holomorphic factorisation.** First observed in two dimensions by [17, 18] and further developed in three dimensions [19–21], the principle of holomorphic factorisation states that 3d supersymmetric indices on different background can be decomposed into the same set of fundamental holomorphic blocks, which are partition functions of the theory on the disk $D_2 \times S^1$. The different blocks are labelled by the corresponding boundary conditions for the fields on the $S^1$ boundary of the disk. These in turn correspond to different massive vacua of the 3d theory. Partition functions on $S_b^3$ (the squashed ellipsoid) and $S^2 \times S^1$ (superconformal index) can then be written as sums over products of the same blocks:

$$\begin{aligned} Z_{S^2\times S^1} &\sim \sum_\alpha \mathbb{B}_\alpha(\xi, y; t, q)\mathbb{B}_\alpha(\bar{\xi}, \bar{y}; \bar{t}, \bar{q}) \\ Z_{S_b^3} &\sim \sum_\alpha \mathbb{B}_\alpha(\xi, y; t, q)\mathbb{B}_\alpha(\tilde{\xi}, \tilde{y}; \tilde{t}, \tilde{q}) \end{aligned} \tag{1.6}$$

[2] We show in section 6 that in the absence of background flux the index is in fact independent of $q$, the angular momentum parameter.

[3] In other words, the Hilbert series of the Higgs branch of the mirror theory.

Here the index $\alpha$ labels the massive vacua of the theory and the relevant conjugation of the variables between the two blocks depends on the particular background.

For 3d $\mathcal{N} = 4$ theories in the presence of background topological, $\tilde{\mathfrak{n}}$, and flavour fluxes, $\mathfrak{n}$, the A and B-twisted indices are expected to have a similar factorisation property. As anticipated in [8], we find the following general factorisation:[4]

$$Z^A_{S^2\times_q S^1} \sim \sum_\alpha \mathbb{B}_\alpha(\xi q^{-\tilde{\mathfrak{n}}}, y q^{-\mathfrak{n}}; q, t)\mathbb{B}_\alpha(\xi q^{\tilde{\mathfrak{n}}}, y q^{\mathfrak{n}}; q^{-1}, t) \tag{1.7}$$

where the blocks are the same as those used in the computation of the supersymmetric index and $\alpha$ labels the massive vacua of the theory as before. In section 2 we demonstrate this factorisation explicitly for the $T[\mathrm{SU}(N)]$ and SQCD$[k, N]$ theories. One may also show that this factorisation reproduces the Bethe ansatz formula [23] for the un-refined twisted index.

The blocks have an expansion in powers of the fugacities for topological charge whose coefficients are naturally interpreted as the contribution of BPS vortices. In section 4.2 we show that the holomorphic block of the $T_\rho[\mathrm{SU}(N)]$ theory, or more accurately the vortex partition function $\mathcal{Z}_V$, which is the non-perturbative contribution to the block, can be interpreted as the generating function of superconformal indices $Z_{\text{S.C.}}$ of $\mathfrak{u}(1,1|2)$ quantum mechanics [9] on the vortex moduli spaces of the $T_\rho[\mathrm{SU}(N)]$ theory. These moduli spaces are the handsaw quiver varieties of Nakajima [24] or equivalently local Laumon spaces $\mathfrak{Q}^\rho_{\vec{d}}$ (defined below), and the superconformal index of the Kähler cone coincides with equivariant Hirzebruch $\chi_t$ genera of the resolution.

$$\begin{aligned} \mathcal{Z}_V[T_\rho[\mathrm{SU}(N)]](q, t, \{y\}, \{\xi\}) &= \sum_{\vec{d}} \left[\prod_{s=1}^{N-1} \left(\frac{\xi_s}{\xi_{s+1}}\right)^{d_s}\right] Z_{\text{S.C.}}(\mathfrak{Q}^\rho_{\vec{d}}, q, t, \{y\}) \\ &= \sum_{\vec{d}} \left[\prod_{s=1}^{N-1} \left(\left(-t^{-\frac{1}{2}}\right)^{(\rho_s+\rho_{s+1})} \frac{\xi_s}{\xi_{s+1}}\right)^{d_s}\right] \chi_t(\mathfrak{Q}^\rho_{\vec{d}}, q, \{y\}) \end{aligned} \tag{1.8}$$

**Laumon space and geometry of the index.** Global Laumon space $\mathcal{Q}_{\vec{d}}$ is a smooth compactification of the space of algebraic maps of degree $\vec{d}$ from $\mathbb{P}^1$ to $\mathcal{B}_N$. $\mathcal{B}_N$ is equivalently the Lagrangian core[5] of the Higgs branch of $T[\mathrm{SU}(N)]$. We also consider the local version of Laumon space $\mathfrak{Q}_{\vec{d}}$ whose definition involves the extra condition that the maps are based (i.e. a marked point on $\mathbb{P}^1$ is mapped to the trivial flag). These coincide with the handsaw quiver varieties of Nakajima [24]. The authors of [25] calculate the equivariant $\chi_t$ genus of $\mathcal{Q}_{\vec{d}}$ using equivariant K-theoretic localisation and find that it factorises in terms of the $\chi_t$ genera of the local Laumon spaces. Turning on the angular momentum refinement $q$ localises the computation to isolated fixed points on the moduli space. Intuitively, the localisation fixes first to the north and south poles of the base $\mathbb{P}^1$ and then to the $S_N$ isolated fixed points on the target space $\mathcal{B}_N$, schematically:

$$\sum_{\alpha} \chi_t(\mathcal{Q}_\alpha) \sim \sum_{S_N} \sum_{\beta,\gamma} \chi_t(\mathfrak{Q}_\beta; q, t) \chi_t(\mathfrak{Q}_\gamma; q^{-1}, t) \tag{1.9}$$

[4] Recent work [22] finds a similar gravity dual factorisation including background fluxes.

[5] For $T_\rho[\mathrm{SU}(N)]$ theories this coincides with the fixed point sub-variety under the group action associated with the fugacity $t$.

The starting point for this paper is the observation that the above formula closely resembles the holomorphic factorisation of the twisted index of $T[\mathrm{SU}(N)]$ into blocks (1.7).

In section 5 we re-interpret the holomorphic block as a generating function of equivariant $\chi_t$ genera of, non-compact, handsaw quiver varieties [24] (or local Laumon space), and describe in detail this correspondence between the physical factorisation and the geometry of global and local Laumon spaces.

We introduce a slight generalisation of global Laumon space in section 4.1 and extend the geometric set up of [25] to understand examples of theories of $T_\rho[\mathrm{SU}(N)]$ type with Chern-Simons levels and background flavour and topological fluxes turned on. The dictionary for these additions is extended as follows:

| | |
|---|---|
| Background topological flux $\tilde{\mathfrak{n}}$ | $\tilde{\mathfrak{n}}^{\text{th}}$ power of line bundle on $\mathcal{Q}_\alpha$<br>Shifting $\xi \to \xi q^{\tilde{\mathfrak{n}}}$ of generating variable |
| Background flavour flux $\mathfrak{n}$ | Shifting $y \to y q^{\mathfrak{n}}$ in the local genera |
| Chern-Simons level $\kappa$ | $\kappa^{\text{th}}$ power of determinant line bundle on $\mathfrak{Q}_\alpha$ |

In section 6 we set the background fluxes to zero. In this case the A-twisted index is expected to coincide with the Coulomb branch Hilbert series of the theory [7]. Indeed, we find the twisted index is the equivariant $\chi_t$ genus of the *compact* global Laumon space. Compactness and isolated fixed points under global symmetries imply independence of the $\chi_t$ genus from fugacities $\{y\}$ and $q$ [26]. The A-twisted index then coincides with the Poincaré polynomial of global Laumon space. The main result of this section is:

$$Z_{\text{H.S.}}[\mathcal{M}_C(T_\rho[\mathrm{SU}(N)])]\,(\xi, t) = \sum_d \xi^d P_t(\mathcal{Q}_d^\rho) = Z^A_{S^2 \times_q S^1}[T_\rho[\mathrm{SU}(N)]]\,(\xi, t) \tag{1.10}$$

This is consistent with the philosophy of Nakajima's construction of the Coulomb branch algebra [10] in terms of the homology of quasimaps to the Higgs branch. In section 5 we turn on background flavour fluxes and angular momentum refinement. Remarkably, we find $q$ independence persists provided we project the index onto the singlet sector under the Cartan subgroup of the global symmetry group:

$$Z_{\text{H.S.}}[\mathcal{M}_C(T_\rho[\mathrm{SU}(N)])]\,(t, \xi; \mathfrak{n}) \sim \int \frac{dy}{y} Z^A_{S^2 \times_A S^1}[T_\rho[\mathrm{SU}(N)]]\,(y, \xi, \mathfrak{n}) \tag{1.11}$$

From the B twist this independence is manifest since the B twist contour integral coincides with the Molien integral (5.14) for the Hilbert series of the Higgs branch of the mirror dual.

Geometrically, this argument translates to a mirror symmetry statement that relates the equivariant Euler character[6] of a line bundle on $\mathcal{M}_C$ to the generating function of $\chi_t$ genera of quasimaps to the Higgs branch $\mathcal{M}_H$:

$$Z_{\text{H.S.}}[\mathcal{M}_C(T_\rho[\text{SU}(N)])]\,(t,\xi;\mathfrak{n}) = \chi_T(\mathcal{M}_C,\mathcal{O}(\mathfrak{n})) \sim \int \frac{dy}{y}\sum_d \xi^d \chi_t(\mathcal{Q}_d;t,y) \tag{1.12}$$

## 2 Factorisation of the A and B twisted indices

In this section we exhibit the $A$ and $B$-twisted index of the $T[\text{SU}(N)]$ theory and the A twisted index of SQCD$[k,N]$. We exhibit the factorisation into holomorphic blocks described in the introduction. We see that the fluxes for the global symmetries through $S^2$ enter only in the holomorphic block as a 'shift' of the corresponding fugacities by the parameter $q$ corresponding to the $\Omega$-deformation/angular momentum refinement. The full details of the calculation are given in appendix A. We also discuss the connection between holomorphic factorisation and the Bethe ansatz formula of the twisted index.

### 2.1 $T[\text{SU}(N)]$

We first factorise the $T[\text{SU}(N)]$ twisted indices. We find the same blocks as those obtained from the factorisation of the squashed ellipsoid partition function [27] in the absence of flux.

**$T[\text{SU}(N)]$ theory.** The $T[\text{SU}(N)]$ theory is specified by the quiver diagram 1 with $N_s = s$ for $s = 1,\ldots,N$ (and $L = N$). It has gauge group $\text{U}(1)\times\ldots\times\text{U}(N-1)$, with an ($\mathcal{N}=4$) vector multiplet for each gauge node, $N$ hypermultiplets in the fundamental of $\text{U}(N-1)$ and a hypermultiplet in the bifundamental of $\text{U}(s)\times\text{U}(s+1)$ for $s = 1,\ldots,N-2$. In addition, it is self mirror dual. For full details of the following calculations, see appendix A.

**A-twist result.** The contour integral for the A-twisted index of $T[\text{SU}(N)]$ is:[7]

$$Z^A[T[\text{SU}(N)]]\left(q,t,\vec{y},\mathfrak{n},\vec{\xi},\tilde{\mathfrak{n}}\right) = \left(\prod_{s=1}^{N-1}\frac{(-1)^s}{s!}\right)\sum_{\{\mathfrak{m}_a^{(s)}\}}\oint_{JK}\prod_{s=1}^{N-1}\prod_{a=1}^{s}\frac{dx_a^{(s)}}{2\pi i x_a^{(s)}} \tag{2.3}$$
$$\times\left(\prod_{s=1}^{N-1}\left(\frac{\xi_s}{\xi_{s+1}}\right)^{\sum_{a=1}^s \mathfrak{m}_a^{(s)}}\right)\left(\prod_{s=1}^{N-1}\prod_{a=1}^{s}(x_a^{(s)})^{\tilde{\mathfrak{n}}_s-\tilde{\mathfrak{n}}_{s+1}}\right)(t-t^{-1})^{-\sum_{s=1}^{N-1}s}$$

[6]In cases where the higher cohomology vanishes, this coincides with the Hilbert series of the Coulomb branch with background flux.

[7]The finite q-Pochhammer symbol is defined as:

$$(a;z)_m = \begin{cases}\prod_{j=0}^{m-1}(1-az^j) & \text{if } m>0\\ 1 & \text{if } m=0\\ \prod_{j=1}^{|m|}(1-az^{-j})^{-1} & \text{if } m<0\end{cases} \tag{2.1}$$

and the infinite q-Pochhamer symbol as:

$$(a;z)_\infty = \begin{cases}\prod_{j=0}^{\infty}(1-az^j) & \text{if } |z|<1\\ \prod_{j=1}^{\infty}(1-az^{-j})^{-1} & \text{if } |z|>1\end{cases}. \tag{2.2}$$

$$\times \left[ \prod_{s=1}^{N-1} \prod_{\substack{a,b=1 \\ a\neq b}}^{s} \frac{\left(\left(\frac{x_a^{(s)}}{x_b^{(s)}}\right)^{\frac{1}{2}}\right)^{(\mathfrak{m}_a^{(s)}-\mathfrak{m}_b^{(s)}-1)}}{\left(\frac{x_a^{(s)}}{x_b^{(s)}} q^{2-\mathfrak{m}_a^{(s)}+\mathfrak{m}_b^{(s)}}; q^2\right)_{\mathfrak{m}_a^{(s)}-\mathfrak{m}_b^{(s)}-1}} \frac{\left(\left(\frac{x_a^{(s)}}{x_b^{(s)}}\right)^{\frac{1}{2}} t^{-1}\right)^{(\mathfrak{m}_a^{(s)}-\mathfrak{m}_b^{(s)}+1)}}{\left(\frac{x_a^{(s)}}{x_b^{(s)}} t^{-2} q^{-\mathfrak{m}_a^{(s)}+\mathfrak{m}_b^{(s)}}; q^2\right)_{\mathfrak{m}_a^{(s)}-\mathfrak{m}_b^{(s)}+1}} \right]$$

$$\times \left[ \prod_{s=1}^{N-1} \prod_{a=1}^{s} \prod_{b=1}^{s+1} \frac{\left(\left(\frac{x_a^{(s)}}{x_b^{(s+1)}}\right)^{\frac{1}{2}} t^{\frac{1}{2}}\right)^{(\mathfrak{m}_a^{(s)}-\mathfrak{m}_b^{(s+1)})}}{\left(\frac{x_a^{(s)}}{x_b^{(s+1)}} t q^{1-\mathfrak{m}_a^{(s)}+\mathfrak{m}_b^{(s+1)}}; q^2\right)_{\mathfrak{m}_a^{(s)}-\mathfrak{m}_b^{(s+1)}}} \frac{\left(\left(\frac{x_b^{(s+1)}}{x_a^{(s)}}\right)^{\frac{1}{2}} t^{\frac{1}{2}}\right)^{(\mathfrak{m}_b^{(s+1)}-\mathfrak{m}_a^{(s)})}}{\left(\frac{x_b^{(s+1)}}{x_a^{(s)}} t q^{1+\mathfrak{m}_a^{(s)}-\mathfrak{m}_b^{(s+1)}}; q^2\right)_{\mathfrak{m}_b^{(s+1)}-\mathfrak{m}_a^{(s)}}} \right]$$

here $\{x_a^{(s)}\}$ denote the gauge fugacities or exponentiated complexified holonomies, $\{\mathfrak{m}_a^{(s)}\}$ the corresponding gauge magnetic fluxes, $\xi_s/\xi_{s+1}$ the F.I. parameter for the $s^{\text{th}}$ gauge node, $\tilde{\mathfrak{n}}_s - \tilde{\mathfrak{n}}_{s+1}$ the corresponding background fluxes through $S^2$, $y_i$ the fugacity for the SU$(N)$ flavour symmetry for the hypermultiplets, and $\mathfrak{n}_i$ the corresponding background fluxes. We have identified $x_a^{(N)} = y_a^{-1}$ and $\mathfrak{m}_a^{(N)} = -\mathfrak{n}_a$. Note that we also identify:

$$\prod_{s=1}^{N} \xi_s = \prod_{i=1}^{N} y_i = 1 \qquad \sum_{s=1}^{N} \tilde{\mathfrak{n}}_s = \sum_{i=1}^{N} \mathfrak{n}_i = 0 \tag{2.4}$$

since in both cases the central symmetry is in fact gauge.

The second line of (2.3) consists of classical contributions around the BPS loci, the third line the 1-loop determinants of the $\mathcal{N}=2$ vector multiplet,[8] and adjoint chiral (together giving the contribution of the $\mathcal{N}=4$ vector multiplet), the fourth line corresponds to the $\mathcal{N}=4$ bifundamental and fundamental hypermultiplets.

As outlined in appendix A we compute the residues of the above integrand at the intersections of hyperplanes specified by the JK residue procedure, and obtain:

$$\mathcal{Z}^A\left[T[\mathrm{SU}(N)]\right]\left(q,t,\vec{y}_i,\mathfrak{n}_i,\vec{\xi}_s,\tilde{\mathfrak{n}}_s\right) = \sum_{\sigma\in S_N} \mathcal{Z}^A_{\text{cl}}\mathcal{Z}^A_{\text{1-loop}}\mathcal{Z}^A_{\text{V}}\mathcal{Z}^A_{\text{aV}}\left(q,t,\vec{y}_{\sigma(i)},\mathfrak{n}_{\sigma(i)},\vec{\xi}_s,\tilde{\mathfrak{n}}_s\right) \tag{2.5}$$

where the perturbative piece is given by:

$$\begin{aligned} \mathcal{Z}^A_{\text{cl}}\mathcal{Z}^A_{\text{1-loop}}\left(q,t,\vec{y},\mathfrak{n},\vec{\xi},\tilde{\mathfrak{n}}\right) &= \prod_{s=1}^{N-1} t^{-s} \left[\prod_{i<j}\frac{\xi_i}{\xi_j}\right] \left[\prod_{s=1}^{N}\left(\xi_s t^{2s}\right)^{-\mathfrak{n}_s}\left(y_s t^{-2s}\right)^{-\tilde{\mathfrak{n}}_s}\right] \\ &\quad \times \left[\prod_{i<j}^{N} \frac{\left(q^2\frac{y_i q^{-\mathfrak{n}_i}}{y_j q^{-\mathfrak{n}_j}};q^2\right)_{\mathfrak{n}_i-\mathfrak{n}_j-1}}{\left(t^2 q^2 \frac{y_i q^{-\mathfrak{n}_i}}{y_j q^{-\mathfrak{n}_j}};q^2\right)_{\mathfrak{n}_i-\mathfrak{n}_j-1}}\right] \end{aligned} \tag{2.6}$$

[8]Here we have used the fact that a $\mathcal{N}=2$ vector multiplet contributes with the same 1-loop determinant as an $\mathcal{N}=2$ chiral multiplet of $R$-charge 2, as in [1].

and the vortex, anti-vortex partition functions are:

$$\mathcal{Z}^A_{\mathrm{V}}\left(q,t,\vec{y},\mathfrak{n},\vec{\xi},\tilde{\mathfrak{n}}\right)=\sum_{\{k_a^{(s)}\}}\prod_{s=1}^{N-1}\left[\left(t^{-2}\frac{\xi_s q^{-\tilde{\mathfrak{n}}_s}}{\xi_{s+1}q^{-\tilde{\mathfrak{n}}_{s+1}}}\right)^{\sum_{a=1}^s k_a^{(s)}}\prod_{\substack{a,b=1\\a\neq b}}^{s}\frac{\left(t^{-2}\frac{y_aq^{-\mathfrak{n}_a}}{y_bq^{-\mathfrak{n}_b}};q^2\right)_{k_a^{(s)}-k_b^{(s)}}}{\left(\frac{y_aq^{-\mathfrak{n}_a}}{y_bq^{-\mathfrak{n}_b}};q^2\right)_{k_a^{(s)}-k_b^{(s)}}}\right.$$
$$\left.\times\prod_{a=1}^{s}\prod_{b=1}^{s+1}\frac{\left(t^2q^2\frac{y_aq^{-\mathfrak{n}_a}}{y_bq^{-\mathfrak{n}_b}};q^2\right)_{k_a^{(s)}-k_b^{(s+1)}}}{\left(q^2\frac{y_aq^{-\mathfrak{n}_a}}{y_bq^{-\mathfrak{n}_b}};q^2\right)_{k_a^{(s)}-k_b^{(s+1)}}}\right] \tag{2.7}$$
$$\mathcal{Z}^A_{\mathrm{aV}}\left(q,t,\vec{y},\mathfrak{n},\vec{\xi},\tilde{\mathfrak{n}}\right)=\mathcal{Z}^A_{\mathrm{V}}\left(q\to q^{-1}\right)$$

Here the summation is over vortex number corresponding to integers $\{k_a^{(s)}\}$ with $a=1,\ldots,s$, satisfying $k_a^{(N)}=0\ \forall a$ and:

$$\begin{array}{rcl} k_1^{(1)}\geq k_1^{(2)}\geq k_1^{(3)}\geq\cdots\geq & k_1^{(N-1)} & \geq 0\\ k_2^{(2)}\geq k_2^{(3)}\geq\cdots\geq & k_2^{(N-1)} & \geq 0\\ k_3^{(3)}\geq\cdots\geq & k_3^{(N-1)} & \geq 0\\ \ddots & \vdots & \\ & k_{N-1}^{(N-1)} & \geq 0\end{array} \tag{2.8}$$

As expected, the localisation calculation yields the twisted index as a sum over Higgs vacua.

One can also factorise the q-Pochhammers in the perturbative piece using the fusion rule:

$$(aq^{-m};q^2)_\infty(aq^m;q^{-2})_\infty=(aq^{-m};q^2)_{m+1} \tag{2.9}$$

defining:

$$B\left(q,t;\vec{Y},\vec{\Xi}\right)=\left[\prod_{i<j}^{N}\frac{\left(q^2\frac{y_iq^{-\mathfrak{n}_i}}{y_jq^{-\mathfrak{n}_j}};q^2\right)_\infty}{\left(t^2q^2\frac{y_iq^{-\mathfrak{n}_i}}{y_jq^{-\mathfrak{n}_j}};q^2\right)_\infty}\right]\mathcal{Z}^A_{\mathrm{V}}\left(q,t,\vec{y},\mathfrak{n},\vec{\xi},\tilde{\mathfrak{n}}\right) \tag{2.10}$$

where we define $Y_i\equiv y_iq^{-\mathfrak{n}_i}$ and $\Xi_i=\xi_iq^{-\tilde{\mathfrak{n}}_i}$, the holomorphic combinations of fugacities and fluxes that the r.h.s. depend on. This coincides with the holomorphic block computed in [28]. Thus we may write:

$$\mathcal{Z}^A\left[T[\mathrm{SU}(N)]\right]\left(t,q,\vec{y},\mathfrak{n},\vec{\xi},\tilde{\mathfrak{n}}\right)=\sum_{\sigma\in S_N}p\left(t,\sigma\vec{y},\sigma\mathfrak{n},\vec{\xi},\tilde{\mathfrak{n}}\right)B\left(q,t;\sigma\vec{Y},\vec{\Xi}\right)B\left(q^{-1},t;\sigma\vec{Y}',\vec{\Xi}'\right) \tag{2.11}$$

where the 'gluing' of blocks corresponds to $Y_i'=y_iq^{\mathfrak{n}_i}$ and $\Xi_i'=\xi_iq^{\tilde{\mathfrak{n}}_i}$, and we define e.g. $\sigma(Y_i)\equiv Y_{\sigma(i)}=y_{\sigma(i)}^{-\mathfrak{n}_{\sigma(i)}}$. The prefactor is:

$$p\left(t,\vec{y},\mathfrak{n},\vec{\xi},\tilde{\mathfrak{n}}\right)=t^{-\frac{N}{2}(N-1)}\left[\prod_{i<j}^{N}\frac{\xi_i}{\xi_j}\right]\left[\prod_{s=1}^{N}\left(\xi_st^{2s}\right)^{-\mathfrak{n}_s}\left(y_st^{-2s}\right)^{-\tilde{\mathfrak{n}}_s}\right] \tag{2.12}$$

The $\xi$ and $y$ monomials in $p\left(t, \vec{y}, \mathfrak{n}, \vec{\xi}, \tilde{\mathfrak{n}}\right)$ can be factorised in terms of $\theta$ functions:[9]

$$\begin{aligned}
&\prod_{s=1}^{N} \left\| \frac{\theta(tq(\xi_s q^{-\tilde{\mathfrak{n}}_s})^{2s}; q^2)}{\theta((\xi_s q^{-\tilde{\mathfrak{n}}_s})^{2s}; q^2)} \frac{\theta(t^{-1}(y_s q^{-\mathfrak{n}_s})^{2s}; q^2)}{\theta((y_s q^{-\mathfrak{n}_s})^{2s}; q^2)} \frac{\theta(\xi_s q^{-\tilde{\mathfrak{n}}_s}; q^2)\theta(y_s q^{-\mathfrak{n}_s}; q^2)}{\theta(\xi_s q^{-\tilde{\mathfrak{n}}_s} y_s q^{-\mathfrak{n}_s}; q^2)} \right\|^2 \\
&\qquad = t^{-N(N+1)/2} \prod_{s=1}^{N} (\xi_s)^{-2s} \left(\xi_s t^{2s}\right)^{-\mathfrak{n}_s} \left(y_s t^{-2s}\right)^{-\tilde{\mathfrak{n}}_s} \qquad (2.13) \\
&\qquad = t^{-N(N+1)/2} \left( \prod_{i<j} \frac{\xi_i}{\xi_j} \right) \prod_{s=1}^{N} \left(\xi_s t^{2s}\right)^{-\mathfrak{n}_s} \left(y_s t^{-2s}\right)^{-\tilde{\mathfrak{n}}_s}
\end{aligned}$$

The fully-factorised block is then defined:

$$\mathbb{B}\left(q, t; \vec{Y}, \vec{\Xi}\right) = \left[ \prod_{s=1}^{N} \frac{\theta(tq(\Xi_s)^{2s}; q^2)}{\theta((\Xi_s)^{2s}; q^2)} \frac{\theta(t^{-1}(Y_s)^{2s}; q^2)}{\theta((Y_s)^{2s}; q^2)} \frac{\theta(\Xi_s; q^2)\theta(Y_s; q^2)}{\theta(\Xi_s Y_s; q^2)} \right] B\left(q, t; \vec{Y}, \vec{\Xi}\right) \tag{2.14}$$

Up to a $t$ dependent pre-factor, the twisted partition function can thus be fully factorised:

$$\mathcal{Z}^A \left[T[\mathrm{SU}(N)]\right] \left(q, t, \vec{y}, \mathfrak{n}, \vec{\xi}, \tilde{\mathfrak{n}}\right) = \sum_{\sigma \in S_N} \mathbb{B}\left(q, t; \sigma\vec{Y}, \vec{\Xi}\right) \mathbb{B}\left(q^{-1}, t; \sigma\vec{Y}', \vec{\Xi}'\right) \tag{2.15}$$

We note that the fully-factorised block is in fact mirror symmetric:

$$\mathbb{B}\left(q, t; \vec{Y}, \vec{\Xi}\right) = \mathbb{B}\left(q, \frac{1}{qt}; \vec{\Xi}, \vec{Y}\right) \tag{2.16}$$

as expected of a mirror self-dual theory. The classical piece of the block given by the ratio of theta functions in (2.14) is clearly separately mirror self-dual. The mirror self-duality of $B\left(q, t; \vec{Y}, \vec{\Xi}\right)$ is given in [28]. The fully factorised block essentially coincides with the one derived in *loc. cit.* after making the replacements (2.10) in that paper of exponentials to theta functions, corresponding to 'resolving' Chern-Simons terms via massive chirals.

[9] The $\theta$ function is defined by $\theta(x;q) \equiv (x;q)_\infty (qx^{-1};q)_\infty$ and the conjugation is $\|\theta(xq^m;q)\|^2 \equiv \theta(xq^m;q)\theta(xq^{-m};q^{-1})$. This identity follows from the fusion rule (3.9). Notice that other factorisations are possible, but they all differ by an elliptic ratio of theta functions which fuses trivially when gluing for the $S_b^3$ partition function, the superconformal index, as well as the twisted index. This ambiguity is discussed in [19] and can be interpreted physically as different ways of 'resolving' Chern-Simons terms via pairs of massive chirals.

**B-twist result.** The $B$ twisted index is given by the integral:

$$Z^B[T[\mathrm{SU}(N)]]\left(q,t,\vec{y},\mathfrak{n},\vec{\xi},\tilde{\mathfrak{n}}\right)=\left(\prod_{s=1}^{N-1}\frac{(-1)^s}{s!}\right)\sum_{\{\mathfrak{m}_a^s\}}\oint_{JK}\prod_{s=1}^{N-1}\prod_{a=1}^{s}\frac{dx_a^{(s)}}{2\pi i x_a^{(s)}} \tag{2.17}$$

$$\times\left(\prod_{s=1}^{N-1}\left(\frac{\xi_s}{\xi_{s+1}}\right)^{\sum_{a=1}^{s}\mathfrak{m}_a^{(s)}}\right)\left(\prod_{s=1}^{N-1}\prod_{a=1}^{s}(x_a^{(s)})^{\tilde{\mathfrak{n}}_s-\tilde{\mathfrak{n}}_{s+1}}\right)(t-t^{-1})^{\sum_{s=1}^{N-1}s}\times\dots$$

$$\times\left[\prod_{s=1}^{N-1}\prod_{\substack{a,b=1\\a\neq b}}^{s}\frac{\left(\left(\frac{x_a^{(s)}}{x_b^{(s)}}\right)^{\frac{1}{2}}\right)^{(\mathfrak{m}_a^{(s)}-\mathfrak{m}_b^{(s)}-1)}}{\left(\frac{x_a^{(s)}}{x_b^{(s)}}q^{2-\mathfrak{m}_a^{(s)}+\mathfrak{m}_b^{(s)}};q^2\right)_{\mathfrak{m}_a^{(s)}-\mathfrak{m}_b^{(s)}-1}}\frac{\left(\left(\frac{x_a^{(s)}}{x_b^{(s)}}\right)^{\frac{1}{2}}t^{-1}\right)^{(\mathfrak{m}_a^{(s)}-\mathfrak{m}_b^{(s)}-1)}}{\left(\frac{x_a^{(s)}}{x_b^{(s)}}t^{-2}q^{2-\mathfrak{m}_a^{(s)}+\mathfrak{m}_b^{(s)}};q^2\right)_{\mathfrak{m}_a^{(s)}-\mathfrak{m}_b^{(s)}-1}}\right]$$

$$\times\left[\prod_{s=1}^{N-1}\prod_{a=1}^{s}\prod_{b=1}^{s+1}\frac{\left(\left(\frac{x_a^{(s)}}{x_b^{(s+1)}}\right)^{\frac{1}{2}}t^{\frac{1}{2}}\right)^{(\mathfrak{m}_a^{(s)}-\mathfrak{m}_b^{(s+1)}+1)}}{\left(\frac{x_a^{(s)}}{x_b^{(s+1)}}tq^{-\mathfrak{m}_a^{(s)}+\mathfrak{m}_b^{(s+1)}};q^2\right)_{\mathfrak{m}_a^{(s)}-\mathfrak{m}_b^{(s+1)}+1}}\frac{\left(\left(\frac{x_b^{(s+1)}}{x_a^{(s)}}\right)^{\frac{1}{2}}t^{\frac{1}{2}}\right)^{(\mathfrak{m}_b^{(s+1)}-\mathfrak{m}_a^{(s)}+1)}}{\left(\frac{x_b^{(s+1)}}{x_a^{(s)}}tq^{\mathfrak{m}_a^{(s)}-\mathfrak{m}_b^{(s+1)}};q^2\right)_{\mathfrak{m}_b^{(s+1)}-\mathfrak{m}_a^{(s)}+1}}\right]$$

with the same identification of $x_a^N=y_a^{-1}$ and $\mathfrak{m}_a^N=-\mathfrak{n}_a$. It can be computed similarly (see appendix A) and is:

$$\mathcal{Z}^B\left[T[\mathrm{SU}(N)]\right]\left(q,t,\vec{y}_i,\mathfrak{n}_i,\vec{\xi}_s,\tilde{\mathfrak{n}}_s\right)=\sum_{\sigma\in S_N}\mathcal{Z}^B_{\mathrm{cl}}\mathcal{Z}^B_{\text{1-loop}}\mathcal{Z}^B_{\mathrm{V}}\mathcal{Z}^B_{\mathrm{aV}}\left(q,t,\vec{y}_{\sigma(i)},\mathfrak{n}_{\sigma(i)},\vec{\xi}_s,\tilde{\mathfrak{n}}_s\right) \tag{2.18}$$

where the perturbative contribution is:

$$\mathcal{Z}^B_{\mathrm{cl}}\mathcal{Z}^B_{\text{1-loop}}\left(q,t,\vec{y},\mathfrak{n},\vec{\xi},\tilde{\mathfrak{n}}\right)=\prod_{s=1}^{N-1}t^s\prod_{i<j}^{N}\left(\frac{y_i}{y_j}\right)\left[\prod_{s=1}^{N}\left(\xi_s t^{-(N-2s+1)}\right)^{-\mathfrak{n}_s}\left(y_s t^{(N-2s+1)}\right)^{-\tilde{\mathfrak{n}}_s}\right]$$
$$\times\left[\prod_{i<j}^{N}\frac{\left(q^2\frac{y_iq^{-\mathfrak{n}_i}}{y_jq^{-\mathfrak{n}_j}};q^2\right)_{\mathfrak{n}_i-\mathfrak{n}_j-1}}{\left(t^2\frac{y_iq^{-\mathfrak{n}_i}}{y_jq^{-\mathfrak{n}_j}};q^2\right)_{\mathfrak{n}_i-\mathfrak{n}_j+1}}\right] \tag{2.19}$$

and the vortex/anti-vortex partition functions are:

$$\mathcal{Z}^B_V\left(q,t,\vec{y},\mathfrak{n},\vec{\xi},\tilde{\mathfrak{n}}\right)=\sum_{\{k_a^{(s)}\}}\prod_{s=1}^{N-1}\left[\left(t^{-2}q^2\frac{\xi_sq^{-\tilde{\mathfrak{n}}_s}}{\xi_{s+1}q^{-\tilde{\mathfrak{n}}_{s+1}}}\right)^{\sum_{a=1}^{s}k_a^{(s)}}\prod_{a\neq b}^{s}\frac{\left(t^{-2}q^2\frac{y_aq^{-\mathfrak{n}_a}}{y_bq^{-\mathfrak{n}_b}};q^2\right)_{k_a^{(s)}-k_b^{(s)}}}{\left(\frac{y_aq^{-\mathfrak{n}_a}}{y_bq^{-\mathfrak{n}_b}};q^2\right)_{k_a^{(s)}-k_b^{(s)}}}\right.$$
$$\left.\times\prod_{a=1}^{s}\prod_{b=1}^{s+1}\frac{\left(t^2\frac{y_aq^{-\mathfrak{n}_a}}{y_bq^{-\mathfrak{n}_b}};q^2\right)_{(k_a^{(s)}-k_b^{(s+1)})}}{\left(q^2\frac{y_aq^{-\mathfrak{n}_a}}{y_bq^{-\mathfrak{n}_b}};q^2\right)_{(k_a^{(s)}-k_b^{(s+1)})}}\right] \tag{2.20}$$

$$\mathcal{Z}^B_{aV}\left(q,t,\vec{y},\mathfrak{n},\vec{\xi},\tilde{\mathfrak{n}}\right)=\mathcal{Z}^B_V\left(q\to q^{-1}\right)$$

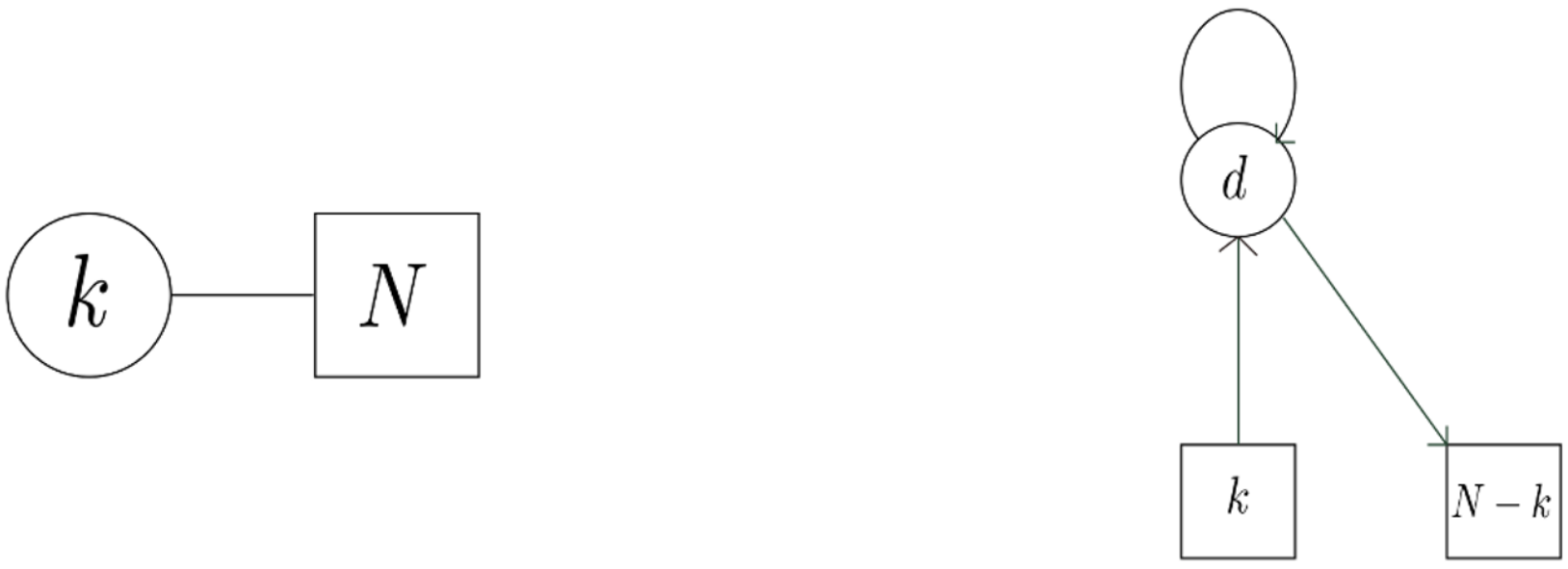

**Figure 2**. The quiver diagram for $\mathcal{N} = 4$ SQCD$[k, N]$, and its vortex moduli space, a simple Handsaw quiver (see section 5).

The sum is over the same set of integers as (2.8). The vortex partition function in the B-twist is the same as in the A-twist up to a relabelling of parameters $t \to t/q$. This is a consequence of different R charge assignments under the two twists corresponding to the chiral multiplets transforming in different powers of the canonical bundle — in the $\Omega$-background this manifests as a shift in the angular momentum grading.[10] In fact we notice that the B-twisted index, up to a $t$-prefactor, can be fully-factorised in terms of the same block under after a $q$ shift of $t$.

$$\mathcal{Z}^B\left[T[\mathrm{SU}(N)]\right]\left(q, t, \vec{y}, \mathfrak{n}, \vec{\xi}, \tilde{\mathfrak{n}}\right) = \sum_{\sigma \in S_N} \mathbb{B}\left(q, tq^{-1}; \sigma\vec{Y}, \vec{\Xi}\right) \mathbb{B}\left(q^{-1}, tq; \sigma\vec{Y}', \vec{\Xi}'\right) \tag{2.21}$$

### 2.2 SQCD$[k, N]$

We also compute the $A$-twisted index for another theory in the class $T_\rho[\mathrm{SU}(N)]$. The quiver diagram is given in figure 2. The details are much the same as $T[\mathrm{SU}(N)]$ and we find:

$$\mathcal{Z}^A\left[SQCD[k, N]\right](q, t, \vec{y}, \mathfrak{n}, \xi, \tilde{\mathfrak{n}}) = \sum_{f \in S_N/(S_k \times S_{N-k})} \mathcal{Z}^{(f)}_{\text{cl}} \mathcal{Z}^{(f)}_{\text{1-loop}} \mathcal{Z}^{(f)}_{\text{V}} \mathcal{Z}^{(f)}_{\text{aV}}(t, q, \vec{y}, \mathfrak{n}, \xi, \tilde{\mathfrak{n}}) \tag{2.22}$$

Here $\xi$ and $\tilde{\mathfrak{n}}$ are the (single) F.I. parameter and flux for the U(1) topological symmetry, and $\vec{y}$ and $\mathfrak{n}$ are $N$-vectors of fugacities and fluxes for the flavour symmetry on the $N$ fundamental $\mathcal{N} = 4$ hypermultiplets. The sum is over the quotient group: $S_N/(S_k \times S_{N-k})$, and is the same for any choice of representative $f$ since the integrand is symmetric in the gauge fugacities $\{x_a\}$, $a = 1, \ldots, k$. The sum is therefore over the $\binom{N}{k}$ Higgs vacua. We assume the matter content has been chosen such that the boundary contributions from the

[10] We thank Mathew Bullimore for pointing this out to us.

localisation procedure vanish. The components are:

$$
\begin{aligned}
\mathcal{Z}_{\text{cl}}^{(f)}\mathcal{Z}_{\text{1-loop}}^{(f)}(t,q,\vec{y},\mathfrak{n},\xi,\tilde{\mathfrak{n}}) &= t^{k^2}\left[\prod_{a=1}^{k}\prod_{i=1}^{N}(-t)^{\mathfrak{n}_{f(a)}-\mathfrak{n}_i-1}\right]\left[\prod_{a=1}^{k}\left(y_{f(a)}^{-1}t^{-1}\right)^{\tilde{\mathfrak{n}}}\xi^{1-\mathfrak{n}_{f(a)}}\right] \\
&\quad\times\left[\prod_{a=1}^{k}\prod_{\substack{i=1\\ i\neq f(b)\forall b=1,\dots,k}}^{N}\frac{\left(q^2\frac{y_{f(a)}q^{-\mathfrak{n}_{f(a)}}}{y_i q^{-\mathfrak{n}_i}}\right)_{\mathfrak{n}_{f(a)}-\mathfrak{n}_i-1}}{\left(t^2q^2\frac{y_{f(a)}q^{-\mathfrak{n}_{f(a)}}}{y_i q^{-\mathfrak{n}_i}};q^2\right)_{\mathfrak{n}_{f(a)}-\mathfrak{n}_i-1}}\right] \\
\mathcal{Z}_{V}^{(f)}(t,q,\vec{y},\mathfrak{n},\xi,\tilde{\mathfrak{n}}) &= \sum_{\{k_a\geq 0\}}\left((-t)^{-N}\xi q^{-\tilde{\mathfrak{n}}}\right)^{\sum_{a=1}^{k}k_a} \\
&\quad\times\prod_{\substack{a,b=1\\ a\neq b}}^{k}\frac{\left(t^{-2}\frac{y_{f(a)}q^{-\mathfrak{n}_{f(a)}}}{y_{f(b)}q^{-\mathfrak{n}_{f(b)}}};q^2\right)_{k_a-k_b}}{\left(\frac{y_{f(a)}q^{-\mathfrak{n}_{f(a)}}}{y_{f(b)}q^{-\mathfrak{n}_{f(b)}}};q^2\right)_{k_a-k_b}}\prod_{a=1}^{k}\prod_{i=1}^{N}\frac{\left(t^2q^2\frac{y_{f(a)}q^{-\mathfrak{n}_{f(a)}}}{y_i q^{-\mathfrak{n}_i}};q^2\right)_{k_a}}{\left(q^2\frac{y_{f(a)}q^{-\mathfrak{n}_{f(a)}}}{y_i q^{-\mathfrak{n}_i}};q^2\right)_{k_a}} \\
\mathcal{Z}_{AV}^{(f)}(t,q,\vec{y},\mathfrak{n},\xi,\tilde{\mathfrak{n}}) &= \mathcal{Z}_{V}^{(f)}\left(q\to q^{-1}\right)
\end{aligned}
\tag{2.23}
$$

This block coincides with the block found in [29] from factorising the superconformal index. The sum is over integers $\{k_a \geq 0\}$, $a = 1,\dots,k$. We note that all these calculations extend straightforwardly to the general $T_\rho[\mathrm{SU}(N)]$ theory.

### 2.3 Bethe ansatz formula

In this subsection we outline how the Bethe ansatz formulation of the twisted index [7, 23] arises from the holomorphic factorisation. This argument is essentially a review of section 9 of [30] for the $\mathcal{N} = 4$ case.

**Block factorisation.** For a generic 3d $\mathcal{N} = 4$ quiver gauge theory, we expect the following factorisation formula for the A-twisted index with background fluxes and angular momentum refinement turned on:

$$
Z^{A}_{S^2\times_q S^1} = \sum_{\alpha}\mathbb{B}_\alpha(\xi q^{\tilde{\mathfrak{n}}}, yq^{\mathfrak{n}}; q, t)\mathbb{B}_\alpha(\xi q^{-\tilde{\mathfrak{n}}}, yq^{-\mathfrak{n}}; q^{-1}, t) \tag{2.24}
$$

where $\mathbb{B}_\alpha$ denotes the fully-factorised block of the theory in a particular vacuum and the sum is taken over massive vacua. The fully-factorised block can be written as a product $\mathbb{B} = Z_{\text{cl.}}Z_{\text{1-loop}}Z_{\text{vortex}}$ where the classical part includes the $\theta$ functions that fuse to give, the generalisation of, the pre-factor of (2.12)[11] and the vortex contributions correspond to a generating function of the $\chi_t$ genera of the vortex moduli space of the theory. The previous subsections verify this formula for the $T[\mathrm{SU}(N)]$ and SQCD$[k, N]$ theories.

[11] Note this pre-factor vanishes in the absence of flux but has a non-trivial contribution when the fugacities are “shifted” by $y \to yq^{\mathfrak{n}}$ and $\xi \to \xi q^{\tilde{\mathfrak{n}}}$.

**$q \to 1$ limit.** We now turn off the angular momentum refinement. This corresponds to taking $\epsilon \to 0$ with $q = e^\epsilon$ in our formulae. For a generic 3d $\mathcal{N} = 4$ quiver gauge theory, the blocks $\mathbb{B}$ have an integral representation determined by the quiver data (see for example appendix A of [28]). In the $\epsilon \to 0$ limit, the block integrals can be evaluated by a saddle point approximation in terms of the effective twisted superpotential $\mathcal{W}$ of the theory where stationary points of $\mathcal{W}$ correspond to the vacua $\alpha$. In this section we work with exponentiated fugacities $y = e^v$, $x = e^u$, $\xi = e^w$ and $t = e^\tau$. The integral form for the blocks gives schematically:[12]

$$\mathbb{B}(\xi, y; q, t) \sim \frac{1}{\sqrt{\epsilon}} \oint du \exp\left\{\frac{1}{2\epsilon}\mathcal{W}(u; w, v, \tau) + \frac{1}{2}\mathcal{W}^0(u; w, v, \tau) + O(\epsilon)\right\} \tag{2.25}$$

Now in the twisted index factorisation the fugacities are shifted in terms of the fluxes so that $w \to w + \epsilon\tilde{\mathfrak{n}}$ and $v \to v + \epsilon\mathfrak{n}$. Making this substitution and Taylor expanding $\mathcal{W}$ we find:

$$\begin{aligned}\mathbb{B}(\xi q^{\tilde{\mathfrak{n}}}, y q^{\mathfrak{n}}; q, t) \sim \frac{1}{\sqrt{\epsilon}} \oint \exp\Big\{&\frac{1}{2\epsilon}\mathcal{W}(u; w, v, \tau) + \frac{1}{2}\mathcal{W}^0(u; w, v, \tau) \\ &+ \frac{1}{2}\mathfrak{n}\partial_v\mathcal{W}(u; w, v, \tau) + \frac{1}{2}\tilde{\mathfrak{n}}\partial_w\mathcal{W}(u; w, v, \tau) + O(\epsilon)\Big\}\end{aligned} \tag{2.26}$$

Now this integral has a set of saddle points, $\alpha$, that correspond to massive vacua of the theory. Each saddle point provides a different integration contour and we see that, to leading and next-to-leading order, each block can be expressed:

$$\begin{aligned}\mathbb{B}_\alpha(\xi q^{\tilde{\mathfrak{n}}}, y q^{\mathfrak{n}}; q, t) \sim \frac{1}{\sqrt{\det(\partial_i\partial_j\mathcal{W}_\alpha^{\text{o.s.}})}} \exp\Big\{&\frac{1}{2\epsilon}\mathcal{W}_\alpha^{\text{o.s.}}(w, v, \tau) + \frac{1}{2}\mathfrak{n}\partial_v\mathcal{W}_\alpha^{\text{o.s.}}(w, v, \tau) \\ &+ \frac{1}{2}\tilde{\mathfrak{n}}\partial_w\mathcal{W}_\alpha^{\text{o.s.}}(w, v, \tau) + \frac{1}{2}\mathcal{W}_\alpha^0(w, v, \tau)\Big\}\end{aligned} \tag{2.27}$$

where $\mathcal{W}_\alpha^{\text{o.s.}}$ denotes the effective twisted superpotential evaluated on a particular vacuum. The twisted index block gluing (2.24) sends $\epsilon \to -\epsilon$ so that the leading divergence cancels. Summing over vacua we find:

$$\begin{aligned}\lim_{q\to 1} Z_{S^2\times_q S^1}[y, \xi, \mathfrak{n}, \tilde{\mathfrak{n}}] = &\sum_\alpha \frac{1}{\det(\partial_i\partial_j\mathcal{W}_\alpha^{\text{o.s.}})} \\ &\times \exp\left(\mathfrak{n}\partial_v\mathcal{W}_\alpha^{\text{o.s.}}(w, v, \tau) + \tilde{\mathfrak{n}}\partial_w\mathcal{W}_\alpha^{\text{o.s.}}(w, v, \tau) + \mathcal{W}_\alpha^0(w, v, \tau)\right)\end{aligned} \tag{2.28}$$

We recognise this as the Bethe ansatz formula of [7].

## 3 Mirror symmetry

3d $\mathcal{N} = 4$ theories enjoy a powerful mirror duality. If theories $T$ and $\check{T}$ are mirror dual, the Coulomb branch of $T$ coincides with the Higgs branch of $\check{T}$ and vice-versa:

$$\mathcal{M}_{H,C}(T) \cong \mathcal{M}_{C,H}(\check{T}) \tag{3.1}$$

[12] $\mathcal{W}_0$ here is essentially the effective dilaton $\Omega$ of [30].

As described in the introduction, the expected twisted index duality is:

$$Z^{A}_{S^2\times_q S^1}[T]\,(t,q,\{z\},\{n\}) \longleftrightarrow Z^{B}_{S^2\times_q S^1}[\check{T}]\left(t^{-1},q,\{\check{z}\},\{\check{n}\}\right) \tag{3.2}$$

where $\{z\}\equiv\{y,\xi\}$, $\{n\}=\{\mathfrak{n},\tilde{\mathfrak{n}}\}$ and $\{\check{z}\}\equiv\{\xi,y\}$, $\{\check{n}\}=\{\tilde{\mathfrak{n}},\mathfrak{n}\}$. Closset and Kim [7] show this duality in some simple examples without the $\Omega$ deformation. In this section we prove this duality, with the $\Omega$ deformation, in two simple Abelian examples: $T[\mathrm{SU}(2)]=\mathrm{SQED}[2]$ and SQED[1], and in the $t\to 1$ limit for the $T[\mathrm{SU}(N)]$ theory where $\mathcal{N}=4$ supersymmetry is restored. For $T_\rho[\mathrm{SU}(N)]$ theories, both $A$ and $B$ twisted indices are given from the JK contour as a sum over Higgs branch vacua, or equivalently over the Weyl group of the flavour symmetry acting on the hypermultiplets. Thus the difficulty in proving mirror symmetry more generally is that the Weyl sum over hypermultiplet flavour symmetry/real masses needs to be exchanged for one over the topological symmetry/F.I. parameters.

### 3.1 SQED[1]

Recall the A twisted partition function for SQED[1] ($k=N=1$ in (2.22) with only topological flux since the flavour symmetry is equivalent to a gauge symmetry in this case):

$$\begin{aligned} Z^{A}_{S^2\times_q S^1}[\mathrm{SQED}[1]] = \xi t^{-\tilde{\mathfrak{n}}}&\left(\sum_{\alpha\geq 0}(\xi t^{-1}q^{\tilde{\mathfrak{n}}})^{\alpha}\frac{(t^2q^{-2};q^{-2})_\beta}{(q^{-2};q^{-2})_\beta}\right)\\ &\times\left(\sum_{\beta\geq 0}(\xi t^{-1}q^{-\mathfrak{n}})^{\beta}\frac{(t^2q^{2};q^{2})_\beta}{(q^{2};q^{2})_\beta}\right)\end{aligned} \tag{3.3}$$

Recalling the $q$-binomial theorem:

$$\sum_{n=0}^{\infty} z^n\frac{(a;q)_n}{(q;q)_n}=\frac{(az;q)_\infty}{(z;q)_\infty} \tag{3.4}$$

as well as the fusion rule (2.9) we can sum the vortex contributions:[13]

$$Z^{A}_{S^2\times_q S^1}[\mathrm{SQED}[1]] = \xi t^{-\tilde{\mathfrak{n}}}\frac{(t\xi q^{-\tilde{\mathfrak{n}}+2};q^2)_{\tilde{\mathfrak{n}}-1}}{(t^{-1}\xi q^{-\tilde{\mathfrak{n}}};q^2)_{\tilde{\mathfrak{n}}+1}} \tag{3.5}$$

A similar computation for the B-twisted index gives:

$$\begin{aligned} Z^{B}_{S^2\times_q S^1}[\mathrm{SQED}[1]] &= y^{-\tilde{\mathfrak{n}}}t^{-\tilde{\mathfrak{n}}}\left(\sum_{\alpha\geq 0}(t^{-1}q^{-1}\xi q^{\tilde{\mathfrak{n}}})^{\alpha}\frac{(t^2;q^{-2})_\alpha}{(q^{-2};q^{-2})_\alpha}\right)\left(\sum_{\beta\geq 0}(t^{-1}q\xi q^{-\tilde{\mathfrak{n}}})^{\beta}\frac{(t^2;q^{2})_\beta}{(q^{2};q^{2})_\beta}\right)\\ &= -y^{-\tilde{\mathfrak{n}}}t^{-\tilde{\mathfrak{n}}}\frac{(t\xi q^{1-\tilde{\mathfrak{n}}};q^2)_{\tilde{\mathfrak{n}}}}{(t^{-1}\xi q^{1-\tilde{\mathfrak{n}}};q^2)_{\tilde{\mathfrak{n}}}}\end{aligned} \tag{3.6}$$

[13]Note that there is no $y$ fugacity dependence since the flavour symmetry is equivalent to the U(1) gauge symmetry.

Comparing with (A.9) and (A.10) we recognise (3.5) and (3.6) as the B and A twisted indices respectively of a free hypermultiplet.[14]

$$\begin{aligned} Z^A_{S^2\times_q S^1}[\text{SQED}[1]] &= Z^B_{S^2\times_q S^1}[\text{Free hyper}] \\ Z^B_{S^2\times_q S^1}[\text{SQED}[1]] &= Z^A_{S^2\times_q S^1}[\text{Free hyper}] \end{aligned} \tag{3.7}$$

where mirror symmetry maps $y \to \xi$, $t \to t^{-1}$ and $\tilde{\mathfrak{n}} \to \mathfrak{n}$.

### 3.2 $T[\mathrm{SU}(2)]$

Since we have noted that the A-twisted index can be block-factorised, proving mirror symmetry (1.5) essentially reduces to the arguments of Nieri and Pasquetti in [8], which we briefly review here. The crucial observation after the introduction of fluxes in the twisted index is that at the level of the blocks, the fluxes appear only as a shift of the fugacities, and the argument is unaffected.

We note first that the vortex partition function (2.7) can be identified as the basic hypergeometric function:

$$\mathcal{Z}^A_{\mathrm{V}}\left(t,q,\vec{y},\mathfrak{n},\vec{\xi},\tilde{\mathfrak{n}}\right) = {}_2\phi_1\left[t^2q^2, t^2q^2\frac{y_1q^{-\mathfrak{n}_1}}{y_2q^{-\mathfrak{n}_2}}; q^2\frac{y_1q^{-\mathfrak{n}_1}}{y_2q^{-\mathfrak{n}_2}}; q^2, t^{-2}\frac{\xi_1q^{-\tilde{\mathfrak{n}}_1}}{\xi_2q^{-\tilde{\mathfrak{n}}_2}}\right] \tag{3.8}$$

This allows us to use known monodromy properties of these functions to prove mirror symmetry. Defining the q-theta function $\Theta(a;z) \equiv (a;z)_\infty(z/a;z)_\infty$, which from (2.9) obeys

$$\Theta(aq^m;q^2)\Theta(aq^{-m},q^{-2}) = (-1)^{m-1}a^{1-m} \tag{3.9}$$

means we may write:

$$\mathcal{Z}^A_{T[\mathrm{SU}(2)]} = t^{-1+(\mathfrak{n}_1-\mathfrak{n}_2)-(\tilde{\mathfrak{n}}_1-\tilde{\mathfrak{n}}_2)}\left(\frac{\xi_1}{\xi_2}\right) y_1^{-\tilde{\mathfrak{n}}_1}y_2^{-\tilde{\mathfrak{n}}_2}\xi_1^{-\mathfrak{n}_1}\xi_2^{-\mathfrak{n}_2}\left[\left\|B_1^I\right\|^2 + \left\|B_2^I\right\|^2\right] \tag{3.10}$$

where e.g. $\left\|B_1^I\right\|^2 = B_1^I(q)B_1^I(q^{-1}) \equiv B_1\tilde{B}_1$ and we rescale the blocks as:

$$\begin{aligned} B_1^I &= \frac{\left(q^2\frac{y_1q^{-\mathfrak{n}_1}}{y_2q^{-\mathfrak{n}_2}};q^2\right)}{\left(q^2t^2\frac{y_1q^{-\mathfrak{n}_1}}{y_2q^{-\mathfrak{n}_2}};q^2\right)}{}_2\phi_1\left[t^2q^2, t^2q^2\frac{y_1q^{-\mathfrak{n}_1}}{y_2q^{-\mathfrak{n}_2}}; q^2\frac{y_1q^{-\mathfrak{n}_1}}{y_2q^{-\mathfrak{n}_2}}; q^2, t^{-2}\frac{\xi_1q^{-\tilde{\mathfrak{n}}_1}}{\xi_2q^{-\tilde{\mathfrak{n}}_2}}\right] \\ B_2^I &= \frac{\Theta\left(\frac{\xi_2q^{-\tilde{\mathfrak{n}}_2}}{\xi_1q^{-\tilde{\mathfrak{n}}_1}};q^2\right)\Theta\left(q^2t^2\frac{y_2q^{-\mathfrak{n}_2}}{y_1q^{-\mathfrak{n}_1}};q^2\right)}{\Theta\left(\frac{\xi_2q^{-\tilde{\mathfrak{n}}_2}}{\xi_1q^{-\tilde{\mathfrak{n}}_1}}\frac{y_1q^{-\mathfrak{n}_1}}{y_2q^{-\mathfrak{n}_2}};q^2\right)\Theta\left(q^2t^2;q^2\right)} \\ &\quad\times\frac{\left(q^2\frac{y_2q^{-\mathfrak{n}_2}}{y_1q^{-\mathfrak{n}_1}};q^2\right)}{\left(q^2t^2\frac{y_2q^{-\mathfrak{n}_2}}{y_1q^{-\mathfrak{n}_1}};q^2\right)}{}_2\phi_1\left[t^2q^2, t^2q^2\frac{y_2q^{-\mathfrak{n}_2}}{y_1q^{-\mathfrak{n}_1}}; q^2\frac{y_2q^{-\mathfrak{n}_2}}{y_1q^{-\mathfrak{n}_1}}; q^2, t^{-2}\frac{\xi_1q^{-\tilde{\mathfrak{n}}_1}}{\xi_2q^{-\tilde{\mathfrak{n}}_2}}\right] \end{aligned} \tag{3.11}$$

[14] This equality is up to some flux dependent sign and we make use of (A.22) to prove the identity.

The blocks which give the Weyl sum over the topological symmetries are defined as:

$$\begin{aligned}B_1^{II} &= \frac{\left(q^2\frac{\xi_1 q^{-\tilde{\mathfrak{n}}_1}}{\xi_2 q^{-\tilde{\mathfrak{n}}_2}};q^2\right)}{\left(t^{-2}\frac{\xi_1 q^{-\tilde{\mathfrak{n}}_1}}{\xi_2 q^{-\tilde{\mathfrak{n}}_2}};q^2\right)}\,{}_2\phi_1\left[t^{-2},t^{-2}\frac{\xi_1 q^{-\tilde{\mathfrak{n}}_1}}{\xi_2 q^{-\tilde{\mathfrak{n}}_2}};q^2\frac{\xi_1 q^{-\tilde{\mathfrak{n}}_1}}{\xi_2 q^{-\tilde{\mathfrak{n}}_2}};q^2,q^2t^2\frac{y_1 q^{-\mathfrak{n}_1}}{y_2 q^{-\mathfrak{n}_2}}\right]\\ B_2^{II} &= \frac{\Theta\left(\frac{y_2 q^{-\mathfrak{n}_2}}{y_1 q^{-\mathfrak{n}_1}};q^2\right)\Theta\left(t^{-2}\frac{\xi_2 q^{-\tilde{\mathfrak{n}}_2}}{\xi_1 q^{-\tilde{\mathfrak{n}}_1}};q^2\right)}{\Theta\left(\frac{y_2 q^{-\mathfrak{n}_2}}{y_1 q^{-\mathfrak{n}_1}}\frac{\xi_1 q^{-\tilde{\mathfrak{n}}_1}}{\xi_2 q^{-\tilde{\mathfrak{n}}_2}};q^2\right)\Theta\left(t^{-2};q^2\right)}\\ &\quad\times\frac{\left(q^2\frac{\xi_2 q^{-\tilde{\mathfrak{n}}_2}}{\xi_1 q^{-\tilde{\mathfrak{n}}_1}};q^2\right)}{\left(t^{-2}\frac{\xi_2 q^{-\tilde{\mathfrak{n}}_2}}{\xi_1 q^{-\tilde{\mathfrak{n}}_1}};q^2\right)}\,{}_2\phi_1\left[t^{-2},t^{-2}\frac{\xi_2 q^{-\tilde{\mathfrak{n}}_2}}{\xi_1 q^{-\tilde{\mathfrak{n}}_1}};q^2\frac{\xi_2 q^{-\tilde{\mathfrak{n}}_2}}{\xi_1 q^{-\tilde{\mathfrak{n}}_1}};q^2,q^2t^2\frac{y_1 q^{-\mathfrak{n}_1}}{y_2 q^{-\mathfrak{n}_2}}\right]\end{aligned} \tag{3.12}$$

The monodromy properties[15] of ${}_2\phi_1$ can the be used to derive:

$$|q|<1\;:\begin{cases}B_1^{II} &= B_1^I\\ B_2^{II} &= B_1^I - B_2^I\end{cases}\qquad\qquad |q|>1\;:\begin{cases}B_1^{II} &= B_1^I + B_2^I\\ B_2^{II} &= -B_2^I\end{cases} \tag{3.13}$$

Note the first of these identities is the statement of mirror symmetry for the holomorphic block. This ensures that, e.g. for $|q|<1$:

$$\left\|B_1^{II}\right\|^2+\left\|B_2^{II}\right\|^2 = B_1^{II}\tilde{B}_1^{II}+B_2^{II}\tilde{B}_2^{II} = B_1^I(\tilde{B}_1^I+\tilde{B}_2^I)+(B_1^I-B_2^I)(-\tilde{B}_2^I) = \left\|B_1^I\right\|^2+\left\|B_2^I\right\|^2 \tag{3.14}$$

This enables us to write:

$$\mathcal{Z}^A_{T[\mathrm{SU}(2)]} = t^{-1+(\mathfrak{n}_1-\mathfrak{n}_2)-(\tilde{\mathfrak{n}}_1-\tilde{\mathfrak{n}}_2)}\left(\frac{\xi_1}{\xi_2}\right)y_1^{-\tilde{\mathfrak{n}}_1}y_2^{-\tilde{\mathfrak{n}}_2}\xi_1^{-\mathfrak{n}_1}\xi_2^{-\mathfrak{n}_2}\left[\left\|B_1^{II}\right\|^2+\left\|B_2^{II}\right\|^2\right] \tag{3.15}$$

Now, under the mirror symmetry transformation on parameters $t\to\frac{1}{t}$, $\{\vec{y},\mathfrak{n}\}\leftrightarrow\{\vec{\xi},\tilde{\mathfrak{n}}\}$, using the fusion rules (2.9), (3.9) and the identification $y_1y_2=\xi_1\xi_2=1$ and $\mathfrak{n}_1+\mathfrak{n}_2=\tilde{\mathfrak{n}}_1+\tilde{\mathfrak{n}}_2=0$, this is precisely the B-twisted index for $T[\mathrm{SU}(2)]$ (2.18).

### 3.3 $T[\mathrm{SU}(N)]$ for $t\to 1$

In this section we prove the mirror symmetry (1.5) of the twisted index in the limit $t\to 1$. First in the A-twist, taking the limit, the vortex partition functions become geometric series, by identifying $k_a^{(s)}$ in (2.8) with $k_a^{(s)}=\sum_{\mu=s}^{N-1} l_a^{(\mu)}$ as in appendix A:

$$\begin{aligned}\lim_{t\to1}\mathcal{Z}^A_{\mathrm{V}} &= \sum_{\{k_a^{(s)}\}}\prod_{s=1}^{N-1}\left(\frac{\xi_s q^{-\tilde{\mathfrak{n}}_s}}{\xi_{s+1}q^{-\tilde{\mathfrak{n}}_{s+1}}}\right)^{\sum_{a=1}^s k_a^{(s)}} = \sum_{\{l_a^{(s)}\geq0\}}\prod_{s=1}^{N-1}\left(\frac{\xi_s q^{-\tilde{\mathfrak{n}}_s}}{\xi_{s+1}q^{-\tilde{\mathfrak{n}}_{s+1}}}\right)^{\sum_{a=1}^s\sum_{\mu=s}^{N-1} l_a^{(\mu)}}\\ &= \sum_{\{l_a^{(s)}\geq0\}}\prod_{s=1}^{N-1}\prod_{a=1}^{s}\prod_{s'=a}^{s}\left(\frac{\xi_{s'} q^{-\tilde{\mathfrak{n}}_{s'}}}{\xi_{s'+1}q^{-\tilde{\mathfrak{n}}_{s'+1}}}\right)^{l_a^{(s)}} = \sum_{\{l_a^{(s)}\geq0\}}\prod_{s=1}^{N-1}\prod_{a=1}^{s}\left(\frac{\xi_a q^{-\tilde{\mathfrak{n}}_a}}{\xi_{s+1}q^{-\tilde{\mathfrak{n}}_{s+1}}}\right)^{l_a^{(s)}}\\ &= \sum_{\{l_{i'}^{j'}\geq0\}}\sum_{i<j}^{N}\left(\frac{\xi_i q^{-\tilde{\mathfrak{n}}_i}}{\xi_j q^{-\tilde{\mathfrak{n}}_j}}\right)^{l_i^{(j)}} = \prod_{i<j}\frac{1}{1-\frac{\xi_i q^{-\tilde{\mathfrak{n}}_i}}{\xi_j q^{-\tilde{\mathfrak{n}}_j}}}\end{aligned} \tag{3.16}$$

[15] Listed in appendix A.4 of [8] which note, are different for $|q|<1$ and $|q|>1$. For example if we choose $|q|<1$ then the latter need to be used for the anti-vortex partition function.

Similarly:

$$\lim_{t\to 1}\mathcal{Z}^A_{\text{aV}}=\prod_{i<j}\frac{1}{1-\frac{\xi_i q^{\tilde{\mathfrak{n}}_i}}{\xi_j q^{\tilde{\mathfrak{n}}_j}}} \tag{3.17}$$

and so the limit for the A-twisted index is:

$$\begin{aligned}\lim_{t\to 1}\mathcal{Z}^A\left[T[\mathrm{SU}(N)]\right]\left(t,q,\vec{y},\mathfrak{n},\vec{\xi},\tilde{\mathfrak{n}}\right)=\sum_{\sigma\in S_N}\Bigg\{&\left[\prod_{i<j}^N\frac{\xi_i}{\xi_j}\right]\left[\prod_{s=1}^N\xi_s^{-\mathfrak{n}_{\sigma(s)}}y_{\sigma(s)}^{-\tilde{\mathfrak{n}}_s}\right]\\&\times\left[\prod_{i<j}^N\frac{1}{1-\frac{\xi_i q^{-\tilde{\mathfrak{n}}_i}}{\xi_j q^{-\tilde{\mathfrak{n}}_j}}}\frac{1}{1-\frac{\xi_i q^{\tilde{\mathfrak{n}}_i}}{\xi_j q^{\tilde{\mathfrak{n}}_j}}}\right]\Bigg\}\end{aligned} \tag{3.18}$$

Now for the B-twist:

$$\begin{aligned}\lim_{t=1}\mathcal{Z}^B_{\text{cl}}\mathcal{Z}^B_{\text{1-loop}}&=\prod_{i<j}^N\left(\frac{y_i}{y_j}\right)\left[\prod_{s=1}^N\xi_s^{-\mathfrak{n}_s}y_s^{-\tilde{\mathfrak{n}}_s}\right]\left[\prod_{i<j}^N\frac{\left(q^2\frac{y_iq^{-\mathfrak{n}_i}}{y_jq^{-\mathfrak{n}_j}};q^2\right)_{\mathfrak{n}_i-\mathfrak{n}_j-1}}{\left(\frac{y_iq^{-\mathfrak{n}_i}}{y_jq^{-\mathfrak{n}_j}};q^2\right)_{\mathfrak{n}_i-\mathfrak{n}_j+1}}\right]\\&=\prod_{i<j}^N\left(\frac{y_i}{y_j}\right)\left[\prod_{s=1}^N\xi_s^{-\mathfrak{n}_s}y_s^{-\tilde{\mathfrak{n}}_s}\right]\left[\prod_{i<j}^N\frac{1}{\left(1-\frac{y_iq^{-\mathfrak{n}_i}}{y_jq^{-\mathfrak{n}_j}}\right)\left(1-\frac{y_iq^{\mathfrak{n}_i}}{y_jq^{\mathfrak{n}_j}}\right)}\right]\end{aligned} \tag{3.19}$$

Note that setting $t=1$ in the vortex partition function/holomorphic block for the B-twist (2.20), from the $b=a$ terms in the last factor, the only non-zero contribution is when all the vortex numbers $\{k_a^{(s)}\}$ are zero.[16] Therefore we have that the vortex partition functions become trivial in this limit.

$$\lim_{t\to 1}\mathcal{Z}^B_V=\lim_{t\to 1}\mathcal{Z}^B_{aV}=1 \tag{3.20}$$

Therefore the limit of the full B-twisted index is:

$$\begin{aligned}&\lim_{t\to 1}\mathcal{Z}^B\left[T[\mathrm{SU}(N)]\right]\left(t,q,\vec{y},\mathfrak{n},\vec{\xi},\tilde{\mathfrak{n}}\right)\\&=\sum_{\sigma\in S_N}\left[\prod_{s=1}^N\xi_s^{-\mathfrak{n}_{\sigma(s)}}y_{\sigma(s)}^{-\tilde{\mathfrak{n}}_s}\right]\left[\prod_{i<j}^N\frac{y_{\sigma(i)}}{y_{\sigma(j)}}\right]\left[\prod_{i<j}^N\frac{1}{\left(1-\frac{y_{\sigma(i)}q^{-n_{\sigma(i)}}}{y_{\sigma(j)}q^{-n_{\sigma(j)}}}\right)\left(1-\frac{y_{\sigma(i)}q^{n_{\sigma(i)}}}{y_{\sigma(j)}q^{n_{\sigma(j)}}}\right)}\right]\\&=\sum_{\sigma\in S_N}\left[\prod_{s=1}^N\xi_s^{-\mathfrak{n}_{\sigma(s)}}y_{\sigma(s)}^{-\tilde{\mathfrak{n}}_s}\right]\left[\prod_{i<j}^N\frac{y_i}{y_j}\right]\left[\prod_{i<j}^N\frac{1}{\left(1-\frac{y_iq^{-n_i}}{y_jq^{-n_j}}\right)\left(1-\frac{y_iq^{n_i}}{y_jq^{n_j}}\right)}\right]\end{aligned} \tag{3.21}$$

The rational factors in both the A and the B twist are obviously mapped to each other under mirror symmetry. Note that:

$$\sum_{\sigma\in S_N}\left[\prod_{s=1}^N\xi_s^{-\mathfrak{n}_{\sigma(s)}}y_{\sigma(s)}^{-\tilde{\mathfrak{n}}_s}\right]=\sum_{\sigma\in S_N}\left[\prod_{s=1}^N\xi_{\sigma^{-1}(s)}^{-\mathfrak{n}_s}y_s^{-\tilde{\mathfrak{n}}_{\sigma^{-1}(s)}}\right]=\sum_{\sigma\in S_N}\left[\prod_{s=1}^N\xi_{\sigma(s)}^{-\mathfrak{n}_s}y_s^{-\tilde{\mathfrak{n}}_{\sigma(s)}}\right] \tag{3.22}$$

[16] Consider the $s=N-1$ terms first with the identification that $k_a^{(N)}=0$.

Therefore we have shown that:

$$\lim_{t\to 1} \mathcal{Z}^A\left[T[\mathrm{SU}(N)]\right]\left(t,q,\vec{y},\mathfrak{n},\vec{\xi},\tilde{\mathfrak{n}}\right) = \lim_{t\to 1} \mathcal{Z}^B\left[T[\mathrm{SU}(N)]\right]\left(t^{-1},q,\vec{\xi},\tilde{\mathfrak{n}},\vec{y},\mathfrak{n}\right) \tag{3.23}$$

## 4 Holomorphic blocks and handsaw quiver varieties

In this section we relate the equivariant $\chi_t$ genera of handsaw quiver varieties to vortex contributions to holomorphic blocks of $T_\rho[\mathrm{SU}(N)]$ theories. In particular, we match the fixed point data in terms of Young tableaux [24] used in the computation of the $\chi_t$ genus to the poles of the holomorphic block integral. We also discuss the geometric interpretation of Chern-Simons levels in this geometrical setup. Finally, we interpret the handsaw quiver variety equivariant $\chi_t$ genus in terms of superconformal quantum mechanics on the $T_\rho[\mathrm{SU}(N)]$ vortex moduli space.

### 4.1 Notation and background

We begin with a brief summary of Laumon space, handsaw quiver varieties and the equivariant $\chi_t$ genus.

**Local Laumon space.** Local Laumon space $\mathfrak{Q}_{\vec{d}}$ is the moduli space of flags of sheaves on $\mathbb{P}^1$. Let $W$ be an $L$ dimensional vector space with basis $\{w_1,\ldots,w_L\}$ and consider:

$$0 \subset \mathcal{W}_1 \subset \ldots \subset \mathcal{W}_{L-1} \subset \mathcal{W}_L = W \otimes \mathcal{O}_{\mathbb{P}^1} \tag{4.1}$$

such that $\mathrm{rank}(\mathcal{W}_k) = k$ and the degree of the sheaves is specified by a vector $\vec{d} = (d_1,\ldots,d_{N-1})$ with $\deg(\mathcal{W}_k) = d_k$. We also impose that $\mathcal{W}_i$ is a vector sub-bundle in a neighbourhood of $\infty \in \mathbb{P}^1$ and the fibre of $\mathcal{W}_i$ at $\infty$ is $\mathrm{span}\{w_1,\ldots w_i\}$. There is a group action on Laumon space $\mathbb{C}^* \times (\mathbb{C}^*)^L$ acting by rotating the $\mathbb{P}^1$ and from the maximal torus of a $\mathrm{GL}(L)$ action on $W$. Ref. [31] describes the fixed points of local Laumon space under this torus action as parametrised by sets of integers $(d_{ij})$ where $0 \le j \le i \le N$ with $d_i = \sum_{j=1}^{i} d_{ij}$ and $d_{kj} \ge d_{ij}$ whenever $i \ge j$. We later match this fixed point data to the holomorphic block poles in section 4.2.

In this paper we also consider a slightly generalised Laumon space which we denote $\mathfrak{Q}^\rho_{\vec{d}}$ where $\rho = (\rho_1,\ldots,\rho_L)$ determines the sheaf ranks via $\mathrm{rank}(\mathcal{W}_k) = \rho_1 + \ldots + \rho_k$. Local Laumon space $\mathfrak{Q}_{\vec{d}}$ is a Kähler variety of dimension $2d_1 + \ldots + 2d_{L-1}$. The generalised space $\mathfrak{Q}^\rho_{\vec{d}}$ has dimension $(\rho_1+\rho_2)d_1 + \ldots + (\rho_{L-1}+\rho_L)d_{L-1}$

**Handsaw quiver varieties.** We recall Nakajima's [24] construction of handsaw quiver varieties. These single-arrow quiver varieties describe the moduli space of vortices in the $T_\rho[\mathrm{SU}(N)]$ theory and are quiver realisations of local Laumon space $\mathfrak{Q}^\rho_{\vec{d}}$. The moduli space of vortices with vortex number $\vec{d} = \{d_1,\ldots,d_{L-1}\}$ with respect to each of the gauge groups in $T_\rho[\mathrm{SU}(N)]$, is specified by the handsaw quiver variety in figure 3. For further details on the vortex interpretation we refer the reader to [32].

The mathematical construction proceeds analogously to the more familiar double-arrow Nakajima quiver varieties [33]. Denote by $V_s \equiv \mathbb{C}^{d_s}$, $s = 1,\ldots,L-1$, the vector space corresponding to the $d_s$ gauge node. Denote by $W_a \equiv \mathbb{C}^{\rho_a}$ the vector space corresponding to the $\rho_a$ flavour node. Also let $V = \bigoplus_{i=1}^{L-1} V_i$, $W = \bigoplus_{i=1}^{L} W_i$.

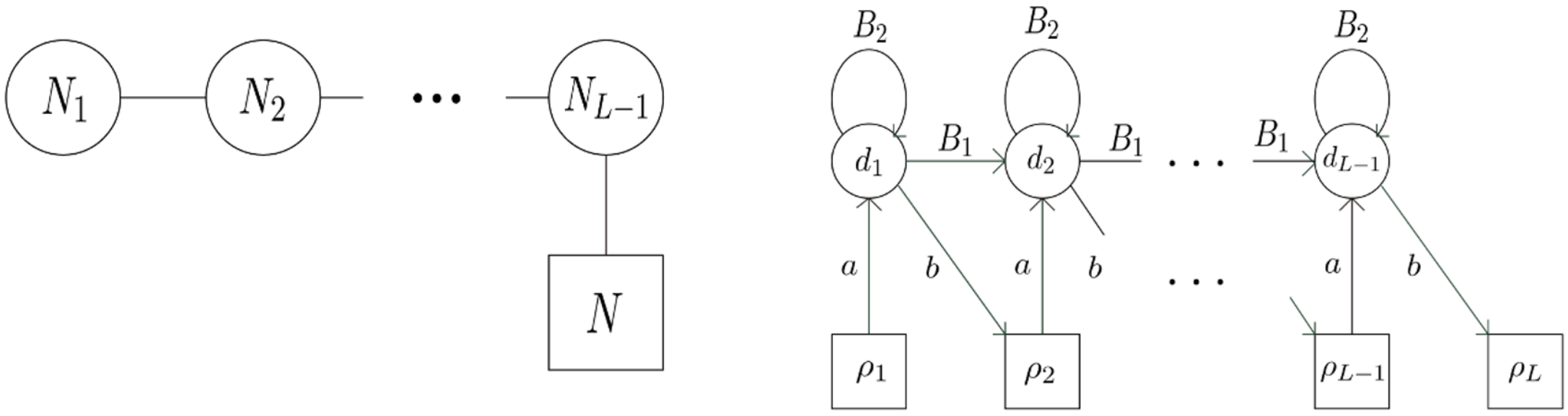

**Figure 3**. The quiver diagram for the 3d $\mathcal{N} = 4$ $T_\rho[\mathrm{SU}(N)]$ theory and its vortex moduli space; the Handsaw quiver variety.

Define:

$$\begin{aligned} B_1 &\in \bigoplus_{i=1}^{L-2} \mathrm{Hom}\,(V_i, V_{i+1}) & \qquad B_2 &\in \bigoplus_{i=1}^{L-1} \mathrm{End}\,(V_i) \\ a &\in \bigoplus_{i=1}^{L-1} \mathrm{Hom}\,(W_i, V_i) & \qquad b &\in \bigoplus_{i=1}^{L-1} \mathrm{Hom}\,(V_i, W_{i+1}) \end{aligned} \tag{4.2}$$

and consider the affine space of all quadruples $(B_1, B_2, a, b)$ of linear data. Defining:

$$\mu(B_1, B_2, a, b) = [B_1, B_2] + ab \in \mathrm{End}(V, V) \tag{4.3}$$

then $\mu^{-1}(0)$ specifies an affine variety with a natural group action of $G = \prod \mathrm{GL}(V_i)$ given by its action on the linear data as:

$$g \in G : (B_1, B_2, a, b) \mapsto (g^{-1}B_1 g, g^{-1}B_2 g, g^{-1}a, bg) \tag{4.4}$$

A point in the space of linear data $(B_1, B_2, a, b)$ is called *stable* if there is no proper graded subspace $S = \bigoplus_{i=1}^{L-1} S_i$ of $V$ stable under $B_1$, $B_2$ and containing $a(W)$, and *costable* if there is no non-zero graded subspace stable under $B_1$, $B_2$ and contained in $\ker(b)$. Nakajima defines the handsaw quiver varieties as:

$$\begin{aligned} \mathscr{L} &= \{(B_1, B_2, a, b) \in \mu^{-1}(0) \,|\, \mathrm{stable}\}/G \\ \mathscr{L}_0 &= \{(B_1, B_2, a, b) \in \mu^{-1}(0)\}/\!\!/G \\ \mathscr{L}_0^{\mathrm{reg}} &= \{(B_1, B_2, a, b) \in \mu^{-1}(0) \,|\, \mathrm{stable\ and\ costable}\}/G \end{aligned} \tag{4.5}$$

Here $/\!\!/$ denotes the affine GIT quotient. There is a projective morphism $\pi : \mathscr{L} \to \mathscr{L}_0$, such that $\pi$ is an isomorphism between $\mathscr{L}_0^{\mathrm{reg}}$ considered as a (possibly empty) open subscheme in both $\mathscr{L}$ and $\mathscr{L}_0$.

Thus, providing $\mathscr{L}_0^{\mathrm{reg}}$ is non-empty, since it is a Zariski open subset of $\mathscr{L}$ and $\mathscr{L}_0$, $\pi$ is a birational morphism and provides a canonical resolution of singularities. The requirement that $\mathscr{L}_0^{\mathrm{reg}}$ is non-empty has a nice physical interpretation. $\mathscr{L}_0^{\mathrm{reg}}$ corresponds to the moduli space of genuine vortices, i.e. those which are not point-like. One can see this from the stratification in [24]:

$$\mathscr{L}_0(\rho,\vec{d}) = \bigsqcup \mathscr{L}_0^{\text{reg}}(\rho,\vec{d}-\vec{d}') \times \text{Sym}^{d'_1}\mathbb{C}\ldots\times \text{Sym}^{d'_{L-1}}\mathbb{C} \tag{4.6}$$

where $\vec{d}' = \{d'_1,\ldots,d'_{L-1}\}$ is such that $d'_s \leq d_s$, analogous to the case for instantons. A sufficient condition for $\mathscr{L}_0^{\text{reg}}$ to be non-empty is to require $\rho_1 \leq \rho_2 \leq \ldots \leq \rho_L$ [24]. Of course, this is the condition imposed on the partition $\rho$ of $N$ specifying the 3d gauge theory $T_\rho[\text{SU}(N)]$ whose vortices are the handsaws with $\dim W_s = \rho_s$, in order for the theory to be 'good'. Of course $T[\text{SU}(N)]$ and its associated vortex moduli spaces, the handsaws with the trivial partition $\rho = (1,\ldots,1)$, obey this condition.

As shown by Nakajima [24], handsaw quiver varieties coincide with local Laumon space $\mathscr{L} = \mathfrak{Q}^\rho_{\vec{d}}$ where the data in the handsaw construction is mapped in the obvious way. The Laumon space of relevance to the $T[\text{SU}(N)]$ theory, and studied by [25], is the case where $\rho = (1,\ldots,1)$ which denote simply $\mathfrak{Q}_{\vec{d}}$.

**Group action on handsaw quiver.** As with local Laumon space there is a torus action on the handsaw quiver. There is an action of $G_w \equiv \prod_{i=1}^L \text{GL}(W_i)$, which acts on the linear data by conjugation, and there is an additional $\mathbb{C}^*$ action given by:

$$(B_1,B_2,a,b) \mapsto (B_1,qB_2,a,qb) \tag{4.7}$$

These commute with equation $\mu = 0$ and the $G$-action and therefore descend to actions on the quotients. Following Nakajima, we fix a decomposition into 1-dimensional subspaces: $W_i = \bigoplus_\alpha W_i^\alpha$ such that $\alpha = 1,\ldots,\dim W_i - \rho_i$. We consider the restriction $G_w$ to the torus $\prod_i \mathbb{T}_i$, $\mathbb{T}_i \subset \text{GL}(W_i)$ preserving these subspaces. We also include the $\mathbb{C}^*$ action to form a larger decomposition-preserving torus action $\mathbb{T}_w \equiv \mathbb{C}^* \times \prod_i \mathbb{T}_i$.

The fixed points of this group action are given by sets of Young-tableaux as described in section 4.2.

**The equivariant $\chi_t$ genus.** The main geometric invariants we study in this paper are the equivariant $\chi_t$ genera of global and local Laumon spaces. Given a group $G$, a space $X$ with a $G$-action and a $G$-equivariant sheaf $E$, the equivariant Euler character is defined by:

$$\chi_G(X;E) \equiv \sum_i (-1)^i \text{ch}_G H^i(X,E) \tag{4.8}$$

Suppose we have an isolated set of fixed points, $X^T$, under a torus $T = (\mathbb{C}^*)^N$ action with associated fugacities $z_1,\ldots,z_N$ then the Euler character can be computed by Grothendieck-Riemann-Roch-Hirzebruch-Atiyah-Singer localisation (the authors used the review [34] for geometric localisation computations):

$$\chi_G(X;E) = \sum_{x\in X^T} \text{ch}_T\left(E_x,q,z\right) \text{PE}\left[\text{ch}_T\left(T^*_x,z\right)\right] \tag{4.9}$$

The $\chi_t$ genus[17] is a sum over equivariant Euler characters where the sheaves are holomorphic j-forms $\Omega^j$:

$$\chi_t(X;t,\{z\}) = \sum_j (-t)^j \chi_G(X,\Omega_X^k) = \sum_{i,j} (-t)^j (-1)^i \mathrm{ch}_G H^i(X,\Omega_X^j) \tag{4.10}$$

In this case the localisation formula becomes:

$$\chi_t(X;t,\{z\}) = \sum_{x\in X^T} \mathrm{PE}\left[(1-t)\mathrm{ch}_T\left(T_x^*,z\right)\right] \tag{4.11}$$

### 4.2 Holomorphic block and $\chi_t$ genera of handsaw quiver varieties

In this section, we give a brief review of $\mathcal{N}=(2,2)$ superconformal quantum mechanics and relate the $\chi_t$ genera of $\mathfrak{Q}^{\rho}_{\vec{d}}$ to superconformal indices of quantum mechanics on the Kähler cones $\mathscr{L}_0$. Further we show that the vortex partition function of the holomorphic block of a $T_\rho[\mathrm{SU}(N)]$ theory is the generating function of superconformal indices/$\chi_t$ genera of its vortex moduli spaces, which are the handsaw quiver varieties.

**Superconformal quantum mechanics.** In recent work [9] by the authors, the superconformal index for supersymmetric $\sigma$ model quantum mechanics with $\mathcal{N}=(2,2)$ superconformal symmetry and target a Kähler cone $X$ is defined.[18] The metric on the Kähler cone is given by:

$$ds^2 = dr^2 + r^2 h_{ij}(\{x\})dx^i dx^j \qquad r\in\mathbb{R}_+ \tag{4.12}$$

where $\{x\}$ are independent of the radial coordinate $r$. The superconformal algebra for this model is $\mathfrak{u}(1,1|2)$. It has bosonic subalgebra:

$$\mathfrak{g}_B = \mathfrak{su}(1,1)\oplus\mathfrak{su}(2)\oplus\mathfrak{u}(1)_{R^I}\oplus\mathfrak{u}(1)_{D^I} \tag{4.13}$$

As is usual, the Hilbert space of the $\sigma$-model is identified with $\Omega(X,\mathbb{C})$, and the symmetry generators above are identified with differential operators on the exterior algebra. Here $\mathfrak{su}(1,1)\simeq\mathfrak{so}(2,1)$ is the conformal algebra, with Cartan generator $D$ realised geometrically as the Lie derivative with respect to the *homothety* $D=r\frac{\partial}{\partial r}$. The $\mathfrak{su}(2)$ subalgebra is a nonabelian R-symmetry, with Cartan generator $J_3$ corresponding to the usual Lefschetz action on forms on a Kähler manifold. The $\mathfrak{u}(1)_{R^I}$ factor with generator $R^I$ is related to the difference $\frac{1}{2}(p-q)$ for forms of bidegree $(p,q)$. The factor $\mathfrak{u}(1)_{D^I}$ lies in the centre of the algebra and the generator $D^I$ corresponds to the Lie derivative with respect to the *Reeb vector field*, defined for Kähler cones as $D^I = I\left(r\frac{\partial}{\partial r}\right)$ ($I$ is the complex structure on the cone). We refer to [9] and references therein for properties of the Reeb vector. An important fact is that there is a 1-1 correspondence between vector fields which act on holomorphic functions on $X$ with positive weight, and Reeb vectors corresponding to Kähler cone metrics [36]. Also as it is central in the superconformal algebra, $\mathfrak{u}(1)_{D^I}$ can mix with global symmetries. Finally the algebra is completed by four supercharges $Q$ of positive dimension and four supercharges $S$ of negative dimension.

[17] Note this is actually the $\chi_{-t}$ genus but we drop the $-$ for notational clarity throughout the rest of this paper.

[18] We use $\mathcal{N}=(2,2)$ to denote the fact that this 1-dimensional $\sigma$ model is the dimensional reduction of the $\mathcal{N}=(2,2)$ $(1+1)$-dimensional $\sigma$ model on a Kähler target, as in [35].

In the aforementioned work, an extensive analysis of BPS representations was carried out, and the superconformal index which recieves contributions only from those states in BPS representations which saturate the unitary bound (corresponding to a positive semi-definite operator $\mathcal{H}$) is defined as:

$$\mathcal{Z}_X^{\text{S.C.}}(t,q,\{z\}) = \text{Tr}\left[(-1)^F e^{-\beta\mathcal{H}} t^{J_3+R^I} q^{D^I} \prod_i z_i^{\mathcal{J}_i}\right] \tag{4.14}$$

where, $\{z_i\}$ are additional fugacities for holomorphic isometries on $X$ generated by $\{\mathcal{J}_i\}$. Note however that a Kahler cone is singular, and the previous definitions in terms of the exterior algebra are not strictly rigorous. To tackle this issue, an equivariant resolution of singularities $\pi : \tilde{X} \to X$ is required, and the S.C. index becomes the equivariant $\chi_t$ genus on the resolved space:[19]

$$\begin{aligned}\mathcal{Z}_X^{\text{S.C.}}(t,q,\{z\}) &= \sum_{p,q=0}^{d_{\mathbb{C}}} (-)^{p+q-d_{\mathbb{C}}}\, t^{p-d_{\mathbb{C}}/2}\, \text{Tr}_{H^q(\tilde{X};\Omega^p_{\tilde{X}})}\left(q^{D^I}\prod_i z_i^{\mathcal{J}_i}\right) \\ &= (-1)^{d_{\mathbb{C}}} t^{-d_{\mathbb{C}}/2}\chi_t\left(\tilde{X};q,\{z\}\right)\end{aligned} \tag{4.15}$$

Much of the work in that paper was to substantiate the validity of this regularisation of the index. In particular, numerous conditions on the Kähler cone were derived to show that (limits of) the index were in fact invariants of the underlying singular Kähler cone, justifying the supposition that it is consistent to assign a spectrum of unitary irreducible representations to quantum mechanics on the cone.

The singular handsaw quiver variety $\mathscr{L}_0$ is in particular a Kähler cone, since it is a fixed point submanifold of a holomorphic isometry acting on the ADHM quiver variety, itself a hyperKähler cone. It has resolution $\pi : \mathfrak{Q}^\rho_{\vec{d}} = \mathscr{L} \to \mathscr{L}_0$. In this section we show that the vortex contributions to the holomorphic block of the $T_\rho[\text{SU}(N)]$ quiver gauge theory are generated by superconformal indices/$\chi_t$ genera on the handsaw quiver varieties, and thus the vortex contribution to the holomorphic block/partition function is related to the count of BPS states in the moduli space quantum mechanics of its vortices. This is analagous to the instanton case where the superconformal index for quantum mechanics on the ADHM quiver variety generates the instanton contribution to the Nekrasov partition function of $\mathcal{N} = 1^*$ supersymmetric SU($N$) Yang-Mills theory on $\Omega$ deformed $\mathbb{C}^2_{q,t} \times S^1$ [37].

**Handsaw superconformal index.** We now proceed to computing the superconformal index of the handsaw quiver variety using the localisation formula (4.11). We recall the fixed point description and character formula from [10]. The $\mathbb{T}_w$ fixed points on $\mathscr{L} = \mathfrak{Q}^\rho_{\vec{d}}$ correspond to tuples of Young diagrams $\vec{Y} = \{Y_{a,\gamma}\}$ corresponding to each $W_a^\gamma$, where the bottom-left corner of $Y_{a,\gamma}$ is shifted such that its $x$-coordinate is $a$, with the restriction that the total number of boxes in the tuple with $x = b$ is $\dim V_b = d_b$. We also need the formula for the character at the tangent space of these fixed points:

$$\mathrm{ch} T^*_{\vec{Y}}\mathscr{L} = \sum_{(a,\gamma),(b,\delta)} \frac{y_{a,\gamma}}{y_{b,\delta}} \Bigg( \sum_{\substack{p\in Y_{a,\gamma}\\ L_{Y_{b,\delta}}(p)=0}} q^{A_{Y_{a,\gamma}}+1} + \sum_{\substack{p\in Y_{b,\delta}\\ L_{Y_{a,\gamma}}(p)=-1}} q^{-A_{Y_{b,\delta}}} \Bigg) \tag{4.16}$$

where $y_{i,\alpha}$ the fugacity corresponding to the action of the sub-torus of $\mathbb{T}_i$ acting on $W_i^\alpha$, and $q$ the aforementiond $\mathbb{C}^*$ action. The leg-length $L_{Y_{b,\delta}}(p)$ of a box $p \in Y_{a,\gamma}$ relative to $Y_{b,\delta}$ is the difference in $x$-coordinate of the right-most box in $Y_{b,\delta}$ in the same row as $p$, minus the $x$-coordinate of $p$. If there are no boxes in the same row, take the difference between $b-1$ and the $x$-coordinate of $p$. The arm-length $A_{Y_{a,\gamma}}(p)$ is the difference in $y$-coordinate of the top box in $Y_{a,\gamma}$ in the same column as $p$, and the $y$-coordinate of $p$.

[19] Note that compared to [9] we have relabelled fugacities as $\tilde{y} \to t$, $\tau \to q$.

The formula for the equivariant $\chi_t$ genus of a general handsaw is then given by:

$$\chi_t(\mathfrak{Q}^\rho_{\vec{d}}) = \chi_t(\mathscr{L}) = \sum_{\vec{Y}} \prod_{(a,\gamma),(b,\delta)} \prod_{\substack{p\in Y_{a,\gamma}\\ L_{Y_{b,\delta}}(s)=0}} \mathrm{PE}\left((1-t)\frac{y_{a,\gamma}}{y_{b,\delta}} q^{A_{Y_{a,\gamma}}+1}\right) \prod_{\substack{p\in Y_{a,\gamma}\\ L_{Y_{b,\delta}}(s)=-1}} \mathrm{PE}\left((1-t)\frac{y_{b,\delta}}{y_{a,\gamma}} q^{-A_{Y_{a,\gamma}}}\right) \tag{4.17}$$

The regularised superconformal index on the singular handsaw $\mathscr{L}_0$, the Kähler cone, is identified as:

$$\mathcal{Z}^{\mathrm{S.C.}}_{\mathscr{L}_0}(t,q,\{y\}) = \left(-t^{-\frac{1}{2}}\right)^{d_{\mathbb{C}}} \chi_t(\mathfrak{Q}^\rho_{\vec{d}})(q,\{y\}) \ , \qquad d_{\mathbb{C}} = \sum_{s=1}^{L-1} d_s(\rho_s + \rho_{s+1}) \tag{4.18}$$

We have identified the $q$ fugacity in (4.17) corresponding to the action (4.7) as the $q$ fugacity in the superconformal index (4.14) grading with respect to a Reeb vector, corresponding to some Kähler cone metric $\mathscr{L}_0$. This is consistent because the action (4.7) grades all holomorphic functions on the handsaw quiver with positive weight, and thus by the aforementioned result of [36] is a valid Reeb vector. This is because the holomorphic functions on the unresolved handsaw correspond to gauge invariant polynomials of the linear data $(B_1, B_2, a, b)$, and thus to gauge invariant paths/cycles on the quiver diagram [38]. Further, from the quiver diagram, all gauge invariant paths pass through either a $B_2$ or $b$, and are thus positively graded under the $q$-action.

**Holomorphic block.** We discussed the holomorphic block of $T[\mathrm{SU}(N)]$ in section 2. The calculation generalises trivially to the case of $T_\rho[\mathrm{SU}(N)]$, essentially just adding an extra index to the fixed points. Physically, the holomorphic block is the partition function of the theory on the disk $D_2 \times_q S^1$ and the vortex contribution is given by an index corresponding to the slow-moving vortex sigma model approximation on the vortex moduli space of the theory. We recall this partition function from section 2 (generalised to the $T_\rho[\mathrm{SU}(N)]$ case):

$$\mathcal{Z}_V[T_\rho[\mathrm{SU}(N)]](q,t,\{y\},\xi) = \sum_{\{k^{(s)}_{a,\gamma}\}} \prod_{s=1}^{L-1} \left( \left(-t^{-\frac{1}{2}}\right)^{(\rho_s+\rho_{s+1})} \frac{\xi_s}{\xi_{s+1}} \right)^{\sum_{a=1}^{s}\sum_{\gamma=1}^{\rho_a} k^{(s)}_{a,\gamma}}$$
$$\times \left( \prod_{\substack{a,b=1,\dots,s\\ \beta=1,\dots,\rho_a\\ \gamma=1,\dots,\rho_b\\ (a,\gamma)\neq(b,\delta)}} \frac{\left(t^{-1}\frac{y_{a,\gamma}}{y_{b,\delta}};q\right)_{k^{(s)}_{a,\gamma}-k^{(s)}_{b,\delta}}}{\left(\frac{y_{a,\gamma}}{y_{b,\delta}};q\right)_{k^{(s)}_{a,\gamma}-k^{(s)}_{b,\delta}}} \right) \left( \prod_{a=1}^{s}\prod_{\gamma=1}^{\rho_b}\prod_{b=1}^{s+1}\prod_{\delta=1}^{\rho_b} \frac{\left(tq\frac{y_{a,\gamma}}{y_{b,\delta}};q\right)_{k^{(s)}_{a,\gamma}-k^{(s+1)}_{b,\delta}}}{\left(q\frac{y_{a,\gamma}}{y_{b,\delta}};q\right)_{k^{(s)}_{a,\gamma}-k^{(s+1)}_{b,\delta}}} \right) \tag{4.19}$$

**N.B.** We have rescaled the parameters:

$$q \to q^2\,, \qquad t \to t^2 \tag{4.20}$$

and we use this definition of $\mathcal{Z}_V(q,t,\{y\},\{\xi\})$ for the rest of the paper.

In the above we sum over all integers such that:

$$\begin{array}{lll} k^{(1)}_{1,\beta} \geq k^{(2)}_{1,\beta} \geq k^{(3)}_{1,\beta} \geq \cdots \geq & k^{(L-1)}_{1,\beta} \geq 0 & \beta=1,\dots,\rho_1 \\ \qquad k^{(2)}_{2,\beta} \geq k^{(3)}_{2,\beta} \geq \cdots \geq & k^{(L-1)}_{2,\beta} \geq 0 & \beta=1,\dots,\rho_2 \\ \qquad\qquad k^{(3)}_{3,\beta} \geq \cdots \geq & k^{(L-1)}_{3,\beta} \geq 0 & \beta=1,\dots,\rho_3 \\ \qquad\qquad\qquad \ddots & & \vdots \\ & k^{(L-1)}_{L-1,\beta} \geq 0 & \beta=1,\dots,\rho_{L-1} \end{array} \tag{4.21}$$

This data specifies a tuple of Young diagrams, where $k^{(s)}_{i,\beta}$ represents the number of boxes with $x$-coordinate $s$ in the Young diagram $Y_{i,\beta}$ in Nakajima's notation. We note these poles correspond directly (in the trivial $\rho$ case) to the local Laumon space description of fixed points described in section 4.1. A given monomial in the F.I. parameters specifies fixed values for $\sum_{i=1}^{s}\sum_{\beta=1}^{\rho_i} k^{(s)}_{i,\beta}$, and we identify:

$$\sum_{i=1}^{s}\sum_{\beta=1}^{\rho_i} k^{(s)}_{i,\beta} = d_s = \mathrm{dim}V_s \tag{4.22}$$

with the condition for the tuple of Young tableau to contribute as a fixed point to a given handsaw quiver $\mathfrak{Q}^{\rho}_{\vec{d}}$ (remember $\rho$ is fixed by the 3d gauge theory $T_\rho[\mathrm{SU}(N)]$.)

We claim that vortex partition function (4.19) generates the superconformal indices of quantum mechanics on its vortex moduli spaces.[20] More precisely:

**Proposition 4.1.**

$$\begin{aligned} \mathcal{Z}_V[T_\rho[\mathrm{SU}(N)]](q,t,\{y\},\{\xi\}) &= \sum_{\vec{d}} \left[ \prod_{s=1}^{N-1} \left(\frac{\xi_s}{\xi_{s+1}}\right)^{d_s} \right] Z_{\mathrm{S.C.}}(\mathfrak{Q}^{\rho}_{\vec{d}},q,t,\{y\}) \\ &= \sum_{\vec{d}} \left[ \prod_{s=1}^{N-1} \left( \left(-t^{-\frac{1}{2}}\right)^{(\rho_s+\rho_{s+1})} \frac{\xi_s}{\xi_{s+1}} \right)^{d_s} \right] \chi_t(\mathfrak{Q}^{\rho}_{\vec{d}},q,\{y\}) \end{aligned} \tag{4.23}$$

[20]The elliptic/4d lift of this statement was recently discussed by the authors of [39].

The proof is reproduced in appendix B. Showing this is an easy but tedious exercise in expanding out and cancelling the $q$-Pochhammer symbols in (4.19). Roughly speaking, the technique is to consider all $q$-Pochhammer symbols depending on pairs of fugacities $y_{a,\gamma}$ and $y_{b,\delta}$ and showing that all terms cancel except for those with leg-lengths 0 and -1, corresponding to the terms which contribute for a given tuple $\vec{Y}$ in (4.17).

We conclude that the disk partition function of the 3d theory is related to a count of BPS states of superconformal quantum mechanics on its vortex moduli spaces.

We note that the relation of the 3d vortex partition function to a refined Witten index of the $\mathcal{N} = (2,2)$ handsaw quiver *gauge* theory, i.e. the 1-dimensional GLSM specified by the handsaw quiver, was anticipated in [40]. The Higgs branch of the 1-dimensional quiver gauge theory is the handsaw quiver variety, and the low energy dynamics is given by the non-linear $\sigma$-model on the Higgs branch. The refined Witten index then coincides with the superconformal index of the non-linear $\sigma$-model.

### 4.3 Chern-Simons levels

In this section we take a brief detour from $\mathcal{N} = 4$ to $\mathcal{N} = 2$ and discuss SQCD$[k,N]$ with a Chern-Simons level $\kappa$. In the geometrical setting we show that Chern-Simons levels arise as line bundles on the local Laumon spaces $\mathfrak{Q}_{\vec{d}}$.

Turning on a CS level introduces a classical term in the twisted index computation which arises after factorising into blocks in the following way (see appendix A for details). The block is:[21]

$$\mathcal{Z}_V[\mathrm{SQCD}[k,N]] = \sum_{\{k_a\geq 0\}} \left((-1)^N t^{-N}\xi\right)^{\sum_{a=1}^k k_a} \prod_{a=1}^{k} y_{f(a)}^{-\kappa d_a} q^{-\kappa d_a(d_a+1)} \\ \times \prod_{\substack{a,b=1\\ a\neq b}}^{k} \frac{\left(t^{-2}\frac{y_{f(a)}}{y_{f(b)}};q^2\right)_{d_a-d_b}}{\left(\frac{y_{f(a)}}{y_{f(b)}};q^2\right)_{d_a-d_b}} \prod_{a=1}^{k}\prod_{i=1}^{N} \frac{\left(t^2q^2\frac{y_{f(a)}}{y_i};q^2\right)_{d_a}}{\left(q^2\frac{y_{f(a)}}{y_i};q^2\right)_{d_a}} \tag{4.24}$$

where $f$ is an injective map $f : \{1,\ldots,k\} \to \{1,\ldots,N\}$.

#### 4.3.1 Line bundles on handsaw quiver varieties

We recall the determinant line bundles on local Laumon space before generalising to the case with non-trivial $\rho$.

**Determinant line bundles.** [41] introduces the determinant line bundles on $\mathfrak{Q}_{\vec{d}}$. In the quiver description as a handsaw these are the familiar tautological line bundles corresponding to each gauge node. We briefly recall the construction.

Points in local Laumon space are flags of sheaves:

$$0 \subset \mathcal{W}_1 \subset \ldots \subset \mathcal{W}_{N-1} \subset \mathcal{W}_N = W \otimes \mathcal{O}_{\mathbb{P}^1} \tag{4.25}$$

To each flag we can associate the fiber of a determinant line bundle $\mathcal{D}_k$ given by the top exterior power of the sections of a particular sheaf $\mathcal{W}_k$, i.e. $\mathcal{D}_k = \det\Gamma(\mathbb{P}^1, \mathcal{W}_k)$. In our

[21] $y$ is rescaled $y \mapsto yt^{-1}$ compared to (A.45) and background fluxes are turned off.

fugacity notation the character of $\mathcal{D}_k$ at a fixed point[22] $(d_{ij})$ is:

$$\mathrm{ch}_T(\mathcal{D}_k|_{(d_{ij})}) = \prod_{j=1}^{k} y_j^{1-d_{kj}} q^{-d_{kj}(d_{kj}-1)} \tag{4.26}$$

where $q$ is the fugacity for the $\mathbb{C}^*$ and $y_1, \ldots, y_N$ are fugacities for $(\mathbb{C}^*)^N$.

**Line bundles for SQCD$[k, N]$.** Now we consider SQCD$[k, N]$. The relevant local map space has $\rho = (k, N-k)$ and the degree is an integer $d \in \mathbb{Z}_{\geq 0}$ (see figure 2), this space is denoted $\mathfrak{Q}_d^{(k,N-k)}$. We slightly generalise the determinant line bundle construction. In this case we have one sub-sheaf:

$$0 \subset \mathcal{W} \subset W \otimes \mathcal{O}_{\mathbb{P}^1} \tag{4.27}$$

where $\mathrm{rk}(\mathcal{W}) = k$, $\deg(\mathcal{W}) = d$ and $\dim(W) = N$. The torus action is $T = \mathbb{C}^* \times (\mathbb{C}^*)^k \times (\mathbb{C}^*)^{N-k}$ with corresponding fugacities $(t, y_1, \ldots, y_k, y_{k+1}, \ldots, y_N)$, the fixed points are labelled by non-negative integers $(d_i)$ such that $d_1 + \ldots + d_k = d$.

We define the determinant line bundle:

$$\mathcal{D} \equiv \det\Gamma(\mathbb{P}^1, \mathcal{W}) \tag{4.28}$$

The corresponding character of the fibre at the fixed point $(d_i)$ is:

$$\mathrm{ch}_T(\mathcal{D}|_{(d_i)}) = \prod_{i=1}^{k} y_i^{(1-d_i)} q^{-d_i(d_i-1)} \tag{4.29}$$

### 4.3.2 Twisted handsaw quiver $\chi_t$ genus

We consider tensor products of the line bundle $\mathcal{D}^{\otimes\kappa}$ and modify the $\chi_t$ genus as follows:[23]

$$\chi_t(\mathfrak{Q}_d^{(k,N-k)}; \mathcal{D}^{\otimes\kappa}) \equiv \sum_j (-1)^j t^j \chi_T(\mathfrak{Q}_d^{(k,N-k)}; \Omega^j \otimes \mathcal{D}^{\otimes\kappa}) \tag{4.30}$$

One can check using the localisation formulae of (4.11) that:

$$\chi_t(\mathfrak{Q}_d^{(k,N-k)}; \mathcal{D}^{\otimes\kappa}) = \sum_{x \in \mathfrak{Q}_d^T} \mathrm{ch}_T(\mathcal{D}_x; t, y)^\kappa \mathrm{PE}[(1-t)\mathrm{ch}_T(T_x^*; t, y)] \tag{4.31}$$

The generating function we consider is:

$$\mathcal{Z}_{\mathfrak{Q}}[\xi, y; q, t; \kappa](\xi, y, q, t) \equiv \sum_{d=0}^{\infty} (q^\kappa \xi (-t)^{-N})^d \chi_t(\mathfrak{Q}_d^{(k,N-k)}; \mathcal{D}^{\otimes\kappa}) \tag{4.32}$$

We observe this matches the holomorphic block (4.24). The $\kappa$-dependent modification to the generating parameter is analogous to the 5d case where the instanton contributions to the 5d $\Omega$ deformed theory arise in the same way.[24] We expect these arguments to generalise to the $T[\mathrm{SU}(N)]$ theory since the Picard group of $\mathfrak{Q}_{\vec{d}}$ is generated by $N-2$ elements [41], and the trivial line bundle so that we have a Chern-Simons level available for each gauge group factor.

[22] the fixed points of $\mathfrak{Q}_{\vec{d}}$ are discussed in section 4.1.

[23] This is analogous to the calculation with a line bundle on *global* Laumon space as in section 5. Ref. [25] describes this invariant as a twisted de Rham complex.

[24] E.g. see equation (2.2) of [42].

## 5 The A twisted index and global Laumon space

In recent work [5, 6] give a geometric interpretation of the $\mathcal{N}=2$ twisted index in terms of the $\chi_t$ genera of vortex moduli spaces — we expand on this geometric interpretation in the $\mathcal{N}=4$ case with the angular momentum refinement, $q$, turned on. In this section we describe how holomorphic factorisation relates to the geometry of global and local Laumon space and the localisation computations of [25].

We elaborate on the correspondence between the Hilbert series and the twisted index in the presence of background flux and angular momentum refinement. The main result of this section is an expression for the Coulomb branch Hilbert series of 3d $\mathcal{N}=4$ theories with background charge, in terms of generating functions of $\chi_t$ genera of Laumon spaces or more generally; holomorphic blocks.

**Global Laumon space and quasimaps.** Global Laumon space $\mathcal{Q}_{\vec{d}}$ is again a moduli space of flags of sheaves on $\mathbb{P}^1$ however compared to local Laumon space we now drop the condition that $\mathcal{W}_i$ is a vector subbundle in a neighbourhood of $\infty \in \mathbb{P}^1$. Local Laumon space is a subset of global Laumon space[25] $\mathfrak{Q}_{\vec{d}} \subset \mathcal{Q}_{\vec{d}}$. We again consider a slightly generalised Laumon space which we denote $\mathfrak{Q}^{\rho}_{\vec{d}} \subset \mathcal{Q}^{\rho}_{\vec{d}}$.

Laumon spaces can be understood as compactifications of spaces of maps,[26] in particular we consider the space of degree $\vec{d}$ algebraic maps $\mathbb{P}^1 \to \mathcal{B}_N$ ($\mathcal{B}_N$ is the complete flag variety which coincides with the Lagrangian core of the Higgs branch of $T[\mathrm{SU}(N)]$) — we denote this space by $\mathcal{Q}^A_{\vec{d}}$. Drinfeld introduced a compactification of this space, denoted $\mathcal{Q}^D_{\vec{d}}$ — this space is compact but may have singularities. Laumon space is then a resolution of singularities $\pi : \mathcal{Q}_{\vec{d}} \to \mathcal{Q}^D_{\vec{d}}$ [44]. Local Laumon space, $\mathfrak{Q}_{\vec{d}}$, arises similarly as a compact resolution of singularities of the space of maps to the complete flags where $\infty$ is mapped to the standard flag.

### 5.1 A twisted index as $\chi_t$ genus of global Laumon space

We recall the geometric localisation calculation of [25]. In that paper they give a formula for the $\chi_t$ genus of global Laumon space:

$$\chi_t(\mathcal{Q}_{\vec{d}}; q, y_1, \ldots, y_N) = \sum_{\substack{\vec{\alpha}+\vec{\beta}=\vec{d} \\ \omega \in S_N}} \chi_t(\mathfrak{Q}_{\vec{\alpha}}; q^{-1}, t, y_1 \ldots, y_N)\chi_t(\mathfrak{Q}_{\vec{\beta}}; q, t, y_1 \ldots, y_N) \prod_{i<j}^{N} \frac{1-ty_i/y_j}{1-y_i/y_j} \tag{5.1}$$

where $\omega \in S_N$ acts by $(y_1, \ldots, y_N) \to (y_{\omega(1)}, \ldots, y_{\omega(N)})$.

[25] See section 4.1 for the definition of local Laumon space.
[26] A useful survey on quasimaps can be found in [43].

Now consider the following generating function of $\chi_t$ genera:[27]

$$\mathcal{Z}_{\mathcal{Q}}[\xi, y; q, t] \equiv \prod_{s=1}^{N-1} \left(\frac{\xi_s}{\xi_{s+1}}\right)^{s(N-s)} \sum_{\vec{d}} \left(t^{-1}\frac{\xi_1}{\xi_2}\right)^{d_1}, \ldots \left(t^{-1}\frac{\xi_{N-1}}{\xi_N}\right)^{d_{N-1}} \chi_t(\mathcal{Q}_{\vec{d}}) \qquad (5.2)$$

Now we recall the A twisted index for the $T[\mathrm{SU}(N)]$ theory (2.11) under the re-scaling $q \to q^{1/2}, t \to t^{1/2}$ — we work with these re-scaled fugacities in the remainder of the paper. With background fluxes turned off:

$$Z^A_{S^2\times_q S^1}[T[\mathrm{SU}(N)]] = \prod_{s=1}^{N-1} \left(\frac{\xi_s}{\xi_{s+1}}\right)^{s(N-s)} \prod_{i<j}^{N} \frac{1 - ty_i/y_j}{1 - y_i/y_j} \mathcal{Z}_{\mathrm{V}} \mathcal{Z}_{\mathrm{aV}} \qquad (5.3)$$

In the previous section we showed we can identify $\mathcal{Z}_{\mathrm{V}}$ and $\mathcal{Z}_{\mathrm{aV}}$ with the generating function of handsaw $\chi_t$ genera i.e.

$$\begin{aligned} \mathcal{Z}_{\mathrm{V}} &= \sum_{\vec{\alpha}} \prod_{s=1}^{N-1} \left(t^{-1}\frac{\xi_s}{\xi_{s+1}}\right)^{\alpha_s} \chi_t(\mathfrak{Q}_{\vec{\alpha}}; q, t) \\ \mathcal{Z}_{\mathrm{aV}} &= \sum_{\vec{\beta}} \prod_{s=1}^{N-1} \left(t^{-1}\frac{\xi_s}{\xi_{s+1}}\right)^{\beta_s} \chi_t(\mathfrak{Q}_{\vec{\beta}}; q^{-1}, t) \end{aligned} \qquad (5.4)$$

So that we have the identification:[28]

| | |
|---|---|
| $Z^A_{S^2\times_q S^1}[T[\mathrm{SU}(N)]]$ | $\mathcal{Z}_{\mathcal{Q}} \sim \sum_d \xi^d \chi_t(\mathcal{Q}_d)$ |
| Higgs branch vacua | $\omega \in S_N$ |
| F.I. parameters | Generating variables of map degree |
| Flavour fugacities | Global symmetries on target |
| $\Omega$ deformation, $q^2$ | $\mathbb{C}^*$ action fugacity, $q$, on Laumon space |
| Adjoint mass, $t^2$ | Homological degree, $t$ |
| Higgs branch core: $\mathfrak{L}[\mathcal{M}_H]$ | Complete flag $\mathcal{B}$ |
| Vortex partition function, $\mathcal{Z}_{\mathrm{V}}$ | $\chi_t$ genera of handsaws, $\sum_\alpha \chi_t(\mathfrak{Q}_\alpha)$ |

With non-trivial $\rho$ the natural generalisation is:

$$\begin{aligned} Z^A_{S^2\times_q S^1}[T_\rho[\mathrm{SU}(N)]] = &\sum_{S_N/S_{\rho_1}\times\ldots\times S_{\rho_L}} \sum_{\vec{\alpha},\vec{\beta}} \prod_{s=1}^{L-1} \left((-t)^{-\frac{1}{2}(\rho_s+\rho_{s+1})}\frac{\xi_s}{\xi_{s+1}}\right)^{\alpha_s+\beta_s} \\ &\times \chi_t(\mathfrak{Q}^\rho_{\vec{\alpha}}; q^{-1}, t)\chi_t(\mathfrak{Q}^\rho_{\vec{\beta}}; q, t) \prod_{h(i)>h(j)} \frac{1 - ty_i/y_j}{1 - y_i/y_j} \end{aligned} \qquad (5.5)$$

where the function $h$ in this equation is a map $h : \{1, \ldots, N\} \to \{1, \ldots, L\}$ defined via $i = \rho_1 + \ldots + \rho_{h(i)}$ for $j = 1, \ldots, \rho_{h(i)}$. The local $\chi_t$ genera can be interpreted as superconformal indices of quantum mechanics on the moduli space of vortices of the theory and the Higgs branch vacua for $T_\rho[\mathrm{SU}(N)]$ are identified with $S_N$ modulo the Levi subgroup specified by $\rho$. It is straightforward to verify this in the SQCD$[k,N]$ case where $\rho=(k,N-k)$ and we find agreement with (2.22). We also make use of this generalisation later in section 6 to compute the Coulomb branch Hilbert series of SQCD$[k,N]$ and find agreement with the known expression.

[27] The work of Bullimore et al. [5, 6] describes the A-twisted index as the equivariant $\chi_t$ genus of the *twisted* quasimap moduli space. We expect that if we had a Laumon-like compactification of twisted quasimaps this generating function with shifted degrees would correspond to the unmodified generating function of the twisted map space compactification. We see this explicitly for the $T[\mathrm{SU}(2)]$ theory in example (6.1).

[28] Up to an overall t pre-factor.

### 5.2 Geometric interpretation of background fluxes

In this section we discuss the inclusion of background flavour, $\mathfrak{n}$, and topological fluxes $\tilde{\mathfrak{n}}$ for the $T[\mathrm{SU}(N)]$ theory.

**Topological flux $\tilde{\mathfrak{n}}$.** Recall, from (2.11), the A twisted index for $T[\mathrm{SU}(N)]$ with topological flux turned on. Using the identifications discussed above we can write this as:[29]

$$\begin{aligned}Z^A_{S^2\times_q S^1}[T[\mathrm{SU}(N)]](\xi,y;q,t;\tilde{\mathfrak{n}}) &= \sum_{\vec{d}}\prod_{s=1}^{N-1}\left(t^{-1}\frac{\xi_s q^{-\tilde{\mathfrak{n}}_s}}{\xi_{s+1}q^{-\tilde{\mathfrak{n}}_{s+1}}}\right)^{d_s} \\ \times \sum_{\substack{\vec{\alpha}+\vec{\beta}=\vec{d}\\ \omega\in S_N}} &\left(\prod_{s=1}^{N-1} q^{-\beta_s(\tilde{\mathfrak{n}}_{s+1}-\tilde{\mathfrak{n}}_s)}\right)\left(\prod_{s=1}^{N} y_s^{\tilde{\mathfrak{n}}_s}\right)\chi_t(\mathfrak{Q}_{\vec{\alpha}};q,t)\chi_t(\mathfrak{Q}_{\vec{\beta}};q^{-1},t)\prod_{i<j}\frac{1-ty_i/y_j}{1-y_i/y_j}\end{aligned} \tag{5.6}$$

where the Weyl group acts by $\omega: y_i\to y_{\omega(i)}$.

Now, given $\lambda\in\mathbb{Z}^N$ one can define a line bundle $\mathcal{O}(\lambda)$ on $\mathcal{Q}_{\vec{d}}$ (more details on this construction can be found in [45]). Ref. [25] consider the equivariant Euler character of the holomorphic forms twisted by this line bundle:

$$\chi_t(\mathcal{Q}_{\vec{d}};\mathcal{O}(\lambda)) \equiv \sum_{i,j}(-1)^{i+j}t^j \mathrm{ch}_T H^i(\mathcal{Q}_{\vec{d}},\Omega^j\otimes\mathcal{O}(\lambda)) \tag{5.7}$$

This object is computed in *loc. cit.* and we can then make the identification:

$$Z^A_{S^2\times_q S^1}[T[\mathrm{SU}(N)]](\xi,y;q,t;\tilde{\mathfrak{n}}) = \sum_{\vec{d}}\prod_{s=1}^{N-1}\left(t^{-1}\frac{\xi_s q^{-\tilde{\mathfrak{n}}_s}}{\xi_{s+1}q^{-\tilde{\mathfrak{n}}_{s+1}}}\right)^{d_s}\chi_t(\mathcal{Q}_{\vec{d}};\mathcal{O}(\lambda=\tilde{\mathfrak{n}})) \tag{5.8}$$

where the topological flux $\tilde{\mathfrak{n}}$ is identified with the line bundle $\lambda$. Comparing with (5.2), the generating function parameters/F.I. parameters are also "shifted" according to $\xi_i\to q^{-\tilde{\mathfrak{n}}_i}\xi_i$.

**Flavour flux $\mathfrak{n}$.** Recall (2.10) where we see that flavour flux enters as a shifting of the flavour fugacities in the $\chi_t$ genus contributions:

$$\chi_t(\mathfrak{Q}_\alpha;q,t,y)\to\chi_t\left(\mathfrak{Q}_\alpha;q,t,yq^{-\mathfrak{n}}\right) \tag{5.9}$$

From the local Laumon space perspective this is a redefinition of the torus action where the action on the base $\mathbb{P}^1$ is mixed with global symmetries on the target $\mathcal{B}_N$.

[29] Up to a $t$ pre-factor: $t^{-N/2}\prod_{s=1}^N t^{1/2(N-2s+1)\tilde{\mathfrak{n}}_s-s/2}$.

Flavour flux also introduces Weyl dependence in the generating function parameters $\xi$ so that different handsaw quivers contribute at different vortex numbers depending on the choice of flavour flux:

$$Z^A_{S^2\times_q S^1}[T[\mathrm{SU}(N)]]\,(\xi,y;q,t;\mathfrak{n}) = \sum_{\vec{d}} \prod_{s=1}^{N-1}\left(t^{-1}\frac{\xi_s}{\xi_{s+1}}\right)^{d_s} \sum_{\substack{\vec{\alpha}+\vec{\beta}=\vec{d}\\ \omega\in S_N}} \left(\prod_{s=1}^{N}\xi_s^{\mathfrak{n}_s}\right)\chi_t\left(\mathfrak{Q}_\alpha;q,t,yq^{-\mathfrak{n}}\right)$$
$$\times\,\chi_t\left(\mathfrak{Q}_\beta;q^{-1},t,yq^{\mathfrak{n}}\right)\prod_{i<j}^{N}\frac{\left(q\frac{y_iq^{-\mathfrak{n}_i}}{y_jq^{-\mathfrak{n}_j}};q\right)_{\mathfrak{n}_i-\mathfrak{n}_j-1}}{\left(tq\frac{y_iq^{-\mathfrak{n}_i}}{y_jq^{-\mathfrak{n}_j}};q\right)_{\mathfrak{n}_i-\mathfrak{n}_j-1}} \tag{5.10}$$

where the Weyl group acts by $\omega: y_i\to y_{\omega(i)}$ and $\omega:\mathfrak{n}_i\to\mathfrak{n}_{\omega(i)}$. It would be interesting to understand the geometrical interpretation of the B twisted index where we expect a mirror dual picture where flavour flux $\mathfrak{n}$ enters as a line bundle on the global vortex moduli space.

### 5.3 Twisted indices and the Hilbert series

In recent work [7] the A and B twisted index, without $\Omega$ deformation $q\to 1$, are shown to coincide with the Higgs branch and Coulomb branch Hilbert series respectively. We discuss how this correspondence is modified when $q\neq 1$ and in the presence of background flavour flux $\mathfrak{n}$.

**Hilbert series.** 3d $\mathcal{N}=4$ theories have a Higgs branch, $\mathcal{M}_H$, and a Coulomb branch, $\mathcal{M}_C$, which are smooth hyperKähler varieties when suitably generic F.I. parameters and real masses are turned on. Physically, the Hilbert series computes chiral operators graded by the global symmetries of the theory. Geometrically, the Hilbert series of either space is defined to the equivariant (with respect to global symmetries) Euler character of the structure sheaf:[30]

$$Z_{\text{H.S.}}[\mathcal{M}] = \chi_G(\mathcal{M};\mathcal{O}_\mathcal{M}) \tag{5.11}$$

In this formalism background baryon number/background magnetic charge enters as a line bundle on the moduli space:

$$Z_{\text{H.S.}}[\mathcal{M};\lambda] = \chi_G(\mathcal{M};\mathcal{O}_\mathcal{M}(\lambda)) \tag{5.12}$$

For more details on the Hilbert series see [46] or for a review [47]. The Hilbert series can then be computed by equivariant localisation as in (4.11).

Often the Higgs branch arises as the resolution of a GIT quotient:

$$\mathcal{M}^0_H = \mathbb{C}[\mu^{-1}(0)]/\!\!/G_{\text{gauge}} \tag{5.13}$$

When the resolution is suitably well-behaved the ring of holomorphic function is preserved under the resolution and the Hilbert series can be computed by purely representation theoretic means, using a classical Molien integral (see for example [48]). The poles of this integral correspond to fixed points on the Higgs branch and background charge is realised as a gauge variable insertion $\sim x^{\mathfrak{n}}$.

[30] In cases there the higher cohomology vanishes and this equivariant character reduces to a character of the zeroth cohomology.

**B-twist as Molien integral.** [7] shows that the B-twist index corresponds to the Molien integral and so to the Higgs branch Hilbert series. We now focus on the $T[\mathrm{SU}(N)]$ theory. Recall the B-twist integral in the zero flux sector:

$$\begin{aligned}\oint \frac{d\xi}{\xi} Z^B[T[\mathrm{SU}(N)]]\,(q,t,y,\tilde{\mathfrak{n}}) = \oint_{S^1} \prod_{s=1}^{N-1}\prod_{a=1}^{s} \frac{dx_a^{(s)}}{2\pi i x_a^{(s)}} \prod_{s=1}^{N-1}\prod_{a=1}^{s} (x_a^{(s)})^{\tilde{\mathfrak{n}}_s-\tilde{\mathfrak{n}}_{s+1}} \\ \times \prod_{s=1}^{N-1}\left(1-\frac{x_a^{(s)}}{x_b^{(s)}}\right)\left(1-t^2\frac{x_a^{(s)}}{x_b^{(s)}}\right) \\ \times \prod_{s=1}^{N-1}\prod_{a=1}^{s}\prod_{b=1}^{s+1} \frac{1}{1-tx_a^{(s)}/x_b^{(s+1)}} \frac{1}{1-tx_b^{(s+1)}/x_a^{(s)}}\end{aligned} \tag{5.14}$$

With the $q$ deformation we note that in the B-twist integrand in the zero $\xi$ sector, $q$ dependence cancels and we land on the Molien integral for the Coulomb branch Hilbert series. Indeed from the evaluation of the B twisted index (2.18) we observe in the $\xi^0$ sector the Hall-Littlewood formula of [16].

Now, using the mirror symmetry of the A and the B twist,[31] we derive the following expression for the Coulomb branch Hilbert series of $T[\mathrm{SU}(N)]$ in terms of the A twisted index holomorphic block expansion:[32]

$$Z_{\mathrm{H.S.}}[\mathcal{M}_C;\mathfrak{n}] = \oint \prod_{s=1}^{N-1} \frac{d\tilde{y}_s}{\tilde{y}_s} \mathbb{B}(q,t;yq^{-\mathfrak{n}},\xi)\mathbb{B}(q^{-1},t;yq^{\mathfrak{n}},\xi) \tag{5.15}$$

In the $T_\rho[\mathrm{SU}(N)]$ case that we focus on in this work, we have shown that the holomorphic block can be given a geometric interpretation in terms of Laumon space. Geometrically, 3d mirror symmetry then relates the equivariant Euler character of a line bundle on the Coulomb branch to the "shifted" $\chi_t$ genus of the moduli space of maps to the Higgs branch. We make this precise in the following section in the case of zero background flux.

## 6 Poincaré polynomial limit

In this section we focus on the A twisted index in the absence of background flavour and topological flux $\mathfrak{n} = \tilde{\mathfrak{n}} = 0$. We show the index is then independent of global symmetries and we provide a geometrical interpretation of this independence. Independence of $q$, the $\Omega$ deformation parameter, means that we are free to send $q \to 0$ and we show that in this limit the $\chi_t$ genera of the non-compact handsaws/local Laumon spaces become the Poincaré polynomials of their compact cores — from the point of view of the index factorisation (2.11) this is a surprising cancellation of $q$ dependence and is unique to the twisted index gluing prescription. It is equally valid to turn off the $\Omega$ deformation, $q \to 1$, where we recover the results of [7] that demonstrate the A twisted index coincides with the Coulomb branch Hilbert series.

[31] We have shown this explicitly for $T[\mathrm{SU}(2)]$ and the $t \to 1$ limit of $T[\mathrm{SU}(N)]$ in section 3 but expect this to hold more generally.

[32] Where $\tilde{y}_s = y_s/y_{s+1}$.

### 6.1 $\chi_t$ genus and Poincaré polynomial

The work of [26] shows when $X$ is compact, symplectic and admits a Hamiltonian circle action (this is true of the Kähler global Laumon space/handsaw quiver variety) with isolated fixed points, the equivariant $\chi_t(X)$ genus is independent of global fugacities and coincides with the ungraded genus. Furthermore [49] explains how in this case the $\chi_t$ genus and Poincaré polynomial essentially coincide: $\chi_t(X) = P_{t^{1/2}}(X)$.[33] This is essentially a consequence of both invariants being computed from the same Białynicki-Birula fixed point formula.

Since the A-twisted index without background flux is identified with a generating function of $\chi_t$ genera (with no line bundle) of global Laumon space, it is then independent of the flavour fugacities $y_i$ and the $\Omega$ parameter $q$, and coincides with the generating function of Poincaré polynomials of global Laumon space.

#### 6.1.1 Poincaré polynomial of handsaw

Given independence of $q$ we now consider the limit $q \to 0$ of the index and take this limit through the holomorphic blocks/$\chi_t$ genera of handsaw quivers. The equivariant $\chi_t$ genus of the handsaw quiver was computed in (4.17).

We note by the arguments of section 4.2 that the action corresponding to $q$ is generated by a valid Reeb vector for some Kähler cone metric on the singular handsaw quiver variety $\mathscr{L}_0$. Laumon space $\mathfrak{Q}_{\vec{d}}$ is a resolution of the singular handsaw. Proposition 5.1 of [9] then tells us the $q \to 0$ limit of the $\chi_t$ genus of Laumon space, or the superconformal index for quantum mechanics on the singular Handsaw, essentially gives the Poincaré polynomial of the preimage of the Kähler cone singularity, and so:[34]

$$\lim_{q\to 0} \mathcal{Z}^{\vec{d}}_{\mathrm{V}}(y,\xi;q,t) = \lim_{q\to 0} \chi_t(\mathfrak{Q}_{\vec{d}}) = P_{t^{1/2}}(\mathfrak{Q}_{\vec{d}}) \tag{6.1}$$

Similarly, for the conjugate block we have:

$$\lim_{q\to 0} \mathcal{Z}^{\vec{d}}_{\mathrm{aV}}(y,\xi;q^{-1},t) = \lim_{q\to 0} \chi_t(\mathfrak{Q}_{\vec{d}}) = t^{\dim(\mathfrak{Q}_d)} P_{t^{-1/2}}(\mathfrak{Q}_{\vec{d}}) \tag{6.2}$$

We now recall Nakajima's generating function over degree of the Poincaré polynomials of handsaw quiver varieties (theorem 4.4 in [24]). In our notation and with GL$(N)$ rather than SL$(N)$ fugacities:

$$\sum_{d} \left(\frac{\xi_1}{\xi_2}\right)^{d_1} \cdots \left(\frac{\xi_{N-1}}{\xi_N}\right)^{d_{N-1}} P_{t^{1/2}}(\mathfrak{Q}^{\rho}_{\vec{d}}) = \prod_{i<j}^{N} \prod_{k=1}^{\rho_i} \frac{1}{1 - t^{\rho_i + \ldots + \rho_{j-1} - k}\xi_i/\xi_j} \tag{6.3}$$

**Example 6.1.** We illustrate the idea of this section with a simple example, the A-twisted index of the abelian $T[\mathrm{SU}(2)]$ theory — from the above arguments this is the generating function of $\chi_t$ genera of global Laumon space with $N = 2$. In this example we can be explicit with the various map spaces. (5.2) becomes:

$$Z^A_{S^2\times_q S^1}[T[\mathrm{SU}(2)]] = \xi \sum_{d=0}^{\infty}(t^{-1}\xi)^d \sum_{\alpha+\beta=d}\sum_{\omega\in S_2}\frac{1-ty_1/y_2}{1-y_1/y_2}\chi_t(\mathfrak{Q}_\alpha)\chi_t(\mathfrak{Q}_\beta) \tag{6.4}$$

[33] The right hand side is still a polynomial in $t$ and not $t^{1/2}$ since the equality implies that the odd homology on $X$ vanishes.

[34] $P_t(\mathfrak{Q})$ here is shorthand for $P_t(\pi^{-1}(0))$, the Poincaré polynomial of the pre-image of the singularity under the resolution $\pi$.

Sending $q \to 0$ through the sum, the $\chi_t$ genera become Weyl group independent and it's straightforward to check:

$$Z^A_{S^2\times_q S^1}[T[\mathrm{SU}(2)]] = \xi(1+t)\left(\sum_{\alpha=0}^{\infty}(\xi t^{-1})^\alpha P_{t^{1/2}}(\mathfrak{Q}_\alpha)\right)\left(\sum_{\beta=0}^{\infty}(\xi t)^\beta P_{t^{-1/2}}(\mathfrak{Q}_\beta)\right) \tag{6.5}$$

For the $T[\mathrm{SU}(2)]$ theory, the relevant geometry is the space of algebraic maps $\mathbb{P}^1 \to \mathbb{P}^1$ — we found [50] particular useful for understanding the map space in this simple example. In coordinates $[z_1 : z_2]$ for the base and $[w_1 : w_2]$ for the target, algebraic maps $\mathbb{P}^1 \to \mathbb{P}^1$ of degree d are specified by 2 homogeneous polynomials $[w_1(z_1, z_2) : w_2(z_1, z_2)]$ each of degree $d$. $w$ and $w'$ define the same map if there exists $\alpha \neq 0$ such that $w' = \alpha w$ and so the coefficients of the polynomials can be compactified to projective space:

$$\mathcal{Q}_d = \mathbb{P}^{2d+1} \tag{6.6}$$

The corresponding handsaw quivers/local map spaces can be shown[35] to be $\mathfrak{Q}_d = \mathbb{C}^d$ and so the Poincaré polynomials of these are trivial $P_{t^{1/2}}(\mathbb{C}^d) = 1$. The Poincaré polynomial of $\mathbb{P}^{2d+1}$ is $\sum_{i=0}^{2d+1} t^i$. Indeed we observe the expected factorisation:

$$\begin{aligned} Z^A_{S^2\times_q S^1}[T[\mathrm{SU}(2)]] &= \xi(1+t)\left(\sum_{\alpha=0}^{\infty}(\xi t^{-1})^\alpha P_t(\mathbb{C}^\alpha)\right)\left(\sum_{\beta=0}^{\infty}(\xi t)^\beta P_{t^{-1}}(\mathbb{C}^\beta)\right) \\ &= \xi(1+t)\sum_{d=0}^{\infty}(\xi t^{-1})^d\left(\sum_{i=0}^{d} t^{2i}\right) \\ &= \sum_{d=0}^{\infty}(\xi t^{-1})^d\left(\xi\sum_{i=0}^{2d+1} t^i\right) = \xi\sum_{d=0}^{\infty}(t^{-1}\xi)^d P_{t^{1/2}}(\mathbb{P}^d) \end{aligned} \tag{6.7}$$

## 6.2 Examples

### 6.2.1 $T[\mathrm{SU}(N)]$

We now consider the $T[\mathrm{SU}(N)]$ theory. From the discussion in the previous subsection and subsection 5.1, the $\chi_t$ genera of global Laumon spaces are identified with Poincaré

[35] To see this, observe the coordinate ring of $\mathscr{L}_0$ is freely generated and so the resolution $\mathscr{L}$ remains as $\mathbb{C}^d = \mathrm{Spec}\,\mathbb{C}[x_1, \ldots, x_s]$. The dimension follows from subsection 4.1.

polynomials and we have:

$$\begin{aligned}Z^A_{S^2\times_q S^1}[T[\mathrm{SU}(N)];t,\xi] &= \prod_{i<j}^N \left(\frac{\xi_i}{\xi_j}\right)\sum_d \left(t^{-1}\frac{\xi_1}{\xi_2}\right)^{d_1}\dots\left(t^{-1}\frac{\xi_{N-1}}{\xi_N}\right)^{d_{N-1}} P_{t^{1/2}}(\mathcal{Q}_d)\\ &= \prod_{i<j}^N \left(\frac{\xi_i}{\xi_j}\right)\sum_{\omega\in S_N}\sum_{\alpha,\beta}\left(\prod_{i<j}^N \frac{1-ty_i/y_j}{1-y_i/y_j}\left(t^{-1}\frac{\xi_1}{\xi_2}\right)^{\alpha_1+\beta_1}\dots\left(t^{-1}\frac{\xi_{N-1}}{\xi_N}\right)^{\alpha_{N-1}+\beta_{N-1}}\right.\\ &\qquad\qquad \left.\times\,\chi_t(\mathfrak{Q}_\alpha;q^{-1},t,y)\chi_t(\mathfrak{Q}_\beta;q,t,y)\right) \end{aligned}\tag{6.8}$$

Taking $q\to 0$ the blocks become Weyl independent (Poincaré polynomials are independent of global symmetries) and the index factorises fully:

$$\begin{aligned}Z^A_{S^2\times_q S^1}[T[\mathrm{SU}(N)];t,\xi] &= \prod_{i<j}^N \left(\frac{\xi_i}{\xi_j}\right)\left(\sum_{\omega\in S_N}\prod_{i<j}\frac{1-ty_i/y_j}{1-y_i/y_j}\right)\\ &\quad\times\left(\sum_\alpha \left(t\frac{\xi_1}{\xi_2}\right)^{\alpha_1}\dots\left(t\frac{\xi_{N-1}}{\xi_N}\right)^{\alpha_{N-1}}P_{t^{-1/2}}(\mathfrak{Q}_\alpha)\right)\\ &\quad\times\left(\sum_\beta \left(t^{-1}\frac{\xi_1}{\xi_2}\right)^{\beta_1}\dots\left(t^{-1}\frac{\xi_{N-1}}{\xi_N}\right)^{\beta_{N-1}}P_{t^{1/2}}(\mathfrak{Q}_\beta)\right)\end{aligned}\tag{6.9}$$

Substituting the generating functions (6.3) in the case $\rho=(1,\dots,1)$ we find:[36]

$$Z^A_{S^2\times_q S^1}[T[\mathrm{SU}(N)];t,\xi] = \prod_{i<j}^N\left(\frac{\xi_i}{\xi_j}\right)\left(\sum_{\omega\in S_N}\prod_{i<j}\frac{1-ty_i/y_j}{1-y_i/y_j}\right)\prod_{i<j}\frac{1}{1-t\xi_i/\xi_j}\frac{1}{1-t^{-1}\xi_i/\xi_j}\tag{6.10}$$

We can re-write this to observe the relation to the Hall-Littlewood formula for the Coulomb branch Hilbert series in the case of zero background GNO charge (equation (3.3) in [46]).

$$Z^A_{S^2\times_q S^1}[T[\mathrm{SU}(N)];t,\xi] = \left(\prod_{i,j=1}^N\frac{1}{1-t\xi_i/\xi_j}\right)(1-t)^N\left(\sum_{\omega\in S_N}\prod_{i<j}\frac{1-ty_i/y_j}{1-y_i/y_j}\right)\tag{6.11}$$

The last factor is in fact $y$ independent due to the following identity from e.g. Macdonald [51]:

$$\sum_{\omega\in S_N}\prod_{i<j}^N\frac{1-ty_i/y_j}{1-y_i/y_j} = \prod_{i=1}^N\frac{1-t^i}{1-t}\tag{6.12}$$

We also note that, reassuringly, this agrees with the generating function of Poincaré polynomials of Laumon space computed by Finkelberg and Kuznetsov [52].

[36] We rescale $\xi_i\to t^i\xi_i$ and $\xi_i\to t^{-i}\xi_i$ in each block respectively.

### 6.2.2 SQCD$[k, N]$

We now consider SQCD$[k, N]$. The Higgs branch vacua are labelled by $\omega \in S_N/W$ where $W$ denotes the Weyl group of the Levi subgroup, which is $W = S_k \times S_{N-k}$ for SQCD$[k, N]$. The relevant handsaw quiver data is $\rho = (k, N-k)$ — see figure 2 for the quiver for the 3d theory and its vortex moduli space.

The localisation computation on the generalised global Laumon space $\mathcal{Q}_d^{(k,N-k)}$ generalises to (as in (5.5)):

$$\begin{aligned} Z^A_{S^2\times_q S^1}[\text{SQCD}[k,N]] &= \sum_d (t^{-\frac{N}{2}}\xi)^d P_{t^{1/2}}(\mathcal{Q}^\rho_d) \\ &= \sum_{S_N/W}\sum_{\alpha,\beta}\left(\prod_{j=1}^{k}\prod_{i=k+1}^{N}\frac{1-ty_i/y_j}{1-y_i/y_j}(\xi t^{-\frac{N}{2}})^\alpha \chi_t(\mathfrak{Q}^\rho_\alpha;q^{-1},t,y)(\xi t^{-\frac{N}{2}})^\beta \chi_t(\mathfrak{Q}^\rho_\beta;q,t,y)\right) \end{aligned} \tag{6.13}$$

Now taking $q \to 0$ we again obtain the factorisation:[37]

$$\stackrel{q\to 0}{=} \left(\sum_{S_N/W}\prod_{j=1}^{k}\prod_{i=k+1}^{N}\frac{1-ty_i/y_j}{1-y_i/y_j}\right)\left(\sum_\alpha (\xi t^{\frac{N}{2}})^\alpha P_{t^{-1/2}}(\mathfrak{Q}^\rho_\alpha)\right)\left(\sum_\beta (\xi t^{-\frac{N}{2}})^\beta P_{t^{1/2}}(\mathfrak{Q}^\rho_\beta)\right) \tag{6.14}$$

The appropriate generating function is (from (6.3)):

$$\sum_d (t^{-\frac{N}{2}}\xi)^d P_t(\mathfrak{Q}_d) = \prod_{i=1}^{k}\frac{1}{1-t^{-\frac{N}{2}+k-i}\xi} \tag{6.15}$$

and it's straightforward to check:

$$\sum_{S_N/W}\prod_{j=1}^{k}\prod_{i=k+1}^{N}\frac{1-ty_i/y_j}{1-y_i/y_j} = \prod_{j=1}^{k}\frac{1-t^{N+1-j}}{1-t^j} \tag{6.16}$$

This gives:

$$Z^A_{S^2\times_q S^1}[\text{SQCD}[k,N]] = \prod_{i=1}^{k}\frac{1-t^{N+1-j}}{(1-t^j)(1-\xi t^{-\frac{N}{2}+i-1})(1-\xi t^{\frac{N}{2}-i+1})} \tag{6.17}$$

We observe this matches the Hilbert series for the Coulomb branch of SQCD$[k, N]$ (equation 5.3 in [16]).

## 7 Further directions

In this work we have interpreted the factorisation of the topologically twisted index of 3d $\mathcal{N} = 4$ theories in terms of geometric invariants of, suitably generalised, Laumon spaces. We have also made proposals of how to include background fluxes and Chern-Simons levels in this geometric setting. In particular, this geometrical setup has allowed us to reinterpret the Coulomb branch Hilbert series with background flux as a particular integral projection (5.15) of the generating function of $\chi_t$ genera of global Laumon spaces in the $T_\rho[\mathrm{SU}(N)]$ case, and holomorphic blocks more generally.

[37] We have used the more general dimension formula: $\dim(\mathfrak{Q}_d^{(k,N-k)}) = dN$ from subsection 4.1.

We have made crucial use of the fact that we have an explicit quiver description of the vortex moduli spaces for the theories considered — namely the handsaw quiver. It would be interesting to understand these moduli spaces and their $\chi_t$ genera moving beyond the class of $T_\rho[\mathrm{SU}(N)]$ theories.

In future work we plan to explore the representation theoretic interpretation of the $\chi_t$ genus/twisted index in more examples. In particular it would be interesting to extend the result (5.15) to the class of affine $\mathcal{N}=4$ quiver gauge theories — since in certain cases [53] these are expected to flow to $\mathcal{N}=2$ Chern-Simons theories with known $\mathrm{AdS}_4$ duals.

## Acknowledgments

The authors would like to thank Mathew Bullimore, Ami Hanany, Chiung Hwang and Alberto Zaffaroni for helpful discussions. This work has been partially supported by STFC consolidated grant ST/P000681/1.

## A Detailed calculations of twisted indices

Here we review the result of the localisation derived in [1] to compute the topologically twisted index of a 3d $\mathcal{N}=2$ theory. We give explicit calculations for the $T[\mathrm{SU}(N)]$ theory, and SQCD$[k,N]$, which generalise straightforwardly to the case of $T_\rho[\mathrm{SU}(N)]$. Given a 3d $\mathcal{N}=2$ theory with a $\mathrm{U}(1)_R$ symmetry and gauge group $G$, the topologically twisted index defined by:

$$Z^{A,B}_{S^2\times_q S^1}(q,t,\{z\},\{n\}) = \mathrm{Tr}_{S^2}\left((-1)^F\, q^{2L_\phi} t^{Q_t} \prod_i z_i^{Q_i}\right) \tag{A.1}$$

can be localised to BPS configurations, resulting in a finite dimensional contour integral equivalent to the Jeffrey-Kirwan residue (using the notation of [7]):

$$\begin{aligned} Z_{S^2\times_q S^1}(q,t,\{z\},\{n\}) &= \frac{1}{|W|}\sum_{\mathfrak{m}} \oint_{\mathrm{JK}} \frac{dx}{2\pi i x} Z_{\mathfrak{m}}(x,q,t,\{z\},\{n\}) \\ &\equiv \frac{(-1)^{\mathrm{rk}\,G}}{|W|}\sum_{\mathfrak{m}}\sum_{x_*\in\mathfrak{M}_{\mathrm{sing}}} \underset{x=x_*}{\text{JK-Res}}\,(Q_{x_*},\eta)\left[Z_{\mathfrak{m}}(x,q,t,\{z\},\{n\})\left(\frac{dx}{x}\right)^{\mathrm{rk}\,G}\right] \\ &\quad + \text{Boundary Contributions} \end{aligned} \tag{A.2}$$

Here $x=e^{iu}$ where $u$ parametrises the set of bosonic zero modes, lying in $H\times\mathfrak{h}$ where $H$ is the maximal torus of $G$ and $\mathfrak{h}$ the corresponding Cartan subalgebra. $\mathfrak{m}$ is the flux of the gauge field through $S^2$, and lives in the co-root lattice $\Gamma_{\mathfrak{h}}$. Here $\{z\}\equiv\{y,\xi\}$ denote (exponentiated) masses and F.I. parameters for the hypermultiplet flavour symmetry and topological symmetry respectively, and $\{n\}\equiv\{\mathfrak{n},\tilde{\mathfrak{n}}\}$ the corresponding fluxes. The integrand $\mathcal{Z}_{\mathrm{int}}$ is composed of the following parts.

- Classical Contributions. The topological symmetry and flux contributes:

$$Z^{\text{top}}_{\text{class}} = x^{\tilde{\mathfrak{n}}}\xi^{\mathfrak{m}} \tag{A.3}$$

A Chern-Simons term contributes:

$$Z^{\text{CS}}_{\text{class}} = x^{\kappa\mathfrak{m}} \tag{A.4}$$

- 1-loop determinants for the 3$d$ $\mathcal{N}=2$ multiplets. An $\mathcal{N}=2$ vector multiplet contributes a 1-loop determinant:

$$Z^{\text{vec}}_{\text{1-loop}} = (-q)^{-\sum_{\alpha>0}|\alpha(\mathfrak{m})|}\prod_{\alpha}\left(1 - x^{\alpha}q^{|\alpha(\mathfrak{m})|}\right) \tag{A.5}$$

and an $\mathcal{N}=2$ chiral multiplet, in a representation $\mathfrak{R}$ of the gauge symmetry, $\mathfrak{R}_{\mathfrak{f}}$ of the flavour symmetry, and $\mathrm{U}(1)_R$ charge $r$ contributes:

$$Z^{\text{chiral}}_{\text{1-loop}} = \prod_{\substack{\rho\in\mathfrak{R}\\ \rho_f\in\mathfrak{R}_f}} \frac{(x^{\rho}y^{\rho_f})^{B/2}}{(x^{\rho}y^{\rho_f}q^{1-B};q^2)_B}\,, \qquad B = \rho(\mathfrak{m}) + \rho_f(\mathfrak{n}) - r + 1 \tag{A.6}$$

We note the identity:

$$(-q)^{-\sum_{\alpha>0}|\alpha(\mathfrak{m})|}\prod_{\alpha}\left(1 - x^{\alpha}q^{|\alpha(\mathfrak{m})|}\right) = \prod_{\alpha}\frac{\left(x^{\alpha/2}\right)^{\alpha(\mathfrak{m})-1}}{\left(x^{\alpha}q^{2-\alpha(\mathfrak{m})};q^2\right)_{\alpha(\mathfrak{m})-1}} \tag{A.7}$$

so that an $\mathcal{N}=2$ vector multiplet contributes in the same way as a chiral of $R$-charge 2 in the adjoint representation of the gauge group. Note that the 1-loop determinant is in fact non-singular.

For a 3d $\mathcal{N}=4$ theory, the superalgebra has a $\mathrm{SU}(2)_H\times\mathrm{SU}(2)_C$ R-symmetry, and one can choose to twist with the either $\mathrm{U}(1)_R = 2\mathrm{U}(1)_{H,C}$, corresponding to the $A$ or $B$ twist. The R-symmetry assignments of the $\mathcal{N}=2$ multiplets making up the $\mathcal{N}=4$ multiplets are fixed by this choice, and can be found in e.g. tables 5 and 6 of [7]. We consider theories containing $\mathcal{N}=4$ vector multiplets, consisting of an $\mathcal{N}=2$ vector multiplet and an $\mathcal{N}=2$ in the adjoint, and $\mathcal{N}=4$ hypermultiplets consisting of $\mathcal{N}=2$ chiral and an $\mathcal{N}=2$ anti-chiral in the same representation, or alternatively a pair of $\mathcal{N}=2$ chirals in conjugate representations. Their 1-loop determinants are products of the $\mathcal{N}=2$ determinants and are:

- A-Twist:

$$\begin{aligned}
Z^{\text{vec}}_{\text{1-loop}} &= \left[(-q)^{-\sum_{\alpha>0}|\alpha(\mathfrak{m})|}\prod_{\alpha}(1-x^{\alpha}q^{|\alpha(\mathfrak{m})|})\right] \\
&\quad\times\left[(t-t^{-1})^{-\operatorname{rk}G}\prod_{\alpha}\frac{\left(x^{\alpha/2}t^{-1}\right)^{\alpha(\mathfrak{m})+1}}{\left(x^{\alpha}t^{-2}q^{-\alpha(\mathfrak{m})};q^2\right)_{\alpha(\mathfrak{m})+1}}\right] \\
&=(t-t^{-1})^{-\operatorname{rk}G}\left[\prod_{\alpha>0}(-1)^{\alpha(\mathfrak{m})}\frac{(1-x^{-\alpha}q^{\alpha(\mathfrak{m})})(1-x^{\alpha}q^{\alpha(\mathfrak{m})})}{q^{\alpha(\mathfrak{m})}}\right] \\
&\quad\times\left[\prod_{\alpha}\frac{\left(x^{\alpha/2}t^{-1}\right)^{\alpha(\mathfrak{m})+1}}{\left(x^{\alpha}t^{-2}q^{-\alpha(\mathfrak{m})};q^2\right)_{\alpha(\mathfrak{m})+1}}\right] \\
&=(t-t^{-1})^{-\operatorname{rk}G}\prod_{\alpha}\frac{\left(x^{\alpha/2}\right)^{\alpha(\mathfrak{m})-1}}{\left(x^{\alpha}q^{2-\alpha(\mathfrak{m})};q^2\right)_{\alpha(\mathfrak{m})-1}}\frac{\left(x^{\alpha/2}t^{-1}\right)^{\alpha(\mathfrak{m})+1}}{\left(x^{\alpha}t^{-2}q^{-\alpha(\mathfrak{m})};q^2\right)_{\alpha(\mathfrak{m})+1}}
\end{aligned}\tag{A.8}$$

Note that the factor $(t-t^{-1})^{-\operatorname{rk}G}$ comes from the chirals in the $\mathcal{N}=2$ adjoint corresponding to the Cartan subalgebra. A hypermultiplet transforming in a representation $\mathfrak{R}$ of the gauge group and $\mathfrak{R}_{\mathfrak{f}}$ of the flavour symmetry contributes:

$$\begin{aligned}
&\mathcal{Z}^{\text{hyper}}_{\text{1-loop}} \\
&=\prod_{\substack{\rho\in\mathfrak{R}\\ \rho_f\in\mathfrak{R}_f}}\frac{\left(x^{\rho/2}y^{\rho_f/2}t^{\frac{1}{2}}\right)^{\rho(\mathfrak{m})+\rho_f(\mathfrak{n})}}{\left(x^{\rho}y^{\rho_f}tq^{1-\rho(\mathfrak{m})-\rho_f(\mathfrak{n})};q^2\right)_{\rho(\mathfrak{m})+\rho_f(\mathfrak{n})}}\frac{\left(x^{-\rho/2}y^{-\rho_f/2}t^{\frac{1}{2}}\right)^{-\rho(\mathfrak{m})-\rho_f(\mathfrak{n})}}{\left(x^{-\rho}y^{-\rho_f}tq^{1+\rho(\mathfrak{m})+\rho_f(\mathfrak{n})};q^2\right)_{-\rho(\mathfrak{m})-\rho_f(\mathfrak{n})}}
\end{aligned}\tag{A.9}$$

- B-twist:

$$\begin{aligned}
Z^{\text{vec}}_{\text{1-loop}} &= \left[(-q)^{-\sum_{\alpha>0}|\alpha(\mathfrak{m})|}\prod_{\alpha}(1-x^{\alpha}q^{|\alpha(\mathfrak{m})|})\right] \\
&\quad\times\left[(t-t^{-1})^{\operatorname{rk}G}\frac{\left(x^{\alpha/2}t^{-1}\right)^{\alpha(\mathfrak{m})-1}}{\left(x^{\alpha}t^{-2}q^{2-\alpha(\mathfrak{m})};q^2\right)_{\alpha(\mathfrak{m})-1}}\right] \\
&=(t-t^{-1})^{\operatorname{rk}G}\prod_{\alpha}\frac{\left(x^{\alpha/2}\right)^{\alpha(\mathfrak{m})-1}}{\left(x^{\alpha}q^{2-\alpha(\mathfrak{m})};q^2\right)_{\alpha(\mathfrak{m})-1}}\frac{\left(x^{\alpha/2}t^{-1}\right)^{\alpha(\mathfrak{m})-1}}{\left(x^{\alpha}t^{-2}q^{2-\alpha(\mathfrak{m})};q^2\right)_{\alpha(\mathfrak{m})-1}}
\end{aligned}\tag{A.10}$$

$$\begin{aligned}
\mathcal{Z}^{\text{hyper}}_{\text{1-loop}} &= \prod_{\substack{\rho\in\mathfrak{R}\\ \rho_f\in\mathfrak{R}_f}}\frac{\left(x^{\rho/2}y^{\rho_f/2}t^{\frac{1}{2}}\right)^{\rho(\mathfrak{m})+\rho_f(\mathfrak{n})+1}}{\left(x^{\rho}y^{\rho_f}tq^{-\rho(\mathfrak{m})-\rho_f(\mathfrak{n})};q^2\right)_{\rho(\mathfrak{m})+\rho_f(\mathfrak{n})+1}} \\
&\quad\times\frac{\left(x^{-\rho/2}y^{-\rho_f/2}t^{\frac{1}{2}}\right)^{-\rho(\mathfrak{m})-\rho_f(\mathfrak{n})+1}}{\left(x^{-\rho}y^{-\rho_f}tq^{\rho(\mathfrak{m})+\rho_f(\mathfrak{n})};q^2\right)_{-\rho(\mathfrak{m})-\rho_f(\mathfrak{n})+1}}
\end{aligned}\tag{A.11}$$

The contour is chosen to include poles, specified by intersections of the hyperplanes corresponding to singularities in the $\mathcal{N}=2$ chiral 1-loop determinants, and can be written in terms of a Jeffrey-Kirwan residue, specified by a parameter $\eta \in \mathfrak{h}^*$. Let $x_*$ be the intersection of $n$ hyperplane singularities of the chiral multiplets, with charges $Q_1, Q_2, \ldots, Q_n$ under the gauge group. If $x^*$ is the intersection of $\mathrm{rk}\, G$ hyperplanes, then the JK residue is the usual residue at the $\mathrm{rk}\, G$-pole if $\eta \in \mathrm{Cone}\,(Q_1, \ldots, Q_{\mathrm{rk}\, G})$, or 0 if $\eta \notin \mathrm{Cone}\,(Q_1, \ldots, Q_{\mathrm{rk}\, G})$. If $u^*$ is the intersection of $n > r$ hyperplanes, a constructive definition of the JK residue is required, for which we refer the reader to [1]. The boundary pieces are given by residues at the asymptotic regions in $H \times \mathfrak{h}$. In the following, we assume that the matter content of the theory has been chosen such that the boundary does not contribute. It was conjectured in [40] that the criterion of whether or not the boundary contributions vanish was whether or not the theories had Higgs vacua (respectively).

### A.1 $T[\mathrm{SU}(N)]$ $A$-twist

Using the above rules, the contour integral for the A-twisted index of $T[\mathrm{SU}(N)]$ is:

$$
\begin{aligned}
Z^A[T[\mathrm{SU}(N)]]\left(q,t,\vec{y},\mathfrak{n},\vec{\xi},\tilde{\mathfrak{n}}\right) &= \left(\prod_{s=1}^{N-1}\frac{(-1)^s}{s!}\right)\sum_{\{\mathfrak{m}_a^{(s)}\}}\oint_{JK}\prod_{s=1}^{N-1}\prod_{a=1}^{s}\frac{dx_a^{(s)}}{2\pi i x_a^{(s)}} \\
&\times\left(\prod_{s=1}^{N-1}\left(\frac{\xi_s}{\xi_{s+1}}\right)^{\sum_{a=1}^{s}\mathfrak{m}_a^{(s)}}\right)\left(\prod_{s=1}^{N-1}\prod_{a=1}^{s}(x_a^{(s)})^{\tilde{\mathfrak{n}}_s-\tilde{\mathfrak{n}}_{s+1}}\right)(t-t^{-1})^{-\sum_{s=1}^{N-1}s} \\
&\times\left[\prod_{s=1}^{N-1}\prod_{\substack{a,b=1\\a\neq b}}^{s}\frac{\left(\left(\frac{x_a^{(s)}}{x_b^{(s)}}\right)^{\frac{1}{2}}\right)^{(\mathfrak{m}_a^{(s)}-\mathfrak{m}_b^{(s)}-1)}}{\left(\frac{x_a^{(s)}}{x_b^{(s)}}q^{2-\mathfrak{m}_a^{(s)}+\mathfrak{m}_b^{(s)}};q^2\right)_{\mathfrak{m}_a^{(s)}-\mathfrak{m}_b^{(s)}-1}}\frac{\left(\left(\frac{x_a^{(s)}}{x_b^{(s)}}\right)^{\frac{1}{2}}t^{-1}\right)^{(\mathfrak{m}_a^{(s)}-\mathfrak{m}_b^{(s)}+1)}}{\left(\frac{x_a^{(s)}}{x_b^{(s)}}t^{-2}q^{-\mathfrak{m}_a^{(s)}+\mathfrak{m}_b^{(s)}};q^2\right)_{\mathfrak{m}_a^{(s)}-\mathfrak{m}_b^{(s)}+1}}\right] \\
&\times\left[\prod_{s=1}^{N-1}\prod_{a=1}^{s}\prod_{b=1}^{s+1}\frac{\left(\left(\frac{x_a^{(s)}}{x_b^{(s+1)}}\right)^{\frac{1}{2}}t^{\frac{1}{2}}\right)^{(\mathfrak{m}_a^{(s)}-\mathfrak{m}_b^{(s+1)})}}{\left(\frac{x_a^{(s)}}{x_b^{(s+1)}}tq^{1-\mathfrak{m}_a^{(s)}+\mathfrak{m}_b^{(s+1)}};q^2\right)_{\mathfrak{m}_a^{(s)}-\mathfrak{m}_b^{(s+1)}}}\frac{\left(\left(\frac{x_b^{(s+1)}}{x_a^{(s)}}\right)^{\frac{1}{2}}t^{\frac{1}{2}}\right)^{(\mathfrak{m}_b^{(s+1)}-\mathfrak{m}_a^{(s)})}}{\left(\frac{x_b^{(s+1)}}{x_a^{(s)}}tq^{1+\mathfrak{m}_a^{(s)}-\mathfrak{m}_b^{(s+1)}};q^2\right)_{\mathfrak{m}_b^{(s+1)}-\mathfrak{m}_a^{(s)}}}\right]
\end{aligned}
\tag{A.12}
$$

Here $\xi_s/\xi_{s+1}$ is the F.I. parameter for the $s^{\mathrm{th}}$ gauge node, $\tilde{\mathfrak{n}}_s - \tilde{\mathfrak{n}}_{s+1}$ the corresponding background through $S^2$, $y_i$ the fugacity for the $\mathrm{SU}(N)$ flavour symmetry for the hypermultiplets, and $\mathfrak{n}_i$ the corresponding background fluxes. We identify $x_a^{(N)} = y_a^{-1}$, $\mathfrak{m}_a^{(N)} = -\mathfrak{n}_a$ and:

$$
\prod_{s=1}^{N}\xi_s = \prod_{i=1}^{N} y_i = 1 \qquad \sum_{s=1}^{N}\tilde{\mathfrak{n}}_s = \sum_{i=1}^{N}\mathfrak{n}_i = 0 \tag{A.13}
$$

Choosing JK parameter $\eta = \vec{1}$, the JK procedure selects a contour enclosing poles corresponding to the intersection of hyperplanes:

$$
x_a^{(s)} = x_{f_{(s)}(a)}^{(s+1)} t^{-1} q^{\mathfrak{m}_a^{(s)}-\mathfrak{m}_{f_{(s)}(a)}^{(s+1)}-1-2p_a^{(s)}} \qquad p_a^{(s)} = 0,\ldots,\mathfrak{m}_a^{(s)}-\mathfrak{m}_{f(a)}^{(s+1)}-1 \tag{A.14}
$$

where $f_{(s)} : I^{(s)} \to I^{(s+1)}$ maps between the set of gauge indices (flavour indices for $I^{(N)}$). That is, we take residues from poles corresponding only to the chiral (and not the anti-chiral) multiplets. We note that although the JK residue procedure allows us to take poles from the $\mathcal{N}=2$ chiral adjoint in the $\mathcal{N}=4$ vector multiplet, in order for $\eta$ to be in the cone of gauge charges, we must also inevitably have to choose a chiral multiplet. Then however, the intersection of hyperplanes specified by choosing such poles coincides with a zero of an anti-chiral 1-loop determinant. Thus such $\operatorname{rk} G$-poles do not contribute to the index.

Now for the residues at (A.14) to be non-vanishing we need $\mathfrak{m}_a^{(s)} - \mathfrak{m}^{(s+1)}_{f_{(s)}(a)} - 1 \geq 0$. We drop the subscript on $f_{(s)}(a)$ to $f(a)$ with the understanding that $a \in I^{(s)}$. Now denoting:

$$\mathfrak{m}_a^{(s)} - \mathfrak{m}^{(s+1)}_{f(a)} - 1 - p_a^{(s)} \equiv \tilde{l}_a^s \qquad p_a^{(s)} \equiv l_a^{(s)} \tag{A.15}$$

where we require $l_a^{(s)}, \tilde{l}_s^{(s)} \geq 0$ for the residues to be non-zero. We can replace the sum over poles by:

$$\sum_{\{\mathfrak{m}\}\,|\,\mathfrak{m}_a^{(s)} - \mathfrak{m}^{(s+1)}_{f_{(s)}(a)} - 1 \geq 0} \quad \sum_{p_a^{(s)}}^{\mathfrak{m}_a^{(s)} - \mathfrak{m}^{(s+1)}_{f_{(s)}(a)} - 1} \quad \to \sum_{\{l_a^{(s)} \geq 0\}} \tag{A.16}$$

We have that:

$$x_a^{(s)} = x^{(s+1)}_{f_{(a)}} t^{-1} q^{\tilde{l}_a^{(s)} - l_a^{(s)}} = \ldots = x^{(N)}_{f^{N-s}(a)} t^{-(N-s)} q^{\sum_{\mu=s}^{N} (\tilde{l}_a^{(s)} - l_a^{(s)})} \tag{A.17}$$

Denoting $f^{N-s}(a) \equiv g_{(s)}(a)$, $k_a^{(s)} \equiv \sum_{\mu=s}^{N} l_a^{(s)}$, $\tilde{k}_a^{(s)} \equiv \sum_{\mu=s}^{N} \tilde{l}_a^{(s)}$ and identifying $x^{(N)}_{f^{N-s}(a)} = y^{-1}_{g_{(s)}(a)}$, we have:

$$x_a^{(s)} = y^{-1}_{g_{(s)}(a)} t^{-(N-s)} q^{\tilde{k}_a^{(s)} - k_a^{(s)}} \qquad k_a^{(s)} + \tilde{k}_a^{(s)} = \mathfrak{m}_a^{(s)} + \mathfrak{n}_{g_{(s)}(a)} - (N-s). \tag{A.18}$$

Now the sum over poles is over $\{k_a^{(s)}\}$ such that $k_a^{(s)} \geq k_a^{(s+1)}$ $\forall\, s$, coinciding with (2.8). Note that we can actually sum over $\{k_a^{(s)} \geq 0\}$ since the residue vanishes if any $l_a^{(s)} < 0$, even if $k_a^{(s)} \geq 0$. The same holds for the $\{\tilde{k}_a^{(s)}\}$.

We now make an observation to drastically simplify the calculation. Notice that the residue contribution to the twisted index vanishes if $f_{(s)}(c) = f_{(s)}(d)$ for any $s$ and $c \neq d$ such that $c, d \in \{1, \ldots, s\}$. This is because if $f_{(s)}(c) = f_{(s)}(d) \Rightarrow g_{(s)}(c) = g_{(s)}(d)$, and if $g_{(s)}(c) = g_{(s)}(d)$ for some $s, c, d$ then the integrand (A.12) evaluated at poles (A.14) is anti-symmetric in $k_c^{(s)}$ and $k_d^{(s)}$, and similarly $\tilde{k}_c^{(s)}$ and $\tilde{k}_d^{(s)}$, thus summing over $\{k_a^{(s)}, \tilde{k}_a^{(s)} \geq 0\}$ the contribution of such $\operatorname{rk} G$-poles vanish. In slightly more detail, the $\mathcal{N}=2$ vector multiplet contribution evaluates to (using (A.7) to write it in a manifestly non-singular

form):

$$\prod_{\alpha>0}(-q)^{-\alpha(\mathfrak{m})}(1-x^{\alpha}q^{\alpha(\mathfrak{m})})(1-x^{-\alpha}q^{\alpha(\mathfrak{m})})$$
$$=\prod_{s=1}^{N-1}\prod_{a<b}^{s}(-)^{\alpha(\mathfrak{m})+1}\left[\left(\frac{x_a^{(s)}}{x_b^{(s)}}q^{\mathfrak{m}_a^{(s)}-\mathfrak{m}_b^{(s)}}\right)^{\frac{1}{2}}-\left(\frac{x_a^{(s)}}{x_b^{(s)}}q^{\mathfrak{m}_a^{(s)}-\mathfrak{m}_b^{(s)}}\right)^{-\frac{1}{2}}\right]$$
$$\times\left[\left(\frac{x_a^{(s)}}{x_b^{(s)}}q^{-\mathfrak{m}_a^{(s)}+\mathfrak{m}_b^{(s)}}\right)^{\frac{1}{2}}-\left(\frac{x_a^{(s)}}{x_b^{(s)}}q^{-\mathfrak{m}_a^{(s)}+\mathfrak{m}_b^{(s)}}\right)^{-\frac{1}{2}}\right] \tag{A.19}$$

Evaluating at (A.18), we obtain:

$$\prod_{s=1}^{N-1}\prod_{a<b}^{s}(-)^{k_a^{(s)}-k_b^{(s)}+\tilde{k}_a^{(s)}-\tilde{k}_b^{(s)}-\mathfrak{n}_{g_{(s)}(a)}+\mathfrak{n}_{g_{(s)}(b)}}$$
$$\times\left[\left(\frac{y_{g_{(s)}(b)}q^{\mathfrak{n}_{g_{(s)}(b)}}}{y_{g_{(s)}(a)}q^{\mathfrak{n}_{g_{(s)}(a)}}}\right)^{\frac{1}{2}}q^{\tilde{k}_a^{(s)}-\tilde{k}_b^{(s)}}-\left(\frac{y_{g_{(s)}(b)}q^{\mathfrak{n}_{g_{(s)}(b)}}}{y_{g_{(s)}(a)}q^{\mathfrak{n}_{g_{(s)}(a)}}}\right)^{-\frac{1}{2}}q^{\tilde{k}_b^{(s)}-\tilde{k}_a^{(s)}}\right]$$
$$\times\left[\left(\frac{y_{g_{(s)}(b)}q^{-\mathfrak{n}_{g_{(s)}(b)}}}{y_{g_{(s)}(a)}q^{-\mathfrak{n}_{g_{(s)}(a)}}}\right)^{\frac{1}{2}}q^{k_b^{(s)}-k_a^{(s)}}-\left(\frac{y_{g_{(s)}(b)}q^{-\mathfrak{n}_{g_{(s)}(b)}}}{y_{g_{(s)}(a)}q^{-\mathfrak{n}_{g_{(s)}(a)}}}\right)^{-\frac{1}{2}}q^{k_a^{(s)}-k_b^{(s)}}\right] \tag{A.20}$$

Setting $g_{(s)}(c)=g_{(s)}(d)$ for some $s,c,d$, the above is antisymmetric in either $k_c^{(s)}$ and $k_d^{(s)}$ or $\tilde{k}_c^{(s)}$ and $\tilde{k}_d^{(s)}$. The contributions from the classical pieces, $\mathcal{N}=4$ hypermultiplets and $\mathcal{N}=2$ adjoint chiral (which with the above makes the contribution to the $\mathcal{N}=4$ vector multiplet) are all symmetric. We spare the reader the proof of these statements, but note these contributions can all be calculated almost identically to below, where we choose a simple $f$ corresponding to a non-vanishing residue.

Thus we may restrict to $f$ injective. The resulting integral is symmetric in the $\{x_a^{(s)}\}$ for each $s$, which cancels the factor of $\prod\frac{1}{s!}$, and there are $N!$ distinct contributions corresponding to permutations of the masses $\{y_i\}$, corresponding to isolated Higgs vacua. We compute the contribution for the identity permutation and obtain the others via a sum over the Weyl Group $S_N$ at the end. That is, we pick $f: I^{(s)}\to I^{(s+1)}$ to be the identity embedding $f: j\in I^{(s)}\mapsto j\in I^{(s+1)}$, and thus evaluate the residue at the poles:

$$x_a^{(s)}=y_a^{-1}t^{-(N-s)}q^{\tilde{k}_a^{(s)}-k_a^{(s)}} \tag{A.21}$$

Which we denote collectively as $x=x_*$. We state the contributions from the different components of the localisation formulae.[38]

[38] Throughout, we use the trivial identity $(a;z)_{m+n}=(a;z)_m(az^m;z)_n$ to separate q-Pochhammers into perturbative, vortex and anti-vortex contributions. Further, we use the identities:

$$(a;z)_{-n}=\frac{1}{(az^{-n};z)_n}$$
$$(a;z)_n=(z^{1-n}/a;z)_n(-a)^nz^{n(n-1)/2}\,. \tag{A.22}$$

- From the $\mathcal{N}=4$ bifundamental and fundamental hypermultiplets:

$$
\begin{aligned}
(2\pi i)^{\mathrm{rk}\,G}\operatorname*{Res}_{x=x_*}&\left[\prod_{s=1}^{N-1}\prod_{a=1}^{s}\frac{1}{2\pi i x_a^{(s)}}\times\text{Fourth line of (A.12)}\right]=\left[\prod_{s=1}^{N-1}\prod_{a=1}^{s}\prod_{b=1}^{s+1}(-t)^{\mathfrak{n}_a-\mathfrak{n}_b-1}\right]\\
&\times\left[\prod_{s=1}^{N-1}\prod_{a=1}^{s}(1-t^2)\right]\left[\prod_{s=1}^{N-1}\prod_{a=1}^{s}\prod_{\substack{b=1\\ b\neq a}}^{s+1}\frac{\left(q^2\frac{y_aq^{-\mathfrak{n}_a}}{y_bq^{-\mathfrak{n}_b}}\right)_{\mathfrak{n}_a-\mathfrak{n}_b-1}}{\left(t^2q^2\frac{y_aq^{-\mathfrak{n}_a}}{y_bq^{-\mathfrak{n}_b}};q^2\right)_{\mathfrak{n}_a-\mathfrak{n}_b-1}}\right]\\
&\times\left[\prod_{s=1}^{N-1}\left(t^{-2}\right)^{\sum_{a=1}^{s}k_a^{(s)}}\prod_{a=1}^{s}\prod_{b=1}^{s+1}\frac{\left(t^2q^2\frac{y_aq^{-\mathfrak{n}_a}}{y_bq^{-\mathfrak{n}_b}};q^2\right)_{k_a^{(s)}-k_b^{(s+1)}}}{\left(q^2\frac{y_aq^{-\mathfrak{n}_a}}{y_bq^{-\mathfrak{n}_b}};q^2\right)_{k_a^{(s)}-k_b^{(s+1)}}}\right]\\
&\times\left[\prod_{s=1}^{N-1}\left(t^{-2}\right)^{\sum_{a=1}^{s}\tilde{k}_a^{(s)}}\prod_{a=1}^{s}\prod_{b=1}^{s+1}\frac{\left(t^2q^{-2}\frac{y_aq^{\mathfrak{n}_a}}{y_bq^{\mathfrak{n}_b}};q^{-2}\right)_{\tilde{k}_a^{(s)}-\tilde{k}_b^{(s+1)}}}{\left(q^{-2}\frac{y_aq^{\mathfrak{n}_a}}{y_bq^{\mathfrak{n}_b}};q^{-2}\right)_{\tilde{k}_a^{(s)}-\tilde{k}_b^{(s+1)}}}\right]
\end{aligned}
\tag{A.23}
$$

- From the $\mathcal{N}=4$ vector multiplet (third line of (A.12) and the $(t-t^{-1})^{\mathrm{rk}\,G}$ factor):

$$
\begin{aligned}
(t-t^{-1})^{-\sum_{s=1}^{N-1}s}\prod_{s=1}^{N-1}\prod_{\substack{a,b=1\\ a\neq b}}^{s}t\,&\frac{\left(t^2q^2\frac{y_aq^{-\mathfrak{n}_a}}{y_bq^{-\mathfrak{n}_b}};q^2\right)_{\mathfrak{n}_a-\mathfrak{n}_b-1}}{\left(q^2\frac{y_aq^{-\mathfrak{n}_a}}{y_bq^{-\mathfrak{n}_b}};q^2\right)_{\mathfrak{n}_a-\mathfrak{n}_b-1}}\\
&\times\frac{\left(t^{-2}\frac{y_aq^{-\mathfrak{n}_a}}{y_bq^{-\mathfrak{n}_b}};q^2\right)_{k_a^{(s)}-k_b^{(s)}}}{\left(\frac{y_aq^{-\mathfrak{n}_a}}{y_bq^{-\mathfrak{n}_b}};q^2\right)_{k_a^{(s)}-k_b^{(s)}}}\frac{\left(t^{-2}\frac{y_aq^{\mathfrak{n}_a}}{y_bq^{\mathfrak{n}_b}};q^{-2}\right)_{\tilde{k}_a^{(s)}-\tilde{k}_b^{(s)}}}{\left(\frac{y_aq^{\mathfrak{n}_a}}{y_bq^{\mathfrak{n}_b}};q^{-2}\right)_{\tilde{k}_a^{(s)}-\tilde{k}_b^{(s)}}}
\end{aligned}
\tag{A.24}
$$

We note that in deriving this formula there are some cancellations between roots $\alpha=(a,b)$ and $-\alpha=(b,a)$.

- The classical contributions in the second line of (A.12) give:

$$
\begin{aligned}
&\left[\prod_{s=1}^{N-1}\left(\frac{\xi_s}{\xi_{s+1}}\right)^{\sum_{a=1}^{s}((N-s)-\mathfrak{n}_a)}\prod_{a=1}^{s}\left(y_a^{-1}t^{-(N-s)}\right)^{\tilde{\mathfrak{n}}_s-\tilde{\mathfrak{n}}_{s+1}}\right]\\
&\times\left[\prod_{s=1}^{N-1}\left(\frac{\xi_sq^{-\tilde{\mathfrak{n}}_s}}{\xi_{s+1}q^{-\tilde{\mathfrak{n}}_{s+1}}}\right)^{\sum_{a=1}^{s}k_a^{(s)}}\right]\left[\prod_{s=1}^{N-1}\left(\frac{\xi_sq^{\tilde{\mathfrak{n}}_s}}{\xi_{s+1}q^{\tilde{\mathfrak{n}}_{s+1}}}\right)^{\sum_{a=1}^{s}\tilde{k}_a^{(s)}}\right]
\end{aligned}
\tag{A.25}
$$

Multiplying (A.23), (A.24) and (A.25) together with the $\prod(-1)^s$ prefactor yields the residues corresponding to the set of poles (A.21), which we can separate into:

$$\mathcal{Z}^A_{\rm cl}\mathcal{Z}^A_{\text{1-loop}}\left(q,t,\vec{y},\mathfrak{n},\vec{\xi},\tilde{\mathfrak{n}}\right)=\prod_{s=1}^{N-1}t^{-s}\left[\prod_{i<j}\frac{\xi_i}{\xi_j}\right]\left[\prod_{s=1}^{N}\left(\xi_s t^{-(N-2s+1)}\right)^{-\mathfrak{n}_s}\left(y_s t^{(N-2s+1)}\right)^{-\tilde{\mathfrak{n}}_s}\right]$$
$$\times\left[\prod_{i<j}^{N}\frac{\left(q^2\frac{y_iq^{-\mathfrak{n}_i}}{y_jq^{-\mathfrak{n}_j}};q^2\right)_{\mathfrak{n}_i-\mathfrak{n}_j-1}}{\left(t^2q^2\frac{y_iq^{-\mathfrak{n}_i}}{y_jq^{-\mathfrak{n}_j}};q^2\right)_{\mathfrak{n}_i-\mathfrak{n}_j-1}}\right] \tag{A.26}$$
$$\mathcal{Z}^A_{\rm V}\left(q,t,\vec{y},\mathfrak{n},\vec{\xi},\tilde{\mathfrak{n}}\right)=\sum_{\{k^{(s)}_a\}}\prod_{s=1}^{N-1}\left[\left(t^{-2}\frac{\xi_sq^{-\tilde{\mathfrak{n}}_s}}{\xi_{s+1}q^{-\tilde{\mathfrak{n}}_{s+1}}}\right)^{\sum_{a=1}^s k^{(s)}_a}\prod_{\substack{a,b=1\\a\neq b}}^{s}\frac{\left(t^{-2}\frac{y_aq^{-\mathfrak{n}_a}}{y_bq^{-\mathfrak{n}_b}};q^2\right)_{k^{(s)}_a-k^{(s)}_b}}{\left(\frac{y_aq^{-\mathfrak{n}_a}}{y_bq^{-\mathfrak{n}_b}};q^2\right)_{k^{(s)}_a-k^{(s)}_b}}\right.$$
$$\left.\times\prod_{a=1}^{s}\prod_{b=1}^{s+1}\frac{\left(t^2q^2\frac{y_aq^{-\mathfrak{n}_a}}{y_bq^{-\mathfrak{n}_b}};q^2\right)_{k^{(s)}_a-k^{(s+1)}_b}}{\left(q^2\frac{y_aq^{-\mathfrak{n}_a}}{y_bq^{-\mathfrak{n}_b}};q^2\right)_{k^{(s)}_a-k^{(s+1)}_b}}\right]$$
$$\mathcal{Z}^A_{\rm aV}\left(q,t,\vec{y},\mathfrak{n},\vec{\xi},\tilde{\mathfrak{n}}\right)=\mathcal{Z}_{\rm V}\left(q\to q^{-1},k\to\tilde{k}\right)$$

where we have used the identifications (A.13) to simply the perturbative prefactor. As described the full index is obtained by summing over the Weyl group:

$$\mathcal{Z}^A\left[T[\mathrm{SU}(N)]\right]\left(q,t,\vec{y}_i,\mathfrak{n}_i,\vec{\xi}_s,\tilde{\mathfrak{n}}_s\right)=\sum_{\sigma\in S_N}\mathcal{Z}^A_{\rm cl}\mathcal{Z}^A_{\text{1-loop}}\mathcal{Z}^A_{\rm V}\mathcal{Z}^A_{\rm aV}\left(q,t,\vec{y}_{\sigma(i)},\mathfrak{n}_{\sigma(i)},\vec{\xi}_s,\tilde{\mathfrak{n}}_s\right) \tag{A.27}$$

The $A$-twisted index for the general $T_\rho[\mathrm{SU}(N)]$ theory can be computed similarly.

## A.2 $T[\mathrm{SU}(N)]$ $B$-twist

The contour integral for the B-twisted index can be computed similarly and is:

$$Z^B[T[\mathrm{SU}(N)]]\left(q,t,\vec{y},\mathfrak{n},\vec{\xi},\tilde{\mathfrak{n}}\right)=\left(\prod_{s=1}^{N-1}\frac{(-1)^s}{s!}\right)\sum_{\{\mathfrak{m}^s_a\}}\oint_{JK}\prod_{s=1}^{N-1}\prod_{a=1}^{s}\frac{dx^{(s)}_a}{2\pi i x^{(s)}_a} \tag{A.28}$$
$$\times\left(\prod_{s=1}^{N-1}\left(\frac{\xi_s}{\xi_{s+1}}\right)^{\sum_{a=1}^s\mathfrak{m}^{(s)}_a}\right)\left(\prod_{s=1}^{N-1}\prod_{a=1}^{s}(x^{(s)}_a)^{\tilde{\mathfrak{n}}_s-\tilde{\mathfrak{n}}_{s+1}}\right)(t-t^{-1})^{\sum_{s=1}^{N-1}s}\times\dots$$
$$\times\left[\prod_{s=1}^{N-1}\prod_{\substack{a,b=1\\a\neq b}}^{s}\frac{\left(\left(\frac{x^{(s)}_a}{x^{(s)}_b}\right)^{\frac12}\right)^{(\mathfrak{m}^{(s)}_a-\mathfrak{m}^{(s)}_b-1)}}{\left(\frac{x^{(s)}_a}{x^{(s)}_b}q^{2-\mathfrak{m}^{(s)}_a+\mathfrak{m}^{(s)}_b};q^2\right)_{\mathfrak{m}^{(s)}_a-\mathfrak{m}^{(s)}_b-1}}\frac{\left(\left(\frac{x^{(s)}_a}{x^{(s)}_b}\right)^{\frac12}t^{-1}\right)^{(\mathfrak{m}^{(s)}_a-\mathfrak{m}^{(s)}_b-1)}}{\left(\frac{x^{(s)}_a}{x^{(s)}_b}t^{-2}q^{2-\mathfrak{m}^{(s)}_a+\mathfrak{m}^{(s)}_b};q^2\right)_{\mathfrak{m}^{(s)}_a-\mathfrak{m}^{(s)}_b-1}}\right]$$
$$\times\left[\prod_{s=1}^{N-1}\prod_{a=1}^{s}\prod_{b=1}^{s+1}\frac{\left(\left(\frac{x^{(s)}_a}{x^{(s+1)}_b}\right)^{\frac12}t^{\frac12}\right)^{(\mathfrak{m}^{(s)}_a-\mathfrak{m}^{(s+1)}_b+1)}}{\left(\frac{x^{(s)}_a}{x^{(s+1)}_b}tq^{-\mathfrak{m}^{(s)}_a+\mathfrak{m}^{(s+1)}_b};q^2\right)_{\mathfrak{m}^{(s)}_a-\mathfrak{m}^{(s+1)}_b+1}}\frac{\left(\left(\frac{x^{(s+1)}_b}{x^{(s)}_a}\right)^{\frac12}t^{\frac12}\right)^{(\mathfrak{m}^{(s+1)}_b-\mathfrak{m}^{(s)}_a+1)}}{\left(\frac{x^{(s+1)}_b}{x^{(s)}_a}tq^{\mathfrak{m}^{(s)}_a-\mathfrak{m}^{(s+1)}_b};q^2\right)_{\mathfrak{m}^{(s+1)}_b-\mathfrak{m}^{(s)}_a+1}}\right]$$

with the same identification of $x_a^N = y_a^{-1}$ and $\mathfrak{m}_a^N = -\mathfrak{n}_a$. Choosing $\eta = \vec{1}$, the JK residue picks basically the same poles as the A-twist. The poles that end up with non-zero residue correspond to intersections of hyperplanes:

$$x_a^{(s)} = x^{(s+1)}_{f_{(s)}(a)} t^{-1} q^{\mathfrak{m}_a^{(s)} - \mathfrak{m}^{(s+1)}_{f_{(s)}(a)} - 2p_a^{(s)}} \qquad p_a^{(s)} = 0, \dots, \mathfrak{m}_a^{(s)} - \mathfrak{m}^{(s+1)}_{f(a)} \tag{A.29}$$

For the residues to be non-vanishing we need $\mathfrak{m}_a^{(s)} - \mathfrak{m}^{(s+1)}_{f(a)} \geq 0$. In the same way as for the A-twist, for non-vanishing contributions $f$ needs to be injective. The integral is symmetric in the $x_a^{(s)}$ for each $s$, cancelling the $1/s!$ out front, and there are $N!$ distinct contributions corresponding to permutations in $S_N$ of the masses $\{y_i\}$, hence we similarly obtain the representation of the B-twist as a sum over Higgs vacua. Again we compute the contribution for the identity permutation and obtain the others via a sum over the Weyl group at the end. Denote:

$$m_a^{(s)} - m_a^{(s+1)} - p_a^{(s)} \equiv \tilde{l_a^{(s)}} \qquad p_a^{(s)} = l_a^{(s)} \tag{A.30}$$

Therefore one can express:

$$x_a^{(s)} = y_a^{-1} t^{-(N-s)} q^{\sum_{\mu=s}^{N-1} \tilde{l}_a^{(\mu)} - \sum_{\mu=s}^{N-1} l_a^{(\mu)}} \equiv y_a^{-1} t^{-(N-s)} q^{\tilde{k}_a^{(s)} - k_a^{(s)}} \tag{A.31}$$

where we have defined:

$$\tilde{k}_a^s \equiv \sum_{\mu=s}^{N-1} \tilde{l}_a^{\mu}\,, \qquad k_a^s \equiv \sum_{\mu=s}^{N-1} l_a^{\mu}\,, \qquad k_a^{(s)} + \tilde{k}_a^{(s)} = \mathfrak{m}_a^{(s)} + \mathfrak{n}_a \tag{A.32}$$

Note that we again have a sum over $\{k_a^{(s)}\}$, with $a = 1, \dots, s$ and $k_a^{(s)} \geq k_a^{(s+1)}$, as in (2.8). The contributions are:

- From the $\mathcal{N} = 4$ bifundamental and fundamental hypermultiplets:

$$\begin{aligned}
(2\pi i)^{\operatorname{rk} G} \operatorname*{Res}_{x = x_*} \left[ \prod_{s=1}^{N-1} \prod_{a=1}^{s} \frac{1}{2\pi i x_a^{(s)}} \times \text{Fourth line of (A.28)} \right] &= \left[ \prod_{s=1}^{N-1} \prod_{a=1}^{s} \prod_{b=1}^{s+1} (-t)^{\mathfrak{n}_a - \mathfrak{n}_b + 1} \right] \\
\times \left[ \prod_{s=1}^{N-1} \frac{1}{(1-t^2)^s} \right] &\left[ \prod_{s=1}^{N-1} \prod_{a=1}^{s} \prod_{\substack{b=1 \\ b \neq a}}^{s+1} \frac{y_a}{y_b} \frac{\left( q^2 \frac{y_a q^{-\mathfrak{n}_a}}{y_b q^{-\mathfrak{n}_b}}; q^2 \right)_{\mathfrak{n}_a - \mathfrak{n}_b - 1}}{\left( t^2 \frac{y_a q^{-\mathfrak{n}_a}}{y_b q^{-\mathfrak{n}_b}}; q^2 \right)_{\mathfrak{n}_a - \mathfrak{n}_b + 1}} \right] \\
\times \Bigg[ \prod_{s=1}^{N-1} (t^{-2} q^2)^{\sum_{a=1}^{s} k_a^{(s)}} &\prod_{a=1}^{s} \prod_{b=1}^{s+1} \frac{\left( t^2 \frac{y_a q^{-\mathfrak{n}_a}}{y_b q^{-\mathfrak{n}_b}}; q^2 \right)_{(k_a^{(s)} - k_b^{(s+1)})}}{\left( q^2 \frac{y_a q^{-\mathfrak{n}_a}}{y_b q^{-\mathfrak{n}_b}}; q^2 \right)_{(k_a^{(s)} - k_b^{(s+1)})}} \Bigg] \\
\times \Bigg[ \prod_{s=1}^{N-1} (t^{-2} q^{-2})^{\sum_{a=1}^{s} \tilde{k}_a^{(s)}} &\prod_{a=1}^{s} \prod_{b=1}^{s+1} \frac{\left( t^2 \frac{y_a q^{\mathfrak{n}_a}}{y_b q^{\mathfrak{n}_b}}; q^{-2} \right)_{(\tilde{k}_a^{(s)} - \tilde{k}_b^{(s+1)})}}{\left( q^{-2} \frac{y_a q^{\mathfrak{n}_a}}{y_b q^{\mathfrak{n}_b}}; q^{-2} \right)_{(\tilde{k}_a^{(s)} - \tilde{k}_b^{(s+1)})}} \Bigg]
\end{aligned} \tag{A.33}$$

- From the $\mathcal{N}=4$ vector multiplet (third line of (A.28) and the $(t-t^{-1})^{\mathrm{rk}\,G}$ factor):

$$\begin{aligned}(t-t^{-1})^{\sum_{s=1}^{N-1} s}\prod_{s=1}^{N-1}\prod_{a\neq b}^{s} t^{-1}\frac{\left(t^2\frac{y_a q^{-\mathfrak{n}_a}}{y_b q^{-\mathfrak{n}_b}};q^2\right)_{\mathfrak{n}_a-\mathfrak{n}_b+1}}{\left(\frac{y_a q^{-\mathfrak{n}_a}}{y_b q^{-\mathfrak{n}_b}};q^2\right)_{\mathfrak{n}_a-\mathfrak{n}_b-1}} \\ \times\frac{\left(t^{-2}q^2\frac{y_a q^{-\mathfrak{n}_a}}{y_b q^{-\mathfrak{n}_b}};q^2\right)_{k_a^{(s)}-k_b^{(s)}}}{\left(\frac{y_a q^{-\mathfrak{n}_a}}{y_b q^{-\mathfrak{n}_b}};q^2\right)_{k_a^{(s)}-k_b^{(s)}}}\frac{\left(t^{-2}q^{-2}\frac{y_a q^{\mathfrak{n}_a}}{y_b q^{\mathfrak{n}_b}};q^{-2}\right)_{\tilde{k}_a^{(s)}-\tilde{k}_b^{(s)}}}{\left(\frac{y_a q^{\mathfrak{n}_a}}{y_b q^{-\mathfrak{n}_b}};q^{-2}\right)_{\tilde{k}_a^{(s)}-\tilde{k}_b^{(s)}}}\end{aligned} \tag{A.34}$$

- The classical contributions in the second line of (A.28) give:

$$\begin{aligned}&\left[\prod_{s=1}^{N-1}\left(\frac{\xi_s}{\xi_{s+1}}\right)^{-\sum_{a=1}^{s}\mathfrak{n}_a}\prod_{a=1}^{s}\left(y_a^{-1}t^{-(N-s)}\right)^{\tilde{\mathfrak{n}}_s-\tilde{\mathfrak{n}}_{s+1}}\right] \\ &\times\left[\prod_{s=1}^{N-1}\left(\frac{\xi_s q^{-\tilde{\mathfrak{n}}_s}}{\xi_{s+1}q^{-\tilde{\mathfrak{n}}_{s+1}}}\right)^{\sum_{a=1}^{s}k_a^{(s)}}\right]\left[\prod_{s=1}^{N-1}\left(\frac{\xi_s q^{\tilde{\mathfrak{n}}_s}}{\xi_{s+1}q^{\tilde{\mathfrak{n}}_{s+1}}}\right)^{\sum_{a=1}^{s}\tilde{k}_a^{(s)}}\right]\end{aligned} \tag{A.35}$$

Multiplying (A.33), (A.34) and (A.35) together with the $\prod(-)^s$ prefactor yields the contribution of the identity permutation, which we can factor into:

$$\begin{aligned}\mathcal{Z}^B_{\mathrm{cl}}\mathcal{Z}^B_{\text{1-loop}}\left(q,t,\vec{y},\mathfrak{n},\vec{\xi},\tilde{\mathfrak{n}}\right)&=\prod_{s=1}^{N-1}t^s\prod_{i<j}^{N}\left(\frac{y_i}{y_j}\right)\left[\prod_{s=1}^{N}\left(\xi_s t^{-(N-2s+1)}\right)^{-\mathfrak{n}_s}\left(y_s t^{(N-2s+1)}\right)^{-\tilde{\mathfrak{n}}_s}\right] \\ &\quad\times\left[\prod_{i<j}^{N}\frac{\left(q^2\frac{y_i q^{-\mathfrak{n}_i}}{y_j q^{-\mathfrak{n}_j}};q^2\right)_{\mathfrak{n}_i-\mathfrak{n}_j-1}}{\left(t^2\frac{y_i q^{-\mathfrak{n}_i}}{y_j q^{-\mathfrak{n}_j}};q^2\right)_{\mathfrak{n}_i-\mathfrak{n}_j+1}}\right] \\ \mathcal{Z}^B_V\left(q,t,\vec{y},\mathfrak{n},\vec{\xi},\tilde{\mathfrak{n}}\right)&=\sum_{k_a^{(s)}\geq 0}\prod_{s=1}^{N-1}\left[\left(t^{-2}q^2\frac{\xi_s q^{-\tilde{\mathfrak{n}}_s}}{\xi_{s+1}q^{-\tilde{\mathfrak{n}}_{s+1}}}\right)^{\sum_{a=1}^{s}k_a^{(s)}}\right. \\ &\quad\times\prod_{a\neq b}^{s}\frac{\left(t^{-2}q^2\frac{y_a q^{-\mathfrak{n}_a}}{y_b q^{-\mathfrak{n}_b}};q^2\right)_{k_a^{(s)}-k_b^{(s)}}}{\left(\frac{y_a q^{-\mathfrak{n}_a}}{y_b q^{-\mathfrak{n}_b}};q^2\right)_{k_a^{(s)}-k_b^{(s)}}} \\ &\quad\left.\times\prod_{a=1}^{s}\prod_{b=1}^{s+1}\frac{\left(t^2\frac{y_a q^{-\mathfrak{n}_a}}{y_b q^{-\mathfrak{n}_b}};q^2\right)_{(k_a^{(s)}-k_b^{(s+1)})}}{\left(q^2\frac{y_a q^{-\mathfrak{n}_a}}{y_b q^{-\mathfrak{n}_b}};q^2\right)_{(k_a^{(s)}-k_b^{(s+1)})}}\right] \\ \mathcal{Z}^B_{aV}\left(q,t,\vec{y},\mathfrak{n},\vec{\xi},\tilde{\mathfrak{n}}\right)&=\mathcal{Z}_V\left(q\to q^{-1},\quad k\to\tilde{k}\right)\end{aligned} \tag{A.36}$$

Notice that the vortex partition function for the B-twist is the same as for the A-twist up to relabelling of parameters. The full B-twisted index is given by summing over the Weyl group:

$$\mathcal{Z}^B\left[T[\mathrm{SU}(N)]\right]\left(q,t,\vec{y}_i,\mathfrak{n}_i,\vec{\xi}_s,\tilde{\mathfrak{n}}_s\right)=\sum_{\sigma\in S_N}\mathcal{Z}^B_{\mathrm{cl}}\mathcal{Z}^B_{\text{1-loop}}\mathcal{Z}^B_{\mathrm{V}}\mathcal{Z}^B_{\mathrm{aV}}\left(q,t,\vec{y}_{\sigma(i)},\mathfrak{n}_{\sigma(i)},\vec{\xi}_s,\tilde{\mathfrak{n}}_s\right) \tag{A.37}$$

### A.3 SQCD$[k,N]$ A twist with Chern-Simons terms

In this section we consider $\mathcal{N}=4$ SQCD.[39] We consider adding a Chern-Simons term $\kappa$, breaking the supersymmetry to $\mathcal{N}=2$. The contour integral is:

$$\begin{aligned}\mathcal{Z}^A\left[\mathrm{SQCD}[k,N]\right](q,t,\vec{y},\mathfrak{n},\xi,\tilde{\mathfrak{n}},\kappa) &= \frac{(-1)^k}{k!}\sum_{\{\mathfrak{m}_a\}}\oint_{JK}\prod_{a=1}^{k}\frac{dx_a}{2\pi i x_a}\left(\prod_{a=1}^{k}(x_a)^{\tilde{\mathfrak{n}}+\kappa\mathfrak{m}_a}\xi^{\mathfrak{m}_a}\right)\\ &\times(t-t^{-1})^{-k}\left[\prod_{\substack{a,b=1\\a\neq b}}^{k}\frac{\left(\left(\frac{x_a}{x_b}\right)^{\frac{1}{2}}\right)^{(\mathfrak{m}_a-\mathfrak{m}_b-1)}}{\left(\frac{x_a}{x_b}q^{2-\mathfrak{m}_a+\mathfrak{m}_b};q^2\right)_{\mathfrak{m}_a-\mathfrak{m}_b-1}}\frac{\left(\left(\frac{x_a}{x_b}\right)^{\frac{1}{2}}t^{-1}\right)^{(\mathfrak{m}_a-\mathfrak{m}_b+1)}}{\left(\frac{x_a}{x_b}t^{-2}q^{-\mathfrak{m}_a+\mathfrak{m}_b};q^2\right)_{\mathfrak{m}_a-\mathfrak{m}_b+1}}\right]\\ &\times\left[\prod_{a=1}^{k}\prod_{i=1}^{N}\frac{\left((x_ay_i)^{\frac{1}{2}}t^{\frac{1}{2}}\right)^{(\mathfrak{m}_a+\mathfrak{n}_i)}}{(x_ay_itq^{1-\mathfrak{m}_a-\mathfrak{n}_i};q^2)_{\mathfrak{m}_a+\mathfrak{n}_i}}\frac{\left((x_ay_i)^{-\frac{1}{2}}t^{\frac{1}{2}}\right)^{(-\mathfrak{m}_a-\mathfrak{n}_i)}}{\left(x_a{}^{-1}y_i^{-1}tq^{1+\mathfrak{m}_a+\mathfrak{n}_i};q^2\right)_{-\mathfrak{m}_a-\mathfrak{n}_i}}\right]\end{aligned}\tag{A.38}$$

Assuming that the boundary terms vanish, the JK residue selects the following poles for $\eta=\vec{1}$, solely from the fundamental $\mathcal{N}=2$ chirals:

$$x_a=y_{f(a)}^{-1}t^{-1}q^{\mathfrak{m}_a+\mathfrak{n}_{f(a)}-1-2p_a}\qquad p_a=0,\ldots,\mathfrak{m}_a+\mathfrak{n}_{f(a)}-1\tag{A.39}$$

where $f:I^{(k)}\to I^{(N)}$ is a map from the set of gauge indices to the set of flavour indices.

As for $T[\mathrm{SU}(N)]$, we do not consider poles of the $\mathcal{N}=2$ adjoint chiral, or $\mathcal{N}=2$ anti-fundamental chirals (equivalently fundamental anti-chirals). Since the notation is simpler here, we explain explicitly why such poles have vanishing residue. Firstly, we do not choose any poles from the anti-chirals, else $\eta=\vec{1}$ will not lie in the cone of gauge charges specifying the $k$-pole. Suppose there was some pole with $x_a=t^2x_bq^{\mathfrak{m}_a-\mathfrak{m}_b-2l}$, $l=0,\ldots,\mathfrak{m}_a-\mathfrak{m}_b$, i.e. corresponding to a charge $k$-vector $(0,\ldots,1,\ldots,-1,\ldots,0)$ if $a<b$. In order for $\eta=\vec{1}$ to lie in the cone of charge vectors, we need to also have a contribution from a chiral with positive charge under gauge symmetry corresponding to $x_b$. If it is another pole from the $\mathcal{N}=4$ vector multiplet, we make the same argument again. Eventually, we arrive at the situation where having chosen some pole with $x_a=t^2x_bq^{\mathfrak{m}_a-\mathfrak{m}_b-2l}$, in order for $\eta=\vec{1}$ to lie in the cone of charges we must choose the pole from the fundamental $x_b=y_{f(b)}^{-1}t^{-1}q^{\mathfrak{m}_b+\mathfrak{n}_{f(b)}-1-2p_b}$, where $p_b=0,\ldots,\mathfrak{m}_b+\mathfrak{n}_{f(b)}-1$ for some $f(b)$. Note that for these to be poles we require $\mathfrak{m}_b+\mathfrak{n}_{f(b)}-1\geq 0$ and $\mathfrak{m}_a-\mathfrak{m}_b\geq 0$. Note that then this implies $x_a=y_{f(b)}^{-1}tq^{\mathfrak{m}_a+\mathfrak{n}_{f(b)}-1-2(p_b+l)}$ where $p_b+l=0,\ldots,\mathfrak{m}_a+\mathfrak{n}_{f(b)}-1$. But this corresponds to a zero in the term $(x_a^{-1}y_{f(b)}^{-1}tq^{1+\mathfrak{m}_a+\mathfrak{n}_{f(b)}};q^2)^{-1}_{-\mathfrak{m}_a-\mathfrak{n}_{f(b)}}$ in the 1-loop determinant of a $\mathcal{N}=2$ anti-chiral.

For a given choice of $f$, the contribution to the A-twisted index is a sum over non-vanishing residues:

$$\sum_{\mathfrak{m}_a+\mathfrak{n}_{f(a)}-1\geq 0}\ \sum_{p_a=0}^{\mathfrak{m}_a+\mathfrak{n}_{f(a)}-1}=\sum_{\tilde{k}_a\geq 0}\sum_{k_a\geq 0}\tag{A.40}$$

[39]Consisting of an $\mathcal{N}=2$ vector multiplet, an $\mathcal{N}=2$ chiral in the adjoint, an $\mathcal{N}=2$ chiral in the fundamental of $\mathrm{U}(k)$, and an $\mathcal{N}=2$ chiral in the antifundamental.

where $\tilde{k}_a \equiv \mathfrak{m}_a + \mathfrak{n}_{f(a)} - 1 - p_a$, and $k_a \equiv p_a$. So we can write the poles as:

$$x_a = y_{f(a)}^{-1} t^{-1} q^{\tilde{k}_a - k_a} \tag{A.41}$$

- The $\mathcal{N} = 4$ fundamental hypermultiplets contribute:

$$\begin{aligned}
&(2\pi i)^k \operatorname*{Res}_{x=x_*} \left[\prod_{a=1}^{k} \frac{1}{2\pi i x_a} \times \text{Third line of (A.38)}\right] = \\
&= (1-t^2)^k \left[\prod_{a=1}^{k} \prod_{\substack{i=1 \\ i\neq f(a)}}^{N} \frac{\left(q^2 \frac{y_{f(a)} q^{-\mathfrak{n}_{f(a)}}}{y_i q^{-\mathfrak{n}_i}}\right)_{\mathfrak{n}_{f(a)}-\mathfrak{n}_i-1}}{\left(t^2 q^2 \frac{y_{f(a)} q^{-\mathfrak{n}_{f(a)}}}{y_i q^{-\mathfrak{n}_i}}; q^2\right)_{\mathfrak{n}_{f(a)}-\mathfrak{n}_i-1}}\right] \\
&\times \left[\prod_{a=1}^{k} \prod_{i=1}^{N} (-t^{-1})^{1+k_a+\tilde{k}_a-\mathfrak{n}_{f(a)}+\mathfrak{n}_i} \frac{\left(t^2 q^2 \frac{y_{f(a)} q^{-\mathfrak{n}_{f(a)}}}{y_i q^{-\mathfrak{n}_i}}; q^2\right)_{k_a} \left(t^2 q^{-2} \frac{y_{f(a)} q^{\mathfrak{n}_{f(a)}}}{y_i q^{\mathfrak{n}_i}}; q^{-2}\right)_{\tilde{k}_a}}{\left(q^2 \frac{y_{f(a)} q^{-\mathfrak{n}_{f(a)}}}{y_i q^{-\mathfrak{n}_i}}; q^2\right)_{k_a} \left(q^{-2} \frac{y_{f(a)} q^{\mathfrak{n}_{f(a)}}}{y_i q^{\mathfrak{n}_i}}; q^{-2}\right)_{\tilde{k}_a}}\right]
\end{aligned} \tag{A.42}$$

- From the $\mathcal{N} = 4$ vector multiplet in the second line of (A.38):

$$\begin{aligned}
&(t - t^{-1})^{-k} \prod_{\substack{a,b=1 \\ a\neq b}}^{k} t \frac{\left(t^2 q^2 \frac{y_{f(a)} q^{-\mathfrak{n}_{f(a)}}}{y_{f(b)} q^{-\mathfrak{n}_{f(b)}}}; q^2\right)_{\mathfrak{n}_{f(a)}-\mathfrak{n}_{f(b)}-1}}{\left(q^2 \frac{y_{f(a)} q^{-\mathfrak{n}_{f(a)}}}{y_{f(b)} q^{-\mathfrak{n}_{f(b)}}}; q^2\right)_{\mathfrak{n}_{f(a)}-\mathfrak{n}_{f(b)}-1}} \\
&\qquad \times \frac{\left(t^{-2} \frac{y_{f(a)} q^{-\mathfrak{n}_{f(a)}}}{y_{f(b)} q^{-\mathfrak{n}_{f(b)}}}; q^2\right)_{k_a^{(s)}-k_b^{(s)}}}{\left(\frac{y_{f(a)} q^{-\mathfrak{n}_{f(a)}}}{y_{f(b)} q^{-\mathfrak{n}_{f(b)}}}; q^2\right)_{k_a^{(s)}-k_b^{(s)}}} \frac{\left(t^{-2} \frac{y_{f(a)} q^{\mathfrak{n}_{f(a)}}}{y_{f(b)} q^{\mathfrak{n}_{f(b)}}}; q^{-2}\right)_{\tilde{k}_a^{(s)}-\tilde{k}_b^{(s)}}}{\left(\frac{y_{f(a)} q^{\mathfrak{n}_{f(a)}}}{y_{f(b)} q^{\mathfrak{n}_{f(b)}}}; q^{-2}\right)_{\tilde{k}_a^{(s)}-\tilde{k}_b^{(s)}}}
\end{aligned} \tag{A.43}$$

- The classical contributions in the first line of (A.38) give:

$$\begin{aligned}
&\prod_{a=1}^{k} \left(y_{f(a)}^{-1} t^{-1}\right)^{\tilde{\mathfrak{n}}+\kappa(1-\mathfrak{n}_{f(a)})} \xi^{1-\mathfrak{n}_{f(a)}} \\
&\quad \times \left(\xi q^{-\tilde{\mathfrak{n}}}\right)^{k_a} \left(\xi q^{\tilde{\mathfrak{n}}}\right)^{\tilde{k}_a} \left(t^{-1} y_{f(a)}^{-1} q^{\mathfrak{n}_{f(a)}}\right)^{\kappa k_a} q^{-\kappa k_a(k_a+1)} \left(t^{-1} y_{f(a)}^{-1} q^{-\mathfrak{n}_{f(a)}}\right)^{\kappa \tilde{k}_a} q^{\kappa \tilde{k}_a(\tilde{k}_a+1)}
\end{aligned} \tag{A.44}$$

Multiplying (A.42), (A.43) and (A.44) together with the $(-)^k$ prefactor (but not the $(1/k!)$ for reasons to be explained) yields the contribution of a choice $f : I^{(k)} \to I^{(N)}$, which

we can factor into:

$$\begin{aligned}
\mathcal{Z}^{(f)}_{\mathrm{cl}}\mathcal{Z}^{(f)}_{\text{1-loop}}(t,q,\vec{y},\mathfrak{n},\xi,\tilde{\mathfrak{n}},\kappa) &= t^{k^2}\left[\prod_{a=1}^{k}\prod_{i=1}^{N}(-t)^{\mathfrak{n}_{f(a)}-\mathfrak{n}_i-1}\right]\left[\prod_{a=1}^{k}\left(y_{f(a)}^{-1}t^{-1}\right)^{\tilde{\mathfrak{n}}_T+\kappa(1-\mathfrak{n}_{f(a)})}\xi^{1-\mathfrak{n}_{f(a)}}\right]\\
&\quad\times\left[\prod_{a=1}^{k}\prod_{\substack{i=1\\ i\neq f(b)\forall b=1,\ldots,k}}^{N}\frac{\left(q^2\frac{y_{f(a)}q^{-\mathfrak{n}_{f(a)}}}{y_i q^{-\mathfrak{n}_i}}\right)_{\mathfrak{n}_{f(a)}-\mathfrak{n}_i-1}}{\left(t^2q^2\frac{y_{f(a)}q^{-\mathfrak{n}_{f(a)}}}{y_i q^{-\mathfrak{n}_i}};q^2\right)_{\mathfrak{n}_{f(a)}-\mathfrak{n}_i-1}}\right]\\
\mathcal{Z}^{(f)}_{V}(t,q,\vec{y},\mathfrak{n},\xi,\tilde{\mathfrak{n}},\kappa) &= \sum_{\{k_a\geq 0\}}\prod_{a=1}^{k}\left(t^{-1}y_{f(a)}^{-1}q^{\mathfrak{n}_{f(a)}}\right)^{\kappa k_a}q^{-\kappa k_a(k_a+1)}\left((-t)^{-N}\xi q^{-\tilde{\mathfrak{n}}}\right)^{\sum_{a=1}^{k}k_a}\\
&\quad\times\prod_{\substack{a,b=1\\ a\neq b}}^{k}\frac{\left(t^{-2}\frac{y_{f(a)}q^{-\mathfrak{n}_{f(a)}}}{y_{f(b)}q^{-\mathfrak{n}_{f(b)}}};q^2\right)_{k_a^{(s)}-k_b^{(s)}}}{\left(\frac{y_{f(a)}q^{-\mathfrak{n}_{f(a)}}}{y_{f(b)}q^{-\mathfrak{n}_{f(b)}}};q^2\right)_{k_a^{(s)}-k_b^{(s)}}}\\
&\quad\times\prod_{a=1}^{k}\prod_{i=1}^{N}\frac{\left(t^2q^2\frac{y_{f(a)}q^{-\mathfrak{n}_{f(a)}}}{y_i q^{-\mathfrak{n}_i}};q^2\right)_{k_a}}{\left(q^2\frac{y_{f(a)}q^{-\mathfrak{n}_{f(a)}}}{y_i q^{-\mathfrak{n}_i}};q^2\right)_{k_a}}\\
\mathcal{Z}^{(f)}_{AV}(t,q,\vec{y},\mathfrak{n},\xi,\tilde{\mathfrak{n}},\kappa) &= \mathcal{Z}^{(f)}_{V}\left(q\to q^{-1}\right)
\end{aligned} \tag{A.45}$$

We note that $f$ needs to be an injective function $f:\{1,\ldots,k\}\to\{1,\ldots,N\}$ in order for the residue to not cancel after summing over vortex number, as in $T[\mathrm{SU}(N)]$. Also since the integration is symmetric in the $k$ fugacities $\{x_a\}$, the sum over $f$ can be converted to one over $S_N/(S_{N-k}\times S_k)$, i.e. over $\binom{N}{k}$ vacua via multiplying by $k!$, cancelling the $1/k!$ prefactor, which is why it was not included earlier. We abuse notation and use $f$ to denote a representative in $S_N/(S_{N-k}\times S_k)$, i.e. a choice of mapping the unordered gauge indices into the unordered flavour indices. Finally, the $A$-twisted index of $\mathcal{N}=4$ SQCD[k,N], with an $\mathcal{N}=2$ CS deformation is:

$$\mathcal{Z}^{A}\left[SQCD[k,N]\right](q,t,\vec{y},\mathfrak{n},\xi,\tilde{\mathfrak{n}},\kappa) = \sum_{f\in S_N/(S_k\times S_{N-k})}\mathcal{Z}^{(f)}_{\mathrm{cl}}\mathcal{Z}^{(f)}_{\text{1-loop}}\mathcal{Z}^{(f)}_{\mathrm{V}}\mathcal{Z}^{(f)}_{\mathrm{aV}}(t,q,\vec{y},\mathfrak{n},\xi,\tilde{\mathfrak{n}},\kappa) \tag{A.46}$$

## B $T_\rho[\mathrm{SU}(N)]$ vortices and supersymmetric quantum mechanics

In this section we prove that the vortex partition function of the $T_\rho[\mathrm{SU}(N)]$ theory (4.19):

$$\mathcal{Z}_V[T_\rho[\mathrm{SU}(N)]](q,t,\{y\},\xi) = \sum_{\{k^{(s)}_{a,\gamma}\}} \prod_{s=1}^{L-1} \left( \left(-t^{-\frac{1}{2}}\right)^{(\rho_s+\rho_{s+1})} \frac{\xi_s}{\xi_{s+1}} \right)^{\sum_{a=1}^{s} \sum_{\gamma=1}^{\rho_a} k^{(s)}_{a,\gamma}} \tag{B.1}$$

$$\times \left( \prod_{\substack{a,b=1,\dots,s\\ \beta=1,\dots,\rho_a\\ \gamma=1,\dots,\rho_b\\ (a,\gamma)\neq(b,\delta)}} \frac{\left(t^{-1}\frac{y_{a,\gamma}}{y_{b,\delta}};q\right)_{k^{(s)}_{a,\gamma}-k^{(s)}_{b,\delta}}}{\left(\frac{y_{a,\gamma}}{y_{b,\delta}};q\right)_{k^{(s)}_{a,\gamma}-k^{(s)}_{b,\delta}}} \right) \left( \prod_{a=1}^{s}\prod_{\gamma=1}^{\rho_b}\prod_{b=1}^{s+1}\prod_{\delta=1}^{\rho_b} \frac{\left(tq\frac{y_{a,\gamma}}{y_{b,\delta}};q\right)_{k^{(s)}_{a,\gamma}-k^{(s+1)}_{b,\delta}}}{\left(q\frac{y_{a,\gamma}}{y_{b,\delta}};q\right)_{k^{(s)}_{a,\gamma}-k^{(s+1)}_{b,\delta}}} \right)$$

(where we have already made the rescaling $q^2 \to q$, and $t^2 \to t$, and the sum is over $\{k^{(s)}_{a,\gamma}\}$ obeying (4.21)) is the generating function of the superconformal indices/ equivariant Hirzebruch $\chi_t$ genera of handsaw quiver varieties $\mathfrak{Q}^{\rho}_{\vec{\alpha}}$, given by:

$$Z_{\text{S.C.}}(\mathfrak{Q}^{\rho}_{\vec{\alpha}},q,t,\{y\}) = (-)^{d_{\mathbb{C}}}(t)^{-d_{\mathbb{C}}/2}\chi_t(\mathfrak{Q}^{\rho}_{\vec{\alpha}};q,y) \tag{B.2}$$

$$= \left(-t^{-\frac{1}{2}}\right)^{d_{\mathbb{C}}} \sum_{\vec{Y}} \prod_{(a,\gamma),(b,\delta)} \prod_{\substack{p\in Y_{a,\gamma}\\ L_{Y_{b,\delta}}(s)=0}} \mathrm{PE}\left((1-t)\frac{y_{a,\gamma}}{y_{b,\delta}}q^{A_{Y_{a,\gamma}}+1}\right) \prod_{\substack{p\in Y_{a,\gamma}\\ L_{Y_{b,\delta}}(s)=-1}} \mathrm{PE}\left((1-t)\frac{y_{b,\delta}}{y_{a,\gamma}}q^{-A_{Y_{a,\gamma}}}\right)$$

Here $\rho$ is an $L$-vector and is the same partition of $N$ specifying the $T_\rho[\mathrm{SU}(N)]$ theory, and $\vec{\alpha}$ is an $L-1$ vector specifying the gauge nodes of the handsaw, which correspond to vortex number physically and map degree in the Laumon space description. The complex dimension of $\mathfrak{Q}^{\rho}_{\vec{\alpha}}$ is given by $d_{\mathbb{C}} = \sum_{s=1}^{L-1}\alpha_s(\rho_s+\rho_{s+1})$. Recall the definition of $\{Y_{a,\gamma}\}$ in section 4.2 as an $N$-tuple of shifted Young tableaux, such that the box in the bottom-left of $Y_{a,\gamma}$ is $a$. That is, we claim:

**Proposition B.1.**

$$\mathcal{Z}_V[T_\rho[\mathrm{SU}(N)]](q,t,\{y\},\{\xi\}) = \sum_{\vec{\alpha}} \left[\prod_{s=1}^{N-1}\left(\frac{\xi_s}{\xi_{s+1}}\right)^{\alpha_s}\right] Z_{\text{S.C.}}(\mathfrak{Q}^{\rho}_{\vec{\alpha}},q,t,\{y\}) \tag{B.3}$$

$$= \sum_{\vec{\alpha}} \left[\prod_{s=1}^{N-1}\left(\left(-t^{-\frac{1}{2}}\right)^{(\rho_s+\rho_{s+1})}\frac{\xi_s}{\xi_{s+1}}\right)^{\alpha_s}\right] \chi_t(\mathfrak{Q}^{\rho}_{\vec{\alpha}},q,\{y\})$$

*Proof.* We identify the data $k^{(s)}_{a,\gamma}$ (4.21) with the height of the column with $x$-coordinate $s$ of the Young tableau $Y_{a,\gamma}$ in Nakajima's notation, and as in section 4.2. Therefore, we identify:

$$\sum_{a=1}^{s}\sum_{\gamma=1}^{\rho_a} k^{(s)}_{a,\gamma} = \alpha_s = \dim V_s \tag{B.4}$$

We note that the contributions in the sum $\sum_{\{k^{(s)}_{a,\gamma}\}}$ in $\mathcal{Z}_V[T_\rho[\mathrm{SU}(N)]](q,t,\{y\},\xi)$ which contribute a fixed monomial $\left[\prod_{s=1}^{N-1}\left(\frac{\xi_s}{\xi_{s+1}}\right)^{\alpha_s}\right]$ in $\xi$, are precisely those obeying the above condition. In addition, identifying (B.4), the powers of $-t^{-\frac{1}{2}}$ match in (B.1) and (B.3).

The proposition holds then if, given some unordered pair of (double) indices $(a,\gamma)$ and $(b,\delta)$, without loss of generality $a\geq b$, and a given set of Young tableaux $\{Y_{a,\gamma}\}$ specified by $\{k^{(s)}_{a,\gamma}\}$ the terms involving the ratio of fugacities $y_{a,\gamma}/y_{b,\delta}$ in the second line of (B.1) match the terms in the plethystic exponentials in (B.2). This is a matching to terms in the character formula corresponding to boxes in two of the tableaux $Y_{a,\gamma}$ and $Y_{b,\delta}$ in the $N$-tuple of tableaux.

The plethystic terms in the superconformal index (B.2) corresponding to Young tableau $Y_{a,\gamma}$ and $Y_{b,\delta}$ are (if $(a,\gamma)\neq(b,\delta)$):

$$\begin{aligned}&\prod_{\substack{p\in Y_{a,\gamma}\\ L_{Y_{b,\delta}}(s)=0}} \frac{1-t\frac{y_{a,\gamma}}{y_{b,\delta}}q^{A_{Y_{a,\gamma}}(s)+1}}{1-\frac{y_{a,\gamma}}{y_{b,\delta}}q^{A_{Y_{a,\gamma}}(s)+1}} \prod_{\substack{p\in Y_{a,\gamma}\\ L_{Y_{b,\delta}}(s)=-1}} \frac{1-t\frac{y_{b,\delta}}{y_{a,\gamma}}q^{-A_{Y_{a,\gamma}}(s)}}{1-\frac{y_{b,\delta}}{y_{a,\gamma}}q^{-A_{Y_{a,\gamma}}(s)}} \prod_{\substack{p\in Y_{b,\delta}\\ L_{Y_{a,\gamma}}(s)=0}} \frac{1-t\frac{y_{b,\delta}}{y_{a,\gamma}}q^{A_{Y_{b,\delta}}(s)+1}}{1-\frac{y_{b,\delta}}{y_{a,\gamma}}q^{A_{Y_{b,\delta}}(s)+1}} \prod_{\substack{p\in Y_{b,\delta}\\ L_{Y_{a,\gamma}}(s)=-1}} \frac{1-t\frac{y_{a,\gamma}}{y_{b,\delta}}q^{-A_{Y_{b,\delta}}(s)}}{1-\frac{y_{a,\gamma}}{y_{b,\delta}}q^{-A_{Y_{b,\delta}}(s)}}\\ &\equiv \qquad \boxed{A} \qquad\qquad\qquad \boxed{B} \qquad\qquad\qquad \boxed{C} \qquad\qquad\qquad \boxed{D}\end{aligned} \tag{B.5}$$

and if $(a,\gamma)=(b,\delta)$ then we just have terms $\boxed{A}$.

The corresponding terms in the 3d vortex partition function (containing the ratio $y_{a,\gamma}/y_{b,\delta}$ are, if $(a,\gamma)=(b,\delta)$ just:

$$\prod_{s=a}^{L-1}\frac{(tq;q)_{k^{(s)}_{a,\gamma}-k^{(s+1)}_{a,\gamma}}}{(q;q)_{k^{(s)}_{a,\gamma}-k^{(s+1)}_{a,\gamma}}} \tag{B.6}$$

which clearly corresponds to the term $\boxed{A}$.

If $(a,\gamma)\neq(b,\delta)$, things are more complicated, and the corresponding terms are:

$$\begin{aligned}&\prod_{s=b}^{L-1}\frac{\left(t^{-1}\frac{y_{a,\gamma}}{y_{b,\delta}};q\right)_{k^{(s)}_{a,\gamma}-k^{(s)}_{b,\delta}}}{\left(\frac{y_{a,\gamma}}{y_{b,\delta}};q\right)_{k^{(s)}_{a,\gamma}-k^{(s)}_{b,\delta}}}\prod_{s=b}^{L-1}\frac{\left(t^{-1}\frac{y_{b,\delta}}{y_{a,\gamma}};q\right)_{k^{(s)}_{b,\delta}-k^{(s)}_{a,\gamma}}}{\left(\frac{y_{b,\delta}}{y_{a,\gamma}};q\right)_{k^{(s)}_{b,\delta}-k^{(s)}_{a,\gamma}}}\\ &\times\prod_{\substack{s=b-1\text{ if }a<b\\ s=b\text{ if }a=b}}^{N-1}\frac{\left(tq\frac{y_{a,\gamma}}{y_{b,\delta}};q\right)_{k^{(s)}_{a,\gamma}-k^{(s+1)}_{b,\delta}}}{\left(q\frac{y_{a,\gamma}}{y_{b,\delta}};q\right)_{k^{(s)}_{a,\gamma}-k^{(s+1)}_{b,\delta}}}\prod_{s=b}^{L-1}\frac{\left(tq\frac{y_{b,\delta}}{y_{a,\gamma}};q\right)_{k^{(s)}_{b,\delta}-k^{(s+1)}_{a,\gamma}}}{\left(q\frac{y_{b,\delta}}{y_{a,\gamma}};q\right)_{k^{(s)}_{b,\delta}-k^{(s+1)}_{a,\gamma}}}\\ &\quad=\prod_{s=b}^{L-1}\frac{\left(t\frac{y_{b,\delta}}{y_{a,\gamma}};q^{-1}\right)_{k^{(s)}_{a,\gamma}-k^{(s)}_{b,\delta}}}{\left(\frac{y_{b,\delta}}{y_{a,\gamma}};q^{-1}\right)_{k^{(s)}_{a,\gamma}-k^{(s)}_{b,\delta}}}\prod_{s=b}^{L-1}\frac{\left(t\frac{y_{a,\gamma}}{y_{b,\delta}};q\right)_{k^{(s)}_{b,\delta}-k^{(s)}_{a,\gamma}}}{\left(\frac{y_{a,\gamma}}{y_{b,\delta}};q^{-1}\right)_{k^{(s)}_{b,\delta}-k^{(s)}_{a,\gamma}}}\\ &\quad\times\prod_{\substack{s=b-1\text{ if }a<b\\ s=b\text{ if }a=b}}^{N-1}\frac{\left(tq\frac{y_{a,\gamma}}{y_{b,\delta}};q\right)_{k^{(s)}_{a,\gamma}-k^{(s+1)}_{b,\delta}}}{\left(q\frac{y_{a,\gamma}}{y_{b,\delta}};q\right)_{k^{(s)}_{a,\gamma}-k^{(s+1)}_{b,\delta}}}\prod_{s=b}^{L-1}\frac{\left(tq\frac{y_{b,\delta}}{y_{a,\gamma}};q\right)_{k^{(s)}_{b,\delta}-k^{(s+1)}_{a,\gamma}}}{\left(q\frac{y_{b,\delta}}{y_{a,\gamma}};q\right)_{k^{(s)}_{b,\delta}-k^{(s+1)}_{a,\gamma}}}\\ &\quad\equiv\boxed{1}\boxed{2}\boxed{3}\boxed{4}\end{aligned} \tag{B.7}$$

where we have used the identity $(a;z)_n=(a^{-1},z^{-1})_n(-a)^n z^{\binom{n}{2}}$. Note that then we have cancellations between $\boxed{1}$ and $\boxed{4}$ and we claim that the remaining terms after cancellations gives precisely $\boxed{B}$ and $\boxed{C}$ in (B.5). We make the same claim for $\boxed{2}$, $\boxed{3}$ and $\boxed{A}$,$\boxed{D}$.

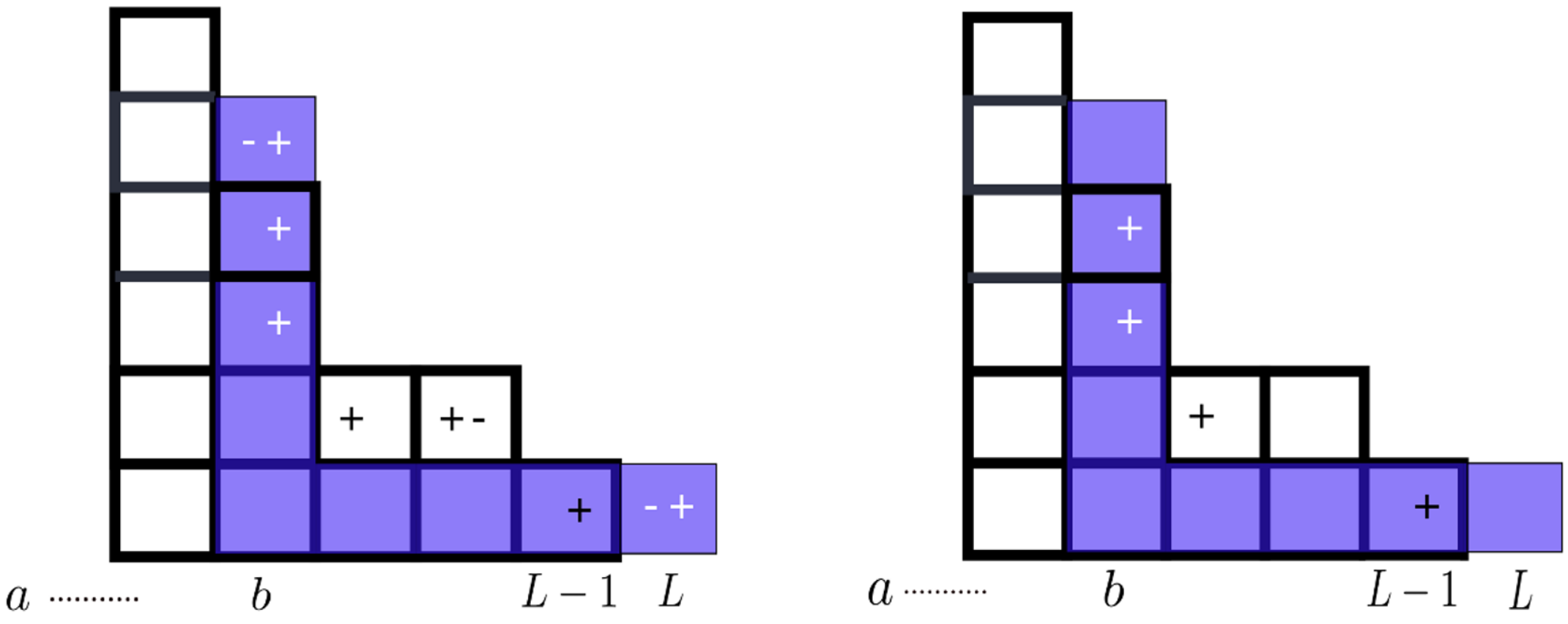

**Figure 4**. Example of Young tableaux $Y_{a,\gamma}$ and $Y_{b,\delta}$.

We may visualise the way the cancellations work in the following way. First draw the Young tableaux $\{Y\}$ in the $xy$ plane, with appropriate shifts in the $x$ direction as described above. At each $s = b, b+1, \ldots, L-1$, evaluate $k^{(s)}_{a,\gamma} - k^{(s)}_{b,\delta}$, corresponding to terms in $\boxed{1}$. If it is 0, we do nothing. If it is positive, draw a '+' in the $k^{(s)}_{a,\gamma} - k^{(s)}_{b,\delta}$ boxes in the $s$-th column in $Y_{a,\gamma}$ above $Y_{b,\delta}$. After cancellation from terms in $\boxed{4}$, these represent boxes $p$ in $Y_{a,\gamma}$ with leg length $-1$ relative to $Y_{b,\delta}$. If it is negative, place a '-' in the $|k^{(s)}_{a,\gamma} - k^{(s)}_{b,\delta}|$ boxes in $Y_{b,\delta}$ above $Y_{a,\gamma}$. The purpose of these '-' is to cancel boxes marked '+' when we consider the terms $\boxed{4}$ corresponding to boxes in $Y_{b,\delta}$ which do not in fact have leg length 0 with respect to $Y_{a,\gamma}$ and should not contribute to (B.5).

Next for each $s$ we evaluate $k^{(s)}_{b,\delta} - k^{(s+1)}_{a,\gamma}$ corresponding to terms in $\boxed{4}$. If it is 0 do nothing. If positive place a mark '+' in the top $k^{(s)}_{b,\delta} - k^{(s+1)}_{a,\gamma}$ boxes in the $s$-th column of $Y_{b,\delta}$. After cancellation from the '-' markings from $Y_{a,\gamma}$ described above, these represent boxes in $Y_{b,\delta}$ which may have leg length 0 with respect to $Y_{a,\gamma}$. If $k^{(s)}_{b,\delta} - k^{(s+1)}_{a,\gamma} < 0$, place a '-' in the top $|k^{(s)}_{b,\delta} - k^{(s+1)}_{a,\gamma}|$ boxes in column $s+1$ of $Y_{a,\gamma}$.

Now for each box in each diagram, if there is both a $+$ and a $-$ marked there, remove both. Note there can be at most one '+' and one '-' in each box. In the aforementioned way, the only boxes remaining will be those marked '+' and it is then easy to see these remaining boxes correspond to the remaining terms in the holomorphic block expression after cancellation. Note that a box cannot have only a '-' marked in it, as all Young tableau have decreasing or constant column height with increasing $s$. The boxes marked in $Y_{a,\gamma}$ correspond to those with length -1 with respect to $Y_{b,\delta}$ and thus terms $\boxed{B}$. Those in $Y_{b,\delta}$ have leg length 0 with respect to $Y_{a,\gamma}$, corresponding to $\boxed{C}$. It is clear also from the definition of the q-Pochhammer Symbols, that remaining '+' marked boxes will contribute consistently with their arm length as in (B.5).

$\square$

See figure 4 for an example illustrating this procedure. $Y_{a,\gamma}$ is drawn plain with bold outline, $Y_{b,\delta}$ shaded. They have been drawn with markings made initially, and then after cancellation. The markings on the left in each box represent those made by considering $\boxed{1}$, those on the right $\boxed{4}$. Markings in black represent those in $Y_{a,\gamma}$, white those in $Y_{b,\delta}$.